\documentclass[12pt]{article}

\pdfoutput=1

\usepackage{graphicx} % Required for inserting images

\usepackage{amsfonts}
\usepackage{amsmath}
\usepackage{epsfig}
\usepackage{latexsym}
\usepackage{graphicx}
\usepackage{color}
\usepackage[dvipsnames]{xcolor}
\usepackage{amssymb}
\usepackage{epstopdf,mathtools,braket}
\numberwithin{equation}{section}
\usepackage{tikz}
\usetikzlibrary{intersections, calc}
\usetikzlibrary {arrows.meta,bending,positioning}
\usetikzlibrary{decorations.pathreplacing,decorations.markings}
\usepackage[hang,small,bf]{caption}
\usepackage[subrefformat=parens]{subcaption}
\allowdisplaybreaks
\usepackage{afterpage}
\usepackage{ulem}
\usepackage{slashed}
\usepackage{url}
\usepackage{bm}
\usepackage{cancel}
\usepackage{hyperref}
\usepackage{titling}
\hypersetup{colorlinks,linkcolor=blue,citecolor=blue ,urlcolor=blue!80}
\usepackage{cite}
\usepackage{comment}
\usepackage{titling}
\usepackage{tikz}
\usetikzlibrary{intersections, calc,positioning,decorations.pathreplacing,decorations.pathmorphing}
\tikzset{>=latex}

\usepackage{comment}

\usepackage{centernot}
\usepackage{mathtools}
\usepackage{stmaryrd}

\def\i{\text{i}}

\def\be{\begin{equation}}
\def\ee{\end{equation}}
\def\bea{\begin{eqnarray}}
\def\eea{\end{eqnarray}}
\def\bequ{\begin{equation}}
\def\eequ{\end{equation}}

\def\Tr{\mbox{Tr}}

\def\i{\text{i}}

\def\Tr{\mathrm{Tr}}

\usepackage{tikz}
\usetikzlibrary{quantikz2}%For quantum circuit
\usetikzlibrary{calc}
\usetikzlibrary{decorations.pathreplacing}
\usetikzlibrary{intersections, calc,positioning,decorations.pathreplacing,decorations.pathmorphing}
\tikzset{>=latex}
\def\i{{\rm i}}

\def\CO{{\cal O}}

\def\BZ{\mathbb{Z}}

\def\b0{\bm{0}_\perp}

\def\L{\mathrm L}

\usepackage{centernot}
\usepackage{mathtools}
\usepackage{stmaryrd}

\usepackage{simplewick}

\usepackage{framed}
 {\endMakeFramed}

\definecolor{Handai-blue}{RGB}{32, 75, 146}

\usepackage{authblk}

\title{
Quantum Computing for Entanglement Asymmetry and the Quantum Mpemba Effect in Field Theory
}

\author[1,2,3,4]{Keisuke Fujii\thanks{Email: fujii@qc.ee.es.osaka-u.ac.jp}}
\author[5]{Harunobu Fujimura\thanks{Email: fujimura@het.phys.sci.osaka-u.ac.jp}}
\author[6,7]{Masazumi Honda\thanks{Email: masazumi.honda@riken.jp}}
\author[6,7]{Duc~Truyen~Le\thanks{Email: ductruyen.le@riken.jp}}

\affil[1]{Graduate School of Engineering Science, The University of Osaka, 1-3 Machikaneyama, Toyonaka, Osaka 560-8531, Japan.}
\affil[2]{Center for Quantum Information and Quantum Biology, The University of Osaka, 1-2 Machikaneyama, Toyonaka, 560-0043, Japan.}
\affil[3]{Graduate School of Informatics, Kyoto University, Sakyo-ku, Kyoto, 606-8501, Japan}
\affil[4]{Center for Quantum Computing, RIKEN, Wako Saitama 351-0198, Japan.}

\affil[5]{Department of Physics, The University of Osaka, Machikaneyama-Cho 1-1, Toyonaka 560-0043, Japan}

\affil[6]{RIKEN Center for Interdisciplinary Theoretical and Mathematical Sciences (iTHEMS), RIKEN, 2-1 Hirosawa, Wako, Saitama 351-0198 Japan}
\affil[7]{Graduate School of Science and Engineering, Saitama University, 255 Shimo-Okubo, Sakura-ku, Saitama 338-8570, Japan}

\begin{document}

\AddToHookNext{shipout/foreground}{%
  \begin{tikzpicture}[remember picture,overlay]
    \node[anchor=north east]
      at ([xshift=-1.5cm,yshift=-1.0cm]current page.north east)
      {OU-HET-1329, RIKEN-iTHEMS-Report-26, STUPP-26-306};
  \end{tikzpicture}%
}

\maketitle

% \KF{Title:\ "A Quantum algorithm for Quantum Mpemba Effect"\\ Subtitle: "The case study of Schwinger model"}\\
% \HF{"A Quantum Algorithm for Quantum Mpemba effect in Quantum Field Theories."}\\
% \HF{"A Scalable Quantum Algorithm for Quantum Mpemba effect."}

\begin{abstract}
We construct a quantum measurement protocol for estimating the 
%second R'enyi
R\'enyi entanglement asymmetry, which quantifies symmetry breaking at the level of a subsystem and provides a probe of the quantum Mpemba effect. The protocol combines the SWAP test with a charge-resolving procedure based on randomized symmetry rotations. For a fixed additive accuracy $\epsilon$, the required number of measurements scales as $\mathcal{O}(\epsilon^{-2})$ without explicit dependence on the total system size, making the protocol particularly attractive for quantum many-body systems and lattice quantum field theories. As a representative application, we study the quantum Mpemba effect in the lattice Schwinger model with a $\theta$ term using classical state-vector simulations. We further estimate the resources required to implement the protocol on noisy intermediate-scale quantum (NISQ) devices, fault-tolerant quantum computers (FTQCs), and a representative early-FTQC architecture. Our results provide a measurement-efficient framework for investigating anomalous symmetry-restoration dynamics in quantum many-body systems and quantum field theories.
\end{abstract}
\thispagestyle{empty}

\newpage
\pagenumbering{arabic}
\setcounter{page}{1}
\tableofcontents
\newpage

%%%%%%%%%%%%%%%%%%%%%%%%%%%%%%%%%%%%%
%%%%%%%%%%%%%%%%%%%%%%%%%%%%%%%%%%%%%
%%%%%%%%%%%%%%%%%%%%%%%%%%%%%%%%%%%%%
\section{Introduction}
\label{sec:introduction}
%%%%%%%%%%%%%%%%%%%%%%%%%%%%%%%%%%%%%
%%%%%%%%%%%%%%%%%%%%%%%%%%%%%%%%%%%%%
%%%%%%%%%%%%%%%%%%%%%%%%%%%%%%%%%%%%%

Nonequilibrium systems can exhibit relaxation phenomena that strongly depart from naive intuition. A celebrated example is the {\it Mpemba effect}, in which a system prepared farther from equilibrium can relax faster than one prepared closer to equilibrium. In its original context, this phenomenon refers to the observation that initially hotter water can freeze faster than initially colder water under certain conditions. 
%Its experimental report was given by  by Mpemba and Osborne \cite{Mpemba_1969}, and the phenomenon has since been investigated both theoretically and experimentally \cite{jeng2005hotwaterfreezefaster,ZLu2017}.
This phenomenon was reported experimentally by Mpemba and Osborne \cite{Mpemba_1969} and has since been investigated both theoretically and experimentally \cite{jeng2005hotwaterfreezefaster,ZLu2017}.

Quantum analogs of the Mpemba effect have attracted considerable attention in recent years \cite{Nava_2019,Ares:2022koq}. In the setting relevant to this work, the quantum Mpemba effect is associated with the dynamical restoration of a symmetry at the level of a subsystem. One considers, for example, an initially symmetry-breaking state evolving under a Hamiltonian that preserves the corresponding global symmetry. A characteristic Mpemba-like behavior then occurs when a state with stronger initial symmetry breaking restores the symmetry faster than a state with weaker initial symmetry breaking. To quantify symmetry breaking on a subsystem, the {\it entanglement asymmetry} was introduced in Ref.~\cite{Ares:2022koq}. Let $\rho_A$ denote the reduced density matrix of a subsystem $A$, and let $\rho_{A,S}$ denote its symmetrized counterpart obtained by averaging over the symmetry transformation, or equivalently by removing coherences between different symmetry sectors. As reviewed in Section~\ref{sec:QC_of_EA}, the entanglement asymmetry is defined by the relative entropy
\begin{align}
    \Delta S_A
    :=
    \Tr_A\left[
    \rho_A
    \left(
    \log\rho_A-\log\rho_{A,S}
    \right)
    \right].
\end{align}
It is non-negative and vanishes when the reduced density matrix is invariant under the symmetry, thereby providing a natural measure of symmetry breaking at the level of a subsystem.

Entanglement asymmetry and the associated quantum Mpemba effect have subsequently been studied in a broad range of settings, including quantum many-body systems \cite{Ares:2023kcz,Ares:2024nkh,Ferro:2023sbn,Murciano:2023qrv,Angelides:2023noe,Rylands:2024fio,Chalas:2024wjz,Liu:2024uqf,Russotto:2024pqg,Bhore:2025nko,Ares:2025vjw,Yamashika:2025vpd,Gibbins:2025bdd,Yamashika:2024hpr,Yamashika:2024mut,Vescovo:2026okg}, high-energy theory \cite{Florio:2025xup,Benini:2025hbj,Lamas:2025eay,Fujimura:2025rnm,Benini:2025lav,Benini:2024xjv,Fossati:2024xtn,Fossati:2024ekt,Kusuki:2024gss,Chen:2023gql,Capizzi:2023yka,Qi:2025duh,Das:2026joo}, and quantum information \cite{Summer:2025wsa,Kusuki:2026end,Moroder:2024jjh,Liu:2024kzv,Turkeshi:2024juo,Ares:2025ljj,Chen:2024lxe, Yamashika:2026wno}. The quantum Mpemba effect has also been observed experimentally \cite{Zhang:2024juc,Joshi:2024sup,Xu:2025wml}, and possible applications have begun to be explored \cite{Westhoff:2025bnz,Yu:2025omg,Zhang:2026vqx}. For recent reviews, see Refs.~\cite{Ares:2025onj,Yu:2025vth}.

Despite this progress, a comprehensive understanding of anomalous symmetry-restoration dynamics remains elusive. In particular, studying the real-time evolution of entanglement asymmetry in quantum field theory is challenging, and comparatively few results on the quantum Mpemba effect in quantum field theories are currently available \cite{Benini:2024xjv,Fujimura:2025rnm,Qi:2025duh}. For critical systems, techniques from conformal field theory can provide analytical access to the dynamics of entanglement asymmetry \cite{Fujimura:2025rnm,Benini:2024xjv}. Extending such analyses to more general interacting quantum field theories, however, remains difficult.

Numerical simulation provides another route to studying nonequilibrium dynamics in quantum field theory. In conventional lattice field theory, the path-integral formulation combined with Monte Carlo sampling has been remarkably successful for equilibrium and Euclidean observables. Real-time evolution, however, leads to highly oscillatory complex weights and hence to a severe sign problem \cite{deForcrand:2009zkb,Gattringer:2016kco,Nagata:2021ugx}. As a consequence, standard importance-sampling Monte Carlo methods are not directly applicable to generic real-time dynamics, precisely the regime required for investigating the quantum Mpemba effect.

Quantum computing offers a conceptually different approach to this difficulty \cite{Bauer:2022hpo,Funcke:2023jbq,DiMeglio:2023nsa}. In the Hamiltonian formulation, real-time evolution is implemented directly as unitary dynamics and does not rely on sampling a positive probability distribution, thereby avoiding the conventional real-time sign problem. Quantum Mpemba dynamics in quantum many-body systems have already been investigated using trapped-ion and superconducting quantum processors \cite{Joshi:2024sup,Xu:2025wml}. Existing experiments have employed techniques such as quantum state tomography and randomized measurements to extract the entanglement asymmetry. For generic many-body states, however, the measurement cost of such approaches can grow rapidly, and in particular exponentially with the size of the subsystem for full state reconstruction or generic multi-copy quantities \cite{Huang:2020tih,Elben:2022jvo,Brydges:2019wut}. Developing measurement protocols with more favorable scaling is therefore important for extending studies of the quantum Mpemba effect to systems with many degrees of freedom.

In this work, we construct a quantum measurement protocol for estimating the second R\'enyi entanglement asymmetry by combining the SWAP test \cite{Nielsen_Chuang_2010} with a charge-resolving procedure based on randomized symmetry rotations. For a fixed additive accuracy $\epsilon$, the number of measurements required by our protocol scales as
$N_{\mathrm{shot}} = \mathcal{O}(\epsilon^{-2}),$
without explicit dependence on the total system size. The circuit resources required to implement the protocol and the real-time evolution still grow with the system size, but the favorable measurement complexity makes the protocol particularly attractive for quantum many-body systems and lattice quantum field theories.

As a representative application, we consider the Schwinger model with a $\theta$ term \cite{Schwinger:1962tn,Schwinger:1962tp,Coleman:1975pw,Coleman:1976uz,Manton:1985jm}, namely quantum electrodynamics in $(1+1)$ dimensions. Using classical state-vector simulations, we demonstrate the application of the protocol in a physically nontrivial quantum-field-theory setting by investigating the quantum Mpemba effect for both spin-product benchmark states and fermion-parity-preserving initial states. We then estimate the resources required to implement the protocol on noisy intermediate-scale quantum (NISQ) devices, fault-tolerant quantum computers (FTQCs), and a representative early-FTQC architecture. Taken together, these analyses connect the proposed measurement protocol to a concrete field-theory application and to quantitative resource estimates for quantum-hardware implementation.

This paper is organized as follows. In Section~\ref{sec:QC_of_EA}, we review the notion of entanglement asymmetry and construct a quantum measurement protocol for estimating the second R\'enyi entanglement asymmetry. In Section~\ref{sec:Schwinger_model}, we demonstrate the application of the protocol to the lattice Schwinger model and investigate the quantum Mpemba effect using classical state-vector simulations. In Section~\ref{sec:resource_estimation}, we estimate the quantum resources required to implement the protocol on noisy, early fault-tolerant, and fully fault-tolerant quantum computers. Finally, we summarize our results and discuss future directions in Section~\ref{sec:summary}.

%%%%%%%%%%%%%%%%%%%%%%%%%%%%%%%%%%%%%
%%%%%%%%%%%%%%%%%%%%%%%%%%%%%%%%%%%%%
%%%%%%%%%%%%%%%%%%%%%%%%%%%%%%%%%%%%%
\section{Quantum Measurement Protocol for Entanglement Asymmetry}
\label{sec:QC_of_EA}
%%%%%%%%%%%%%%%%%%%%%%%%%%%%%%%%%%%%%
%%%%%%%%%%%%%%%%%%%%%%%%%%%%%%%%%%%%%
%%%%%%%%%%%%%%%%%%%%%%%%%%%%%%%%%%%%%
In this section, we review the notion of entanglement asymmetry \cite{Ares:2022koq} and construct a quantum measurement protocol for estimating the R\'enyi entanglement asymmetry. Combining the SWAP test \cite{Nielsen_Chuang_2010} with a charge-resolving procedure based on randomized symmetry rotations \cite{Bandyopadhyay:2023gwz}, we show that the R\'enyi entanglement asymmetry can be estimated with favorable measurement complexity.

%%%%%%%%%%%%%%%%%%%%%%%%%%%%%%%%%%%%%
%%%%%%%%%%%%%%%%%%%%%%%%%%%%%%%%%%%%%
\subsection{Entanglement Asymmetry}
\label{sec:def EA}
%%%%%%%%%%%%%%%%%%%%%%%%%%%%%%%%%%%%%
%%%%%%%%%%%%%%%%%%%%%%%%%%%%%%%%%%%%%
\begin{figure}[t]
    \centering
    \begin{tikzpicture}%[scale=1,baseline={([yshift=-.5ex]current bounding box.center)}]
    %coordinates
    \coordinate (p0) at (0,0);
    \coordinate (p1) at (1,0);
    \coordinate (p2) at (2,0);
    \coordinate (p3) at (3,0);
    \coordinate (p4) at (4,0);
    \coordinate (p5) at (5,0);
    \coordinate (p6) at (6,0);
    \coordinate (p7) at (7,0);
    \coordinate (p8) at (8,0);
    %points
    \pgfmathsetmacro{\r}{0.1} %radius of circle
    \fill [red!80] (p0) circle [radius=\r];
    \fill [red!80] (p1) circle [radius=\r];
    \fill [red!80] (p2) circle [radius=\r];
    \fill [red!80] (p3) circle [radius=\r];
    \fill (p4) circle [radius=\r];
    \fill (p5) circle [radius=\r];
    \fill (p6) circle [radius=\r];
    \draw (p7) node {\scalebox{1.2}{$\cdots$}};
    \fill (p8) circle [radius=\r];
    %brace
    \draw [red!80, decorate, decoration={brace, amplitude = 10pt, raise = 8pt}, line width = 1pt] (p0) -- (p3);
    \draw[red!80] ($($(p0)!0.5!(p3)$)+(0,1)$) node {\scalebox{1.2}{$A$}};
    \draw [decorate, decoration={brace, amplitude = 10pt, raise = 8pt}, line width = 1pt] (p4) -- (p8);
    \draw ($($(p4)!0.5!(p8)$)+(0,1)$) node {\scalebox{1.2}{$B$}};
    \end{tikzpicture}
    \caption{A schematic illustration of a one-dimensional quantum system divided into two subsystems, $A$ and $B$. Each dot represents a quantum degree of freedom. The subsystem $A$ is indicated by red dots, while the subsystem $B$ is indicated by black dots.}
    \label{fig:subsystem}
\end{figure}
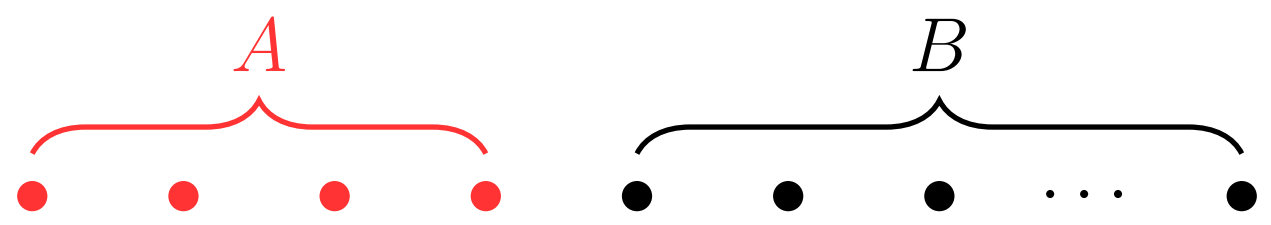

Let us consider a quantum system which can be divided into the two subsystems $A$ and $B$ such that the total Hilbert space $\mathcal{H}$ is factorized as $\mathcal{H}=\mathcal{H}_A \otimes \mathcal{H}_B$ as illustrated in Fig.~\ref{fig:subsystem}. 
Suppose that the system has a charge operator $Q$ which generates a $U(1)$ action\footnote{
The $U(1)$ action is typically taken to be a global $U(1)$ symmetry of the system but it does not have to be a symmetry of the system in general. Therefore, the discussion in Section~\ref{sec:def EA} holds even if the $U(1)$ action is not a symmetry while the lattice Schwinger model, our main example in Section~\ref{sec:Schwinger_model}, possesses a $U(1)$ symmetry. Also, although this paper focuses on the case of a $U(1)$ action, our formulation can be straightforwardly extended to general group action.  
} on the Hilbert space $\mathcal{H}$ and decomposes as
%is factorized as
\begin{align}
Q = Q_A\otimes \mathbf{1}_B + \mathbf{1}_A \otimes Q_B ,
\end{align}
where $Q_{A,B}$ acts on the subsystem Hilbert space $\mathcal{H}_{A,B}$.
%We assume that the theory possesses a global $U(1)$ symmetry, whose charge operator takes the form $Q = Q_A\otimes \mathbf{1}_B + \mathbf{1}_A \otimes Q_B$, where $Q_{A}$ and $Q_{B}$ act on the Hilbert space of the subsystem $A$ and $B$, respectively\footnote{Although we restrict our attention to a $U(1)$ symmetry, our formulation can be straightforwardly extended to other symmetry groups.}. 

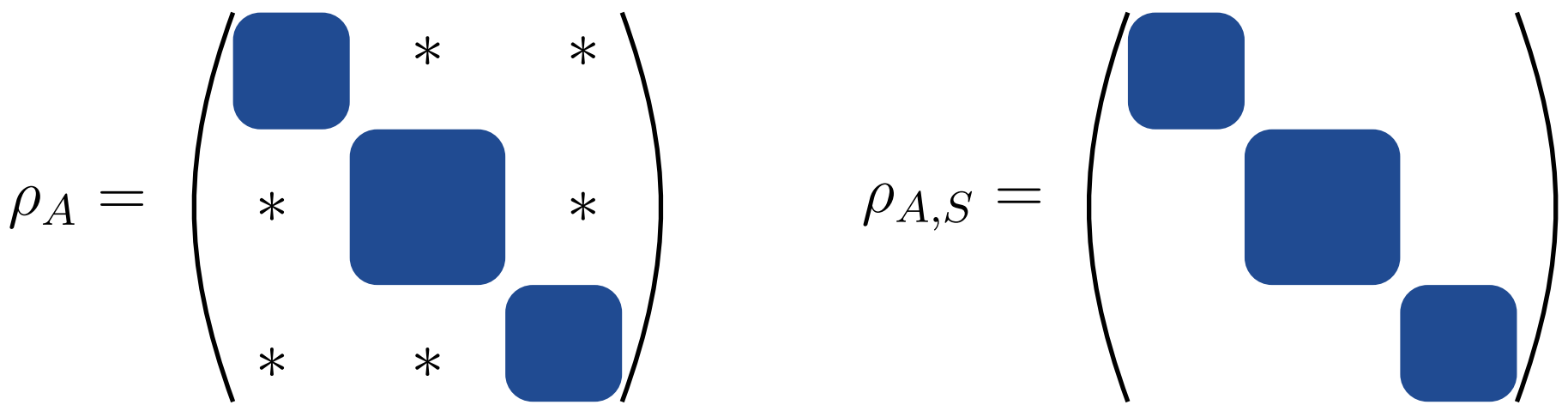
\begin{figure}[t]
        \centering
    \begin{minipage}[b]{0.4\columnwidth}
        \centering
        \begin{tikzpicture}
            %parameters
            \pgfmathsetmacro{\L}{3} %length of matrix
            %coordinates
            \coordinate (left_bottom) at (0,0); 
            \coordinate (left_top) at (0,\L);
            \coordinate (right_bottom) at (\L,0);
            \coordinate (right_top) at (\L,\L); %size of matrix
            \coordinate (d1) at (\L*0,\L); %block1
            \coordinate (d2) at (\L*0.3,\L*0.7);
            \coordinate (d3) at (\L*0.3,\L*0.7); %block2
            \coordinate (d4) at (\L*0.7,\L*0.3);
            \coordinate (d5) at (\L*0.7,\L*0.3); %block3
            \coordinate (d6) at (\L*1,\L*0);
            \coordinate (rho_A) at ($($(left_bottom)!0.5!(left_top)$)+(-\L*0.4,0)$);
            \coordinate (off1) at (\L*0.5,\L*0.9); %off-diagonal element
            \coordinate (off2) at (\L*0.9,\L*0.9); 
            \coordinate (off3) at (\L*0.9,\L*0.5); 
            \coordinate (off4) at (\L*0.1,\L*0.1); 
            \coordinate (off5) at (\L*0.1,\L*0.5); 
            \coordinate (off6) at (\L*0.5,\L*0.1); 
            %\fill (left_bottom) circle [radius=0.1] (left_top) circle [radius=0.1] (right_bottom) circle [radius=0.1] (right_top) circle [radius=0.1];
            %lines
            \draw[line width = 1pt] (left_bottom) to [out=110, in=250] (left_top);
            \draw[line width = 1pt] (right_bottom) to [out=70, in=290] (right_top);
            %blocks
            \fill[Handai-blue, rounded corners=6pt] (d1) rectangle (d2);
            \fill[Handai-blue, rounded corners=6pt] (d3) rectangle (d4);
            \fill[Handai-blue, rounded corners=6pt] (d5) rectangle (d6);
            % \fill[cyan!30, rounded corners=6pt] (d1) rectangle (d2);
            % \fill[cyan!30, rounded corners=6pt] (d3) rectangle (d4);
            % \fill[cyan!30, rounded corners=6pt] (d5) rectangle (d6);
            %nodes
            \draw (rho_A) node {\scalebox{1.2}{$\rho_A =$}};
            \draw (off1) node {\scalebox{1.2}{$*$}};
            \draw (off2) node {\scalebox{1.2}{$*$}};
            \draw (off3) node {\scalebox{1.2}{$*$}};
            \draw (off4) node {\scalebox{1.2}{$*$}};
            \draw (off5) node {\scalebox{1.2}{$*$}};
            \draw (off6) node {\scalebox{1.2}{$*$}};
        \end{tikzpicture}
    \end{minipage}
    \begin{minipage}[b]{0.4\columnwidth}
        \centering
        \begin{tikzpicture}
            %parameters
            \pgfmathsetmacro{\L}{3} %length of matrix
            %coordinates
            \coordinate (left_bottom) at (0,0); 
            \coordinate (left_top) at (0,\L);
            \coordinate (right_bottom) at (\L,0);
            \coordinate (right_top) at (\L,\L); %size of matrix
            \coordinate (d1) at (\L*0,\L); %block1
            \coordinate (d2) at (\L*0.3,\L*0.7);
            \coordinate (d3) at (\L*0.3,\L*0.7); %block2
            \coordinate (d4) at (\L*0.7,\L*0.3);
            \coordinate (d5) at (\L*0.7,\L*0.3); %block3
            \coordinate (d6) at (\L*1,\L*0);
            \coordinate (rho_AS) at ($($(left_bottom)!0.5!(left_top)$)+(-\L*0.45,0)$);
            %\fill (left_bottom) circle [radius=0.1] (left_top) circle [radius=0.1] (right_bottom) circle [radius=0.1] (right_top) circle [radius=0.1];
            %lines
            \draw[line width = 1pt] (left_bottom) to [out=110, in=250] (left_top);
            \draw[line width = 1pt] (right_bottom) to [out=70, in=290] (right_top);
            %blocks
            \fill[Handai-blue, rounded corners=6pt] (d1) rectangle (d2);
            \fill[Handai-blue, rounded corners=6pt] (d3) rectangle (d4);
            \fill[Handai-blue, rounded corners=6pt] (d5) rectangle (d6);
            % \fill[cyan!30, rounded corners=6pt] (d1) rectangle (d2);
            % \fill[cyan!30, rounded corners=6pt] (d3) rectangle (d4);
            % \fill[cyan!30, rounded corners=6pt] (d5) rectangle (d6);
            %nodes
            \draw (rho_AS) node {\scalebox{1.2}{$\rho_{A,S} =$}};
        \end{tikzpicture}
    \end{minipage}
    \caption{
    Schematic illustration of the reduced density matrix $\rho_A$ (left) and the symmetrized reduced density matrix $\rho_{A,S}$ (right), shown in the eigenbasis of symmetry charge $Q_A$. The shaded boxes 
    \begin{tikzpicture}[scale=1,baseline={([yshift=-.5ex]current bounding box.center)}]
        \fill[Handai-blue, rounded corners=4pt] (0,0) rectangle (0.5,0.5);
        % \fill[cyan!30, rounded corners=4pt] (0,0) rectangle (0.5,0.5);
    \end{tikzpicture}
    represent the block-diagonal elements and the symbol $*$ denotes nonzero off-diagonal elements. The presence of nonzero off-diagonal charge blocks
    %elements 
    indicates symmetry breaking in subsystem $A$. In contrast, $\rho_{A,S}$ is defined by eliminating the off-diagonal charge blocks
    %elements 
    and is therefore always block-diagonal.
    }
\label{fig:reduced_density_matrix}
\end{figure}
%%%%%%%%%%%%%%%%%%%%%%%%%%%%%%%%%%%%%

Now let us consider a state of the total system associated with the density matrix $\rho$ and ask whether or not the subsystem $A$ is symmetric under the $U(1)$ action.\footnote{
Below we often refer to the $U(1)$ action as ``symmetry" for simplicity of explanation while it is not necessarily a symmetry as noted in footnote 1.} 
%Given a dendity matrix $\rho$ of the total system, the reduced density matrix on $A$ is defined as 
To see this, we first introduce the reduced density matrix on $A$ as
\begin{align}
\rho_A := \Tr_B \left[ \rho \right] ,
%\rho_A := \Tr_B \left[ \ket{\Psi}\bra{\Psi} \right].
\end{align}
where $\Tr_B$ denotes the trace over $\mathcal{H}_B$.
%We focus on the symmetry of the subsytem $A$. 
Then a symmetry-preserved state and a symmetry-broken state on $A$ are defined as
%A symmetry-preserved state and a symmetry-broken state are defined by
\begin{align}\label{eq:def_symmetry}
    \left\{ \quad
    \begin{aligned}
        \left[\rho_A,  Q_A \right] 
        &=
        0\quad \text{(symmetry preserved)},\\
        \left[\rho_A,  Q_A\right] 
        &\neq
        0\quad \text{(symmetry broken)}
    \end{aligned}
    \right.\ .
\end{align}
When the symmetry is preserved on the subsystem $A$, the reduced density matrix $\rho_A$ commutes with the charge operator $Q_A$ and is therefore block-diagonal in the eigenbasis of charge $Q_A$. In contrast, when the symmetry is broken on the subsystem $A$, $\rho_A$ contains off-diagonal charge blocks,
%elements, 
as illustrated in the left panel of Fig.~\ref{fig:reduced_density_matrix}.\footnote{
The navy blue used in this paper is ``Handai Blue,'' the signature color of The University of Osaka.
}
To define the entanglement asymmetry, we introduce the symmetrized reduced density matrix $\rho_{A,S}$\cite{Ares:2022koq}, which is obtained by projecting $\rho_A$ onto the eigenspace of charge $Q_A$,
\begin{align}
    \rho_{A,S}
    &:=
    \sum_{q_A\in \BZ} \Pi_{q_A} \rho_A \Pi_{q_A}\ ,
\end{align}
where $\Pi_{q_A}$ denotes the projector onto the eigenspace of $Q_A$ with the eigenvalue $q_A$ and we take the normalization s.t.~$q_A\in \BZ$. 
%where $\Pi_{q_A}$ denotes the projector onto the eigenspace of $Q_A$ with eigenvalue $q_A\in \BZ$. 
Schematically, the symmetrized reduced density matrix $\rho_{A,S}$ is obtained by eliminating the off-diagonal charge blocks
%elements 
of reduced density matrix $\rho_A$, as illustrated in the right panel of Fig.~\ref{fig:reduced_density_matrix}. By construction, $\rho_{A,S}$ is always block-diagonal and commutes with $Q_A$. With this definition, the criterion for symmetry preservation or breaking in \eqref{eq:def_symmetry} can be equivalently expressed as
\begin{align}\label{eq:def_symmetry_2}
    \left\{ \quad
    \begin{aligned}
        \rho_A
        &=
        \rho_{A,S}\quad \text{(symmetry preserved)},\\
        \rho_A
        &\neq
        \rho_{A,S}\quad \text{(symmetry broken)}
    \end{aligned}
    \right.\ .
\end{align}
To quantify the degree of symmetry breaking on subsystem $A$, we introduce the entanglement asymmetry $\Delta S_{A}$ as the relative entropy between $\rho_A$ and $\rho_{A,S}$ \cite{Ares:2022koq},
\begin{align}
\begin{aligned}
    \Delta S_{A}
    &:=  \Tr_{A} \left[ \rho_A ( \log \rho_A  -\log \rho_{A,S}) \right]  \ .
\end{aligned}
\label{eq:EA_def}
\end{align}
The entanglement asymmetry satisfies the following properties\cite{Kullback:1951zyt}.
\begin{align}
    \left\{\ 
    \begin{aligned}
        \Delta S_{A} = 0\  &\iff \  \rho_A = \rho_{A,S} \iff \ \text{(symmetry preserved)}\\
        \Delta S_{A} > 0\  &\iff \  \rho_A \neq \rho_{A,S} \iff \ \text{(symmetry broken)}
    \end{aligned}
    \right.\ .
    \label{eq:EA_property}
\end{align}
Namely, the entanglement asymmetry vanishes for symmetry-preserved states and takes a positive value for symmetry-broken states. From this property, $\Delta S_{A}$ can be regarded as a quantitative measure of symmetry breaking at the level of the subsystem $A$. 

In general, analyzing the entanglement asymmetry is challenging due to the presence of the logarithm
%logarithm
of the reduced density matrix in its definition. It is therefore convenient to introduce the R\'enyi entanglement asymmetry,
\begin{align}
    \Delta S_{A}^{(n)}
    :=
    \frac{1}{1-n} 
    \left( 
    \log 
    \Tr_{A} \left[ \rho_{A,S}^n \right]
    -
    \log 
    \Tr_{A} \left[ \rho_{A}^n \right]
    \right)\ ,
    \label{eq:REA_def}
\end{align}
where $n$ is a positive integer. The entanglement asymmetry defined in \eqref{eq:EA_def} is recovered by taking the limit $n \to 1$ of the R\'enyi entanglement asymmetry,
\begin{align}
    \lim_{n\to1} \Delta S_{A}^{(n)} 
    =
    \Delta S_{A}\ .
\end{align}

%%%%%%%%%%%%%%%%%%%%%%%%%%%%%%%%%%%%%%%%%%%%%%%%%%%%%
%%%%%%%%%%%%%%%%%%%%%%%%%%%%%%%%%%%%%%%%%%%%%%%%%%%%%
\subsection{SWAP Test with Charge-Resolving Procedure}
%%%%%%%%%%%%%%%%%%%%%%%%%%%%%%%%%%%%%%%%%%%%%%%%%%%%%
%%%%%%%%%%%%%%%%%%%%%%%%%%%%%%%%%%%%%%%%%%%%%%%%%%%%%
Recently, the R\'enyi entanglement asymmetry has been investigated on NISQ devices \cite{Joshi:2024sup,Xu:2025wml}, but existing experimental approaches based on tomography or randomized measurements can require a number of measurements that grows rapidly with the subsystem size. This scaling is significant when we consider a quantum system with a large number of degrees of freedom, such as quantum field theories. In this subsection, we describe a quantum measurement protocol for estimating the R\'enyi entanglement asymmetry by combining the SWAP test with a charge-resolving procedure based on randomized symmetry rotations. For simplicity, we focus on the second R\'enyi entanglement asymmetry $(n = 2)$.\footnote{
This restriction is made only for simplicity. Our protocol can be straightforwardly extended to higher-order R\'enyi entanglement asymmetries $(n = 3,4, \ldots)$.
}
\begin{align}\label{eq:2nd_REA}
    \Delta S_{A}^{(2)}
    &=
    \log  \Tr_A
    \left[
        \rho_A^2
    \right]
    -
    \log  \Tr_A
    \left[
        \rho_{A,S}^2
    \right]\ ,
\end{align}
where the reduced density matrix $\rho_A $ is defined by
\begin{align}
    \rho_A = \Tr_B \left[ \ket{\Psi} \bra{\Psi}\right],  \qquad (\ket{\Psi}\in\mathcal{H}).
\end{align}
In the following, we explain how to compute $\Tr_A \left[ \rho_A^2 \right]$ and $\Tr_A \left[ \rho_{A,S}^2 \right]$ separately.

%%%%%%%%%%%%%%%%%%%%%%%%%%%%%%%%%%%%%%%%%%%%%
\subsubsection{Computation of $\Tr_A \left[\rho_A^2 \right]$}
%%%%%%%%%%%%%%%%%%%%%%%%%%%%%%%%%%%%%%%%%%%%%
\begin{figure}[t]
    \centering
    \begin{quantikz}[row sep=0.4cm, column sep=0.4cm]
    \lstick{\text{ancilla}\ $\ket{0}$\ } &\gate{H}     & \ctrl{2} & \gate{H} & \meter{} \\
    \lstick[wires=2]{$\ket{\Psi}$}
    \push{B\ }                           & \qwbundle{} &          & \qw      & \qw \\
    \push{\textcolor{red!80}{A}\ }       & \qwbundle{}   & \swap{2} & \qw      & \qw \\
    \lstick[wires=2]{$\ket{\Psi}$}
    \push{B\ }                           & \qwbundle{} &          & \qw      & \qw \\
    \push{\textcolor{red!80}{A}\ }       & \qwbundle{} & \targX{} & \qw      & \qw
    \end{quantikz}
    \caption{Quantum circuit for the SWAP test used to estimate the purity $\Tr_A \left[ \rho_A^2 \right]$. The quantum circuit consists of two copies of the system and one ancilla qubit, and applies the SWAP operator to subsystem $A$.}
    \label{fig:SWAP_test}
\end{figure}
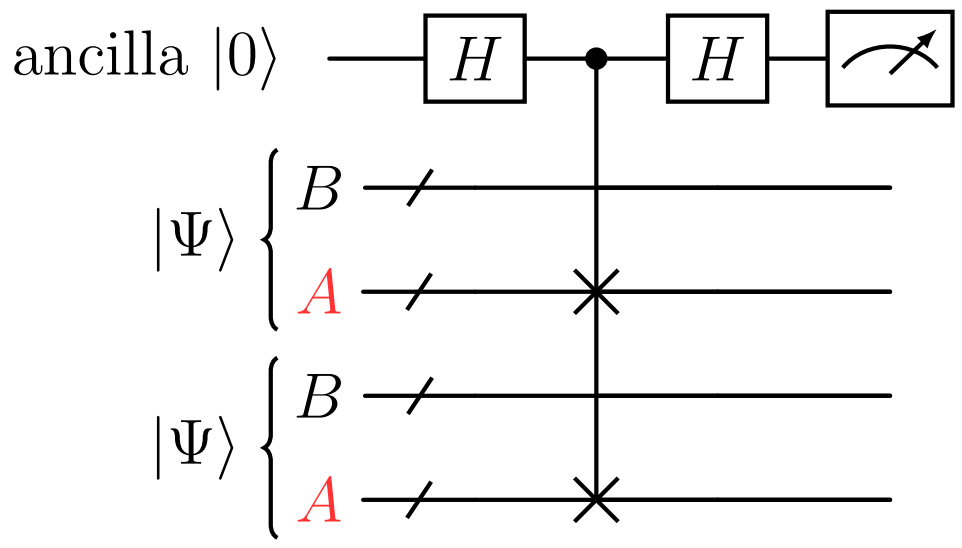
%%%
%The first term to be evaluated is the purity $\Tr_A \left[ \rho_A^2 \right]$, which can be expressed as
%The first term to be evaluated in \eqref{eq:2nd_REA} is the so-called purity $\Tr_A \left[ \rho_A^2 \right]$.
To evaluate the first term in \eqref{eq:2nd_REA}, we estimate the so-called purity $\Tr_A \left[ \rho_A^2 \right]$.
There are various known ways to compute the purity as well as the higher R\'enyi
%Renyi 
entropies \cite{Hastings:2010zka,Johri:2017fqx,Acharya:2017oqk,Brydges:2019wut,Subramanian:2021xxc}.
Below we employ the SWAP trick \cite{Hastings:2010zka,Johri:2017fqx} although other purity-estimation methods could also be used.
It is known that the purity can be expressed as 
\begin{align}\label{eq:purity}
    \Tr_A
    \left[
        \rho_A^2
    \right]
    &=
    \bra{\Psi} \otimes \bra{\Psi}\  \text{SWAP}_A\  \ket{\Psi} \otimes \ket{\Psi}\ ,
\end{align}
where the operator $\text{SWAP}_A $ on $\mathcal{H} \otimes \mathcal{H}$ denotes the SWAP operation between the two $\mathcal{H}_A$-parts of $\mathcal{H} \otimes \mathcal{H}$ defined as
%the action of $\text{SWAP}_A$ is defined by 
%$\text{SWAP}_A \ket{\psi_i}_A \otimes \ket{\psi_j}_A = \ket{\psi_j}_A \otimes \ket{\psi_i}_A $. 
\begin{align}
\text{SWAP}_A \ket{\psi_1}_A \otimes \ket{\phi_1}_B \otimes \ket{\psi_2}_A \otimes \ket{\phi_2}_B
= \ket{\psi_2}_A \otimes \ket{\phi_1}_B \otimes \ket{\psi_1}_A \otimes \ket{\phi_2}_B .
\end{align}
%In this expression, 
In the above expression, the purity is given by the expectation value of the SWAP operator acting on subsystem $A$ in the duplicated system. 
The purity can be estimated using the quantum circuit for the SWAP test shown in Fig.~\ref{fig:SWAP_test}. The purity $\Tr_A \left[ \rho_A^2 \right]$ is obtained from the expectation value of the Pauli-$Z$ operator measured on the ancilla qubit,
\begin{align}\label{eq:purity_ancilla}
    \Tr_A
    \left[
        \rho_A^2
    \right]
    &=
    \langle Z \rangle_{\text{ancilla}}\ ,
\end{align}
with a statistical error $\epsilon$ given by
\begin{align}
    \epsilon
    &:=
    \sqrt{\frac{\langle Z^2 \rangle_{\text{ancilla}} -\langle Z \rangle_{\text{ancilla}}^2 }{N_{\text{shot}}}} 
%    =
%    \sqrt{\frac{1 -\left( \Tr_A \left[ \rho_A^2 \right] \right)^2 }{N_{\text{shot}}}}     < \frac{1}{\sqrt{N_{\text{shot}}}}\ ,
\end{align}
where $N_{\text{shot}}$ denotes the number of measurements. 
From this formula, it is clear that the statistical error $\epsilon$ is bounded from above as
\begin{align}\label{eq:statistical_error1}
    \epsilon     \leq \frac{1}{\sqrt{N_{\text{shot}}}}\ .
\end{align}

%%%%%%%%%%%%%%%%%%%%%%%%%%%%%%%%%%%%%%%%%%%
\subsubsection{Computation of $\Tr_A \left[\rho_{A,S}^2 \right]$}
%%%%%%%%%%%%%%%%%%%%%%%%%%%%%%%%%%%%%%%%%%%
The second term in \eqref{eq:2nd_REA} involves the evaluation of the purity for $\rho_{A,S}$.
%$\Tr_A \left[ \rho_{A,S}^2 \right]$. 
Although the SWAP test can in principle be applied to the quantum state $\rho_{A,S}$, the main challenge is how to realize $\rho_{A,S}$ starting from the original reduced density matrix $\rho_A$. 
%In general, the number of off-diagonal elements in $\rho_A$ scales as $\CO \left( N_A^2 \right)$, where $N_A$ denotes the size of subsystem $A$. 
Classically, one may obtain $\rho_{A,S}$ by explicitly projecting $\rho_A$ onto the charge eigenspaces by naive matrix multiplication. 
However, this procedure requires a computational cost of $\CO \left( 2^{3N_A} \right)$ where $N_A$ denotes the size of subsystem $A$, and thus becomes inefficient for large systems. Therefore, it is useful to realize $\rho_{A,S}$ directly within a quantum circuit. 

For a finite subsystem, the symmetrized reduced density matrix $\rho_{A,S}$ can be written as
\begin{align}\label{eq:rho_A_S_projector}
    \rho_{A,S}
    &= \sum_{q_A =q_{\rm min} }^{q_{\rm max}} \Pi_{q_A} \rho_A \Pi_{q_A} ,
    %\qquad ({\rm for}\ N_A\ {\rm qubit\ system}) ,
\end{align}
where $q_{\rm min}$ and $q_{\rm max}$ are minimal and maximal charges of the subsystem $A$, respectively.
Importantly, one can rewrite this as\footnote{
This can be derived by noting $\rho_{A,S}=\int_{-\pi}^\pi \frac{d\alpha}{2\pi} e^{i\alpha Q_A}\rho_A e^{-i\alpha Q_A}$ and using the fact that $q_A$ is in the finite range $[q_{\rm min},q_{\rm max}]$.
} 
\begin{align}\label{eq:rho_A_S_U(1)}
    \rho_{A,S}
    &=
    \frac{1}{2^{n}}
    \sum_{x=0}^{2^n-1} 
    \left( U_A\right)^x \rho_A \left( U_A^{\dagger} \right)^x ,
%    \ \qquad ({\rm for}\ N_A\ {\rm qubit\ system}) ,
\end{align}
where $U_A$ is a unitary operator defined as
\begin{align}
    U_A=e^{2\pi \i \frac{ Q_A }{2^n} }\  ,
    \label{eq:charge_operator}
\end{align}
and $n$ is an integer satisfying\footnote{
Strictly speaking, using an integer $M> q_{\rm max} -q_{\rm min}$, $\rho_{A,S}$ can also be written as
\[
\rho_{A,S}=   \frac{1}{M}\sum_{x=0}^{M-1} 
    e^{2\pi \i \frac{x Q_A }{M} } \rho_A  e^{-2\pi \i \frac{x Q_A }{M} } .
\]  
Here we are taking $M=2^n$ for convenience.
Also, $q_{\rm max} -q_{\rm min}$ is typically $\mathcal{O}(N_A )$ and it suffices to take $n =\mathcal{O}(\log{N_A})$.
} 
\begin{align}
    n > \log_2{ (q_{\rm max} -q_{\rm min}) } .
\end{align}

\begin{figure}[t]
    \centering
    \begin{quantikz}%[row sep=0.2cm, column sep=0.2cm]
        \lstick{$\rho_A $} 
%        & \qwbundle{} & \gate{U_A^{x}}
        & \qwbundle{} & \gate{\left( U_A \right)^{x}} \gategroup[wires=1, steps=1, style={dashed, style=Handai-blue, rounded corners=5pt},label style={label position=below, style=Handai-blue, anchor=north, yshift=-0.2cm} ]{$x$: random integer} & &  \rstick{\textcolor{Handai-blue}{$\rho_{A,S} $}}
    \end{quantikz}
    \caption{
    Realization of the symmetrized state $\rho_{A,S}$ in a quantum circuit. We apply the unitary $\left( U_A\right)^x$ where $U_A$ is given by \eqref{eq:charge_operator} and an integer $x \in \{0,1,\cdots , 2^{n}-1\}$ is chosen uniformly at random for each experimental run.
    } 
    \label{fig:QC_for_rho_A_S}
\end{figure}
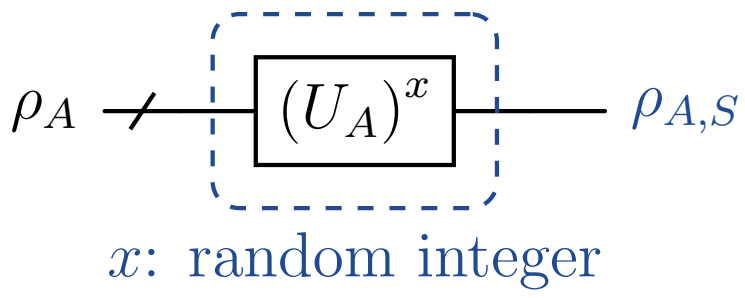

From \eqref{eq:rho_A_S_U(1)}, $\rho_{A,S}$ can be regarded as a classical mixture of the states $(U_A)^x \rho_A (U_A^\dagger)^x$ with equal probability. Figure~\ref{fig:QC_for_rho_A_S} shows a schematic implementation of this procedure. In each experimental run, an integer $x\in\{0,1,\ldots,2^n-1\}$ is chosen uniformly at random, and the unitary $(U_A)^x$ is applied to subsystem $A$.\footnote{
This construction can also be understood from the viewpoint of quantum phase estimation.
\begin{align}
    \begin{quantikz}[row sep=0.2cm, column sep=0.2cm]
        \lstick{\textcolor{red!80}{$\rho_A $}} 
        & \qwbundle{} &\qw &\qw \gategroup[wires=4, steps=8, style={dashed, style=Handai-blue, rounded corners=10pt},label style={label position=below, style=Handai-blue, anchor=north, yshift=-0.2cm} ]{Quantum phase estimation}  & \gate{U_A^{2^0}} & \ldots\ & \gate{U_A^{2^{n-1}}} & &  & &\qw &\qw &\qw 
        \rstick{\textcolor{red!80}{$\rho_{A,S} $}}\\
        \lstick[wires = 3]{$\ket{0}^{\otimes n}$}
            &\qw &\qw & \gate{H} &  \ctrl{-1} &  & & & \gate[wires=3]{\mathrm{QFT}^\dagger}& & \meter[wires=3]{}\\
            \push{\vdots} \\
            &\qw &\qw & \gate{H} & & & \ctrl{-3} & & & & 
    \end{quantikz}
    \nonumber
\end{align}
Coupling an ancilla register to powers of $U_A$ resolves the charge sectors and removes coherences between different charges when the ancilla degrees of freedom are discarded. Since only the resulting symmetrized state is required here, neither the inverse quantum Fourier transformation nor the measurement of the ancilla register is necessary. The randomized implementation in Fig.~\ref{fig:QC_for_rho_A_S} provides a simpler realization of the same symmetry-twirling channel.
}

%%%
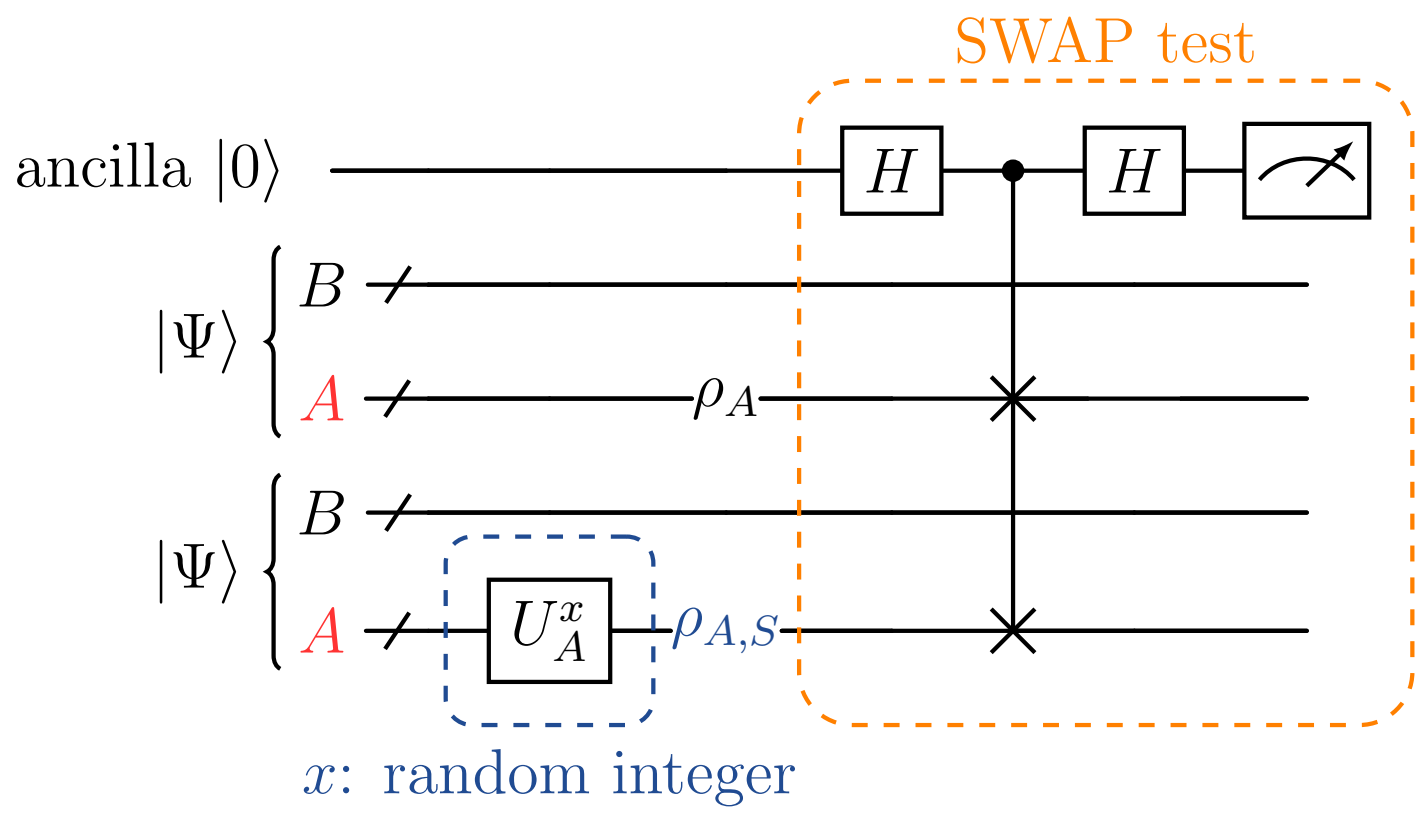
\begin{figure}[t]
    \centering
    \begin{quantikz}[row sep=0.3cm, column sep=0.4cm]
        \lstick{\text{ancilla}\ $\ket{0}$\ } &\qw &\qw &\qw &\gate{H}\gategroup[wires=5, steps=4, style={dashed, style=orange, rounded corners=10pt},label style={label, style=orange} ]{SWAP test} &\ctrl{2} &\gate{H} &\meter{} \\
        \lstick[wires=2]{$\ket{\Psi}$}
        \push{B\ } &\qwbundle{} &\qw &\qw &\qw &\qw &\qw &\qw \\
        \push{\textcolor{red!80}{A}\ } &\qwbundle{} &\qw &\push{\rho_{A}}\qw &\qw &\swap{2} &\hphantomgate{}\qw &\qw\\
        \lstick[wires=2]{$\ket{\Psi}$}
        \push{B\ } &\qwbundle{} &\qw &\qw &\qw &\qw &\qw &\qw \\
        \push{\textcolor{red!80}{A}\ } &\qwbundle{} &\gate{U_A^{x}}\gategroup[wires=1, steps=1, style={dashed, style=Handai-blue, rounded corners=5pt},label style={label position=below, style=Handai-blue, anchor=north, yshift=-0.2cm} ]{$x$: random integer} &\push{\textcolor{Handai-blue}{\rho_{A,S}}}\qw &\qw &\targX{} &\qw &\qw
    \end{quantikz}
    \caption{
    Quantum circuit for estimating $\Tr_A [ \rho_{A,S}^2 ]$, obtained by combining the quantum circuit in Fig.~\ref{fig:QC_for_rho_A_S} with the SWAP test. Using the identity $\Tr_A [ \rho_{A,S}^2 ] = \Tr_A [ \rho_{A} \rho_{A,S} ]$, it is sufficient to apply the unitary operator $U_A^{x}$ on subsystem $A$ of one of the two copies of the system.
    }
    \label{fig:QC_for_P_Q}
\end{figure}
%%%
We now combine the quantum circuit for preparing $\rho_{A,S}$ with the SWAP test. The resulting quantum circuit for estimating the purity $\Tr_A \left[ \rho_{A,S}^2 \right]$ is shown in  Fig.~\ref{fig:QC_for_P_Q}.
The purity is obtained from the expectation value of measurements on the ancilla qubit,
\begin{align} \label{eq:purity_Q_ancilla}
    \Tr_A
    \left[
        \rho_{A,S}^2
    \right]
    &=
    \langle Z \rangle_{S,\text{ancilla}} \ ,
%    \langle Z \rangle_{\text{ancilla}}'\ ,
\end{align}
where
%$\langle Z \rangle_{\text{ancilla}}'$ 
%$\langle Z \rangle_{S,\text{ancilla}}$ denotes the expectation value of the ancilla qubit,
$\langle \cdots \rangle_{S,\text{ancilla}}$ denotes the expectation value on the ancilla qubit, 
with $x \in \{ 0,1,\cdots 2^{n}-1 \}$ chosen randomly for each experimental run. 
The statistical error $\epsilon_S$
%$\epsilon'$ 
of this expectation value is analogous to that of the standard SWAP test,
\begin{align}
%    \epsilon'
    \epsilon_S
    &=
    \sqrt{\frac{1 -\left( \Tr_A \left[ \rho_{A,S}^2 \right] \right)^2}{N_{\text{shot}}}} .
%    < \frac{1}{\sqrt{N_{\text{shot}}}}\ ,
\end{align}
It is also clear that $\epsilon_S$ is bounded as
\begin{align}\label{eq:statistical_error2}
    \epsilon_S \leq \frac{1}{\sqrt{N_{\text{shot}}}}\ .
\end{align}

%%%%%%%%%%%%%%%%%%%%%%%%%%%%%%%%%%%%%%%%%%%%
%%%%%%%%%%%%%%%%%%%%%%%%%%%%%%%%%%%%%%%%%%%%
\subsection{Overview of Our Protocol}
\label{sec:Overview_of_protocol}
%%%%%%%%%%%%%%%%%%%%%%%%%%%%%%%%%%%%%%%%%%%%
%%%%%%%%%%%%%%%%%%%%%%%%%%%%%%%%%%%%%%%%%%%%
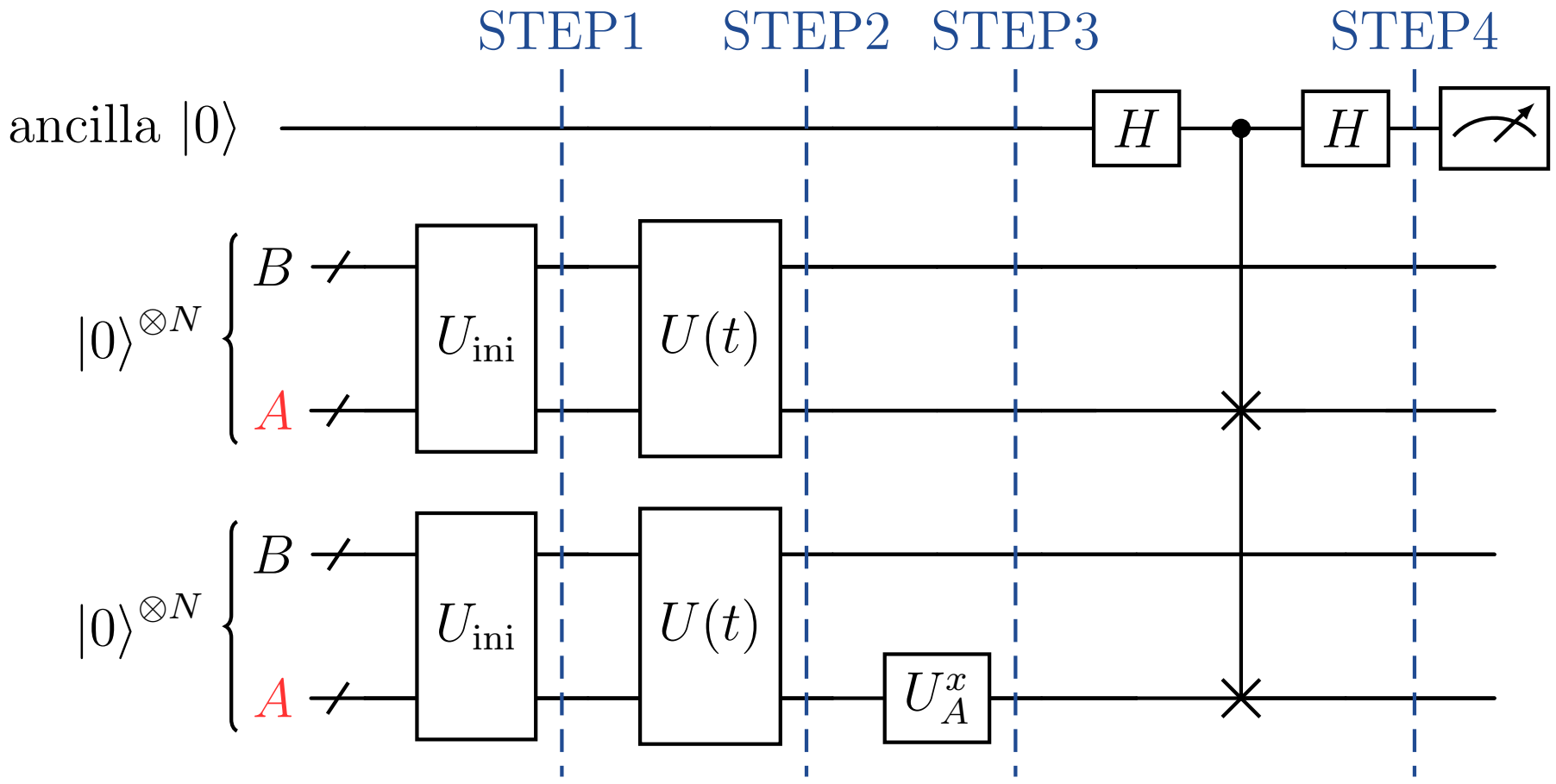
\begin{figure}[t]
    \centering
    \begin{quantikz}[row sep=0.4cm, column sep=0.4cm]
        \lstick{\text{ancilla}\ $\ket{0}$\ }
            &\qw &\qw &\qw  &\qw &\qw &\qw\slice[style=Handai-blue]{\textcolor{Handai-blue}{STEP3}} &\qw &\gate{H} &\ctrl{2} &\gate{H}\slice[style=Handai-blue]{\textcolor{Handai-blue}{STEP4}} &\meter{} \\
        \lstick[wires=2]{$\ket{0}^{\otimes N}$}
        \push{B\ }
            &\qwbundle{} &\gate[wires=2]{U_{\text{ini}}}\slice[style=Handai-blue]{\textcolor{Handai-blue}{STEP1}} &\qw  &\gate[wires=2]{U(t)}\slice[style=Handai-blue]{\textcolor{Handai-blue}{STEP2}}&\qw &\qw &\qw &\qw &\qw &\qw &\qw  \\
        \push{\textcolor{red!80}{A}\ }
            &\qwbundle{} &\qw &\qw &\qw &\qw &\qw &\qw &\hphantomgate{}\qw &\swap{2}&\hphantomgate{}\qw &\qw\\
        \lstick[wires=2]{$\ket{0}^{\otimes N}$}
        \push{B\ }
            &\qwbundle{} &\gate[wires=2]{U_{\text{ini}}} &\qw & \gate[wires=2]{U(t)}&\qw &\qw &\qw &\qw &\qw &\qw &\qw \\
        \push{\textcolor{red!80}{A}\ }
            &\qwbundle{} &\qw &\qw &\qw &\qw &\gate{U_A^{x}}&\qw &\qw &\targX{} &\qw &\qw
    \end{quantikz}
    \caption{
    Quantum circuit for investigating the quantum Mpemba effect. The symmetrized purity $\Tr_A\left[ \rho_{A,S}^2 \right]$ is obtained from the expectation value of measurements on the ancilla qubit. To evaluate the purity $\Tr_A\left[ \rho_{A}^2 \right]$, we perform the SWAP test while omitting STEP 3.
    }
    \label{fig:QC_Overview}
\end{figure}

In the previous subsection, we presented a quantum measurement protocol for estimating the second R\'enyi entanglement asymmetry. One of the key advantages of our approach is that real-time dynamics can be naturally implemented within a quantum circuit. In this subsection, we summarize our protocol and outline how it can be used to investigate the quantum Mpemba effect. The complete quantum circuit of the protocol is shown in Fig.~\ref{fig:QC_Overview}.
\begin{itemize}
    \item [\textbf{STEP~1:}]
        Prepare an initial state $\ket{\psi_{\text{ini}}} \otimes \ket{\psi_{\text{ini}}}$ that explicitly breaks the symmetry,
        \begin{align}
            \ket{\psi_{\text{ini}}}
            &=
            U_{\text{ini}}\ \ket{0}^{\otimes N}\ ,\nonumber
        \end{align}
        where $N$ is the number of qubits in the original system.
    \item [\textbf{STEP~2:}]
        Perform a real-time evolution on both duplicated systems,
        \begin{align}
            \ket{\psi(t) }
            &=
            U(t) \ket{\psi_{\text{ini}}}\ ,\ U(t) = e^{-\i H t}\nonumber
        \end{align}
        where $H$ is a symmetric Hamiltonian.
    \item [\textbf{STEP~3:}]
        Randomly choose an integer $x\in \{0,1,\cdots, 2^n-1\}$ and apply the unitary operator $U_A^{x}$ to subsystem $A$ of one of the two copies of the system.
    \item [\textbf{STEP~4:}]
        Apply the SWAP test circuit and measure the ancilla qubit.
\end{itemize}
By repeating STEPS 1--4 
%STEP 1--4 for 
$N_{\text{shot}}$ times, we obtain the expectation value $\langle Z \rangle_{S,\text{ancilla}}$. 
To obtain the expectation value $\langle Z \rangle_{\text{ancilla}}$ in \eqref{eq:purity_ancilla}, we omit STEP 3 and perform the same procedure. The second R\'enyi entanglement asymmetry is then estimated as
\begin{align}
    \Delta S_{A}^{(2)}
    &=
    \log \frac{\langle Z \rangle_{\text{ancilla}}}{\langle Z \rangle_{S,\text{ancilla}}} \ .
    %\ ,
%    \log \frac{\langle Z \rangle_{\text{ancilla}}}{\langle Z \rangle_{\text{ancilla}}'}\ ,
    \label{EA_meas}
\end{align}
The corresponding statistical errors of $\langle Z \rangle_{\text{ancilla}}$ and $\langle Z \rangle_{S,\text{ancilla}}$ are bounded
%given 
by \eqref{eq:statistical_error1} and \eqref{eq:statistical_error2}, respectively. Thus, estimating either purity to a fixed additive accuracy requires $N_{\text{shot}}=\mathcal{O}(\epsilon^{-2})$ measurements. Importantly, this measurement complexity has no explicit dependence on the total system size. Therefore, the protocol provides a favorable measurement strategy for studying the R\'enyi entanglement asymmetry in systems with a large number of degrees of freedom, including quantum field theories.

In practice, additional errors arise from the implementation of the time-evolution operator $U(t)$ using a finite sequence of elementary gates, namely the Trotterization error. The behavior of this error depends on the specific form of the Hamiltonian. In the next section, we consider the Schwinger model as a representative quantum field theory and demonstrate the application of the protocol using classical state-vector simulations.

%%%%%%%%%%%%%%%%%%%%%%%%%%%%%%%%%%%%
%%%%%%%%%%%%%%%%%%%%%%%%%%%%%%%%%%%%
%%%%%%%%%%%%%%%%%%%%%%%%%%%%%%%%%%%%
\section{Application to the Lattice Schwinger Model}
\label{sec:Schwinger_model}
%%%%%%%%%%%%%%%%%%%%%%%%%%%%%%%%%%%%
%%%%%%%%%%%%%%%%%%%%%%%%%%%%%%%%%%%%
In the previous section, we introduced a quantum measurement protocol for estimating the R\'enyi entanglement asymmetry and explained how it can be used to investigate the quantum Mpemba effect in quantum field theories. In this section, we demonstrate the application of the protocol in a concrete interacting quantum field theory, namely the Schwinger model \cite{Schwinger:1962tn,Schwinger:1962tp}. Using classical state-vector simulations, we investigate the real-time dynamics of the second R\'enyi entanglement asymmetry and examine the quantum Mpemba effect for several classes of symmetry-breaking initial states. This analysis provides a concrete example of the nonequilibrium field-theory dynamics that can be probed using the measurement protocol developed in the previous section.

%%%%%%%%%%%%%%%%%%%%%%%%%%%%%%%%%%%%
%%%%%%%%%%%%%%%%%%%%%%%%%%%%%%%%%%%%
\subsection{Lattice Schwinger Model as Qubits}
%%%%%%%%%%%%%%%%%%%%%%%%%%%%%%%%%%%%
%%%%%%%%%%%%%%%%%%%%%%%%%%%%%%%%%%%%
Here we review the lattice Schwinger model formulated in a suitable way for digital quantum simulation.
The Schwinger model is $(1+1)$-dimensional quantum electrodynamics with a single Dirac fermion (electron) $\psi$ coupled to a $U(1)$ gauge field $A_{\mu}$. 
%Its Lagrangian 
The Lagrangian of the continuum Schwinger model is given by
\begin{align}\label{eq:Lagrangian_Schwinger}
    \mathcal{L}
    &=
    - \frac{1}{4} F_{\mu \nu}F^{\mu \nu} 
    + \i \overline{\psi}\gamma^{\mu} \left( \partial_{\mu} + \i g A_{\mu} \right)\psi 
    - m \overline{\psi} e^{\i \theta \gamma^5} \psi\ ,
\end{align}
where $F_{\mu \nu} = \partial_{\mu}A_{\nu} - \partial_{\nu}A_{\mu}$ and we take the convention $\gamma^0 = \sigma^3$, $\gamma^1 = \i \sigma^2$, $\gamma^5 = \gamma^0 \gamma^1$. 
%where $\gamma^0 = \sigma^3$, $\gamma^1 = \i \sigma^2$, $\gamma^5 = \gamma^0 \gamma^1$, and $F_{\mu \nu} = \partial_{\mu}A_{\nu} - \partial_{\nu}A_{\mu}$. 
The parameters $g$, $m$, and $\theta$ denote the gauge coupling, the fermion mass, and the topological angle, respectively.\footnote{
The original definition of the topological angle $\theta$ is the coefficient of the topological $\theta$-term $\frac{g \theta}{4\pi} \epsilon_{\mu \nu} F^{\mu \nu}$ in the Lagrangian.
The Lagrangian \eqref{eq:Lagrangian_Schwinger} is obtained by making the chiral transformation $\psi \to e^{\i \frac{\theta}{2} \gamma^5}\psi$ to absorb the $\theta$-term via the transformation of the path integral measure \cite{Fujikawa:1979ay}.
The two Lagrangians are physically equivalent in the continuum theories on spacetime without boundaries but they are not completely equivalent when we put the theories on 
%lattice or/and spacetime
a lattice and/or on spacetime with boundaries.
Here we have chosen to work with \eqref{eq:Lagrangian_Schwinger} for our simulation.
}

In numerical simulations, we discretize the Schwinger model on a spatial lattice in order to regularize the infinite-dimensional degrees of freedom. 
Let us put the theory on a one-dimensional lattice with $N$ sites and lattice spacing $a$.
Adopting the temporal gauge $A_0 = 0$ and employing the Kogut-Susskind formulation \cite{Kogut:1974ag, Susskind:1976jm}, the lattice Hamiltonian with open boundary conditions is given by \cite{Hamer:1997dx, Chakraborty:2020uhf, Honda:2021aum}
%.
\begin{align}
\begin{aligned}\label{eq:Hamiltonian_KS}
    H
    &=  J \sum_{n=1}^{N-1} L_n^2
    -\i \sum_{n=1}^{N-1} 
    \left( w - (-1)^n\frac{m_{\text{lat}}}{2} \sin \theta \right) 
    \left[ \chi_n^{\dagger} U_n \chi_{n+1} - \chi_{n+1}^{\dagger} U_n^{\dagger} \chi_{n} \right] \\
    &~~~ +m_{\text{lat}} \cos \theta \sum_{n=1}^{N} (-1)^n \chi_n^{\dagger} \chi_n     
    \ , 
\end{aligned}
\end{align}
where the parameters are given by\footnote{
It was shown in \cite{Dempsey:2022nys} that one can 
%achieve a faster continuum limit 
improve convergence to the continuum limit by relating the mass parameters in the lattice and continuum theories without the $\theta$-term as $m_{\text{lat}} = m - \frac{g^2 a}{8}$. Here, we extend their result with the $\theta$-term. 
} 
%$w = \frac{1}{2a}$, $J=\frac{g^2 a}{2}$, and $m_{\text{lat}} = m - \frac{g^2 a}{8\cos \theta}$.
\begin{align}
w = \frac{1}{2a},\quad J=\frac{g^2 a}{2},\quad m_{\text{lat}} = m - \frac{g^2 a}{8\cos \theta} .
\end{align}
%Here, $a$ denotes the lattice spacing. 
The gauge operators are related to the continuum fields as $U_n \leftrightarrow e^{\i a g A^1(x)}$, $L_n \leftrightarrow \Pi(x) / g$, where $\Pi(x)$ denotes the conjugate momentum of $A^1$. The correspondence between the staggered fermion $\chi_n$ and the components of the Dirac fermion $\psi = \left( \psi_u(x), \psi_d (x) \right)^{\mathsf{T}}$ is given by
\begin{align}
    \frac{\chi_n}{\sqrt{a}}
    \ \longleftrightarrow \left\{\ 
    \begin{aligned}
        &\psi_u(x)\quad n: \text{even}\\
        &\psi_d(x)\quad n: \text{odd}
    \end{aligned}
    \right.\ .
\end{align}
Physical states must satisfy the lattice Gauss-law constraint,
\begin{align}\label{eq:Gauss_law}
    L_n - L_{n-1}
    &=
    \chi_n^{\dagger} \chi_n - \frac{1-(-1)^n}{2}\ .
\end{align}

The lattice Hamiltonian in \eqref{eq:Hamiltonian_KS} is expressed in terms of fermionic and gauge variables. To implement the quantum simulation on a qubit-based architecture, it is necessary to rewrite the lattice Schwinger model in terms of spin variables. To this end, we perform the Jordan--Wigner transformation \cite{Jordan:1928wi} to map the staggered fermions onto spin degrees of freedom,
\begin{align}\label{eq:Jordan_Wigner}
    \chi_n
    &=
    \left[ \prod_{\ell<n}\left(-\i Z_{\ell}\right) \right]\frac{X_n - \i Y_n}{2}\ ,
\end{align}
where $X_n$, $Y_n$, and $Z_n$ denote the Pauli operators acting on the $n$-th site. By utilizing the lattice Gauss-law constraint in \eqref{eq:Gauss_law} and imposing the open boundary condition $L_0 = 0$, 
%$L_0 = 0$ from the open boundary condition, 
the Hamiltonian can be rewritten solely in terms of spin variables \cite{Hamer:1997dx, Chakraborty:2020uhf, Honda:2021aum},
\begin{align}\label{eq:Hamiltonian_spin}
    H = H_{ZZ} + H_{Z} + H_{XY} \ ,
\end{align}
where
\begin{align}\label{eq:def_H_Z_H_ZZ_H_XY}
    \begin{aligned}
        H_{ZZ}
        &=
        \frac{J}{2} \sum_{1\le n < m \le N-1} (N-m) Z_n Z_{m} \ ,\\
        H_{Z}
        &=
        \frac{m_{\text{lat}}\cos \theta }{2}
        \sum_{n=1}^{N} (-1)^n Z_n
        - \frac{J}{2} \sum_{n=1}^{N-1} (n \text{ mod }2) \sum_{m = 1}^n Z_{m}\ ,\\
        H_{XY}
        &=
        \frac{1}{2} \sum_{n=1}^{N-1} 
        \left( w - (-1)^n \frac{m_{\text{lat}}}{2} \sin \theta \right)
        \left( X_n X_{n+1} + Y_n Y_{n+1} \right)\ .
    \end{aligned}
\end{align}
In this expression, we neglect constant terms that do not affect the quantum dynamics.    
%\footnote{Strictly speaking, we redefine the fermion as $\chi_n \to \left[\prod_{\ell<n} U_{\ell}\right] \chi_n$ to eliminate the gauge variables.}
The first term, $H_{ZZ}$, contains nonlocal interactions that arise from resolution of the Gauss-law constraint. 

%%%%%%%%%%%%%%%%%%%%%%%%%%%%%%%%%%%%
%%%%%%%%%%%%%%%%%%%%%%%%%%%%%%%%%%%%
\subsection{Setup of numerical simulation}
%%%%%%%%%%%%%%%%%%%%%%%%%%%%%%%%%%%%
%%%%%%%%%%%%%%%%%%%%%%%%%%%%%%%%%%%%
We now specify the simulation setup used to demonstrate the application of the measurement protocol introduced in Section~\ref{sec:QC_of_EA}. The purpose of the numerical analysis is not to simulate the SWAP-test circuit itself, but to evaluate the quantities that would be measured by the protocol during the real-time evolution of the lattice Schwinger model. We consider the Hamiltonian defined in \eqref{eq:Hamiltonian_spin} and \eqref{eq:def_H_Z_H_ZZ_H_XY}, with a total system of $N$ qubits and subsystem $A$ chosen as the leftmost $N_A$ qubits, forming a single contiguous interval, as illustrated in Fig.~\ref{fig:subsystem}. The Hamiltonian \eqref{eq:Hamiltonian_spin} has a $U(1)$ symmetry generated by the charge operator\footnote{
In the language of quantum field theory, this symmetry corresponds to a remnant global symmetry after gauge fixing.
}
\begin{align}\label{eq:charge_def}
    Q
    =
    \sum_{n=1}^N \frac{1-Z_n}{2}\ .
\end{align}
Since the summand gives zero on $\ket{0}_n$ and one on $\ket{1}_n$, the charge $Q$ counts the number of occupied spin sites in the computational basis.

To study the restoration of $U(1)$ symmetry, we consider two families of initial states as follows\footnote{
These states are gauge invariant since we have already solved the Gauss law at the operator level. 
Nevertheless, one might wonder whether or not the operators acting on $\ket{\Omega}$ are gauge invariant.
The apparent contradiction is resolved by distinguishing the operator representations before and after solving the Gauss law:
the operator representations in \eqref{eq:initial_state} are valid only after solving the Gauss law.
In the original gauge theory language, $Y_n$ in \eqref{eq:initial_state} corresponds to a gauge-invariant dressed operator with an electric-flux string extending to the right boundary.
This construction relies on leaving the right boundary flux unrestricted.
Note that this construction is specific to open boundary conditions and these states generally violate gauge invariance for periodic boundary conditions.
}
%\MH{I'll comment on gauge invarience of the states}
\begin{align}
\left\{\quad
\begin{aligned}\label{eq:initial_state}
    \ket{\psi_1(\phi)}
    &=
    e^{-\i \frac{\phi}{2}\sum_{n=1}^{N} Y_n } \ket{\Omega}\\
    \ket{\psi_2(\phi)}
    &=
    e^{-\i \frac{\phi}{2}\sum_{n=1}^{N-1} Y_n Y_{n+1} } \ket{\Omega}
\end{aligned}
\right.\ ,
\end{align}
where $\ket{\Omega}$ is some reference state and the parameter $\phi$ controls the degree of initial symmetry breaking.\footnote{
We assume that the reference state $\ket{\Omega}$ does not break the symmetry.
}
For the product reference states considered below,
%The 
the first family $\ket{\psi_1(\phi)}$ consists of a simple class of spin-product states and is useful for benchmarking, diagnosing the relaxation dynamics and developing a simple interpretation of the symmetry restoration. However, since the single-site rotation changes the fermion parity, these states do not satisfy the fermion-parity superselection rule and should not be regarded as physical fermionic states \cite{Wick:1952nb, Vidal:2021xmv}. On the other hand, the second 
%state 
family $\ket{\psi_2(\phi)}$ satisfies the fermion-parity superselection rule when the reference state is chosen to have a definite fermion parity because the generator of this rotation is even under fermion parity. This fermion-parity-preserving state therefore provides a physically admissible class of initial states for the fermionic model.

For the real-time evolution, we approximate the unitary operator $U(t) = e^{-\i H t}$ by a second-order Suzuki-Trotter decomposition\cite{Trotter1959, Suzuki1976}. For a single time step $\delta t$, we use
\begin{equation}\label{eq:Trotterization}
    U(\delta t) 
    =
    e^{-\i \frac{\delta t}{2} \left( H_{ZZ} + H_Z \right)}
    e^{-\i \frac{\delta t}{2} H_{\text{odd}} }
    e^{-\i \delta t H_{\text{even}} }
    e^{-\i \frac{\delta t}{2} H_{\text{odd}} }
    e^{-\i \frac{\delta t}{2} \left( H_{ZZ} + H_Z \right)}
    + \mathcal{O}\left( \delta t^3 \right),
\end{equation}
where $H_{ZZ}$ and $H_{Z}$ are defined in \eqref{eq:def_H_Z_H_ZZ_H_XY}. $H_{\text{even}}$ and $H_{\text{odd}}$ are given by
\begin{align}
    &H_{\text{even/odd}}
    = 
    \frac{1}{2} \sum_{n=\text{even/odd}} 
    \left(w - (-1)^n \frac{m_{\text{lat}}}{2} \sin \theta \right)
    \left( X_n X_{n+1} + Y_n Y_{n+1}\right) \ .
    %\ ,
    \label{eq:H_even_odd}
\end{align}
Each unitary operator appearing in \eqref{eq:Trotterization} can be decomposed into elementary gates in a straightforward way, as described in Appendix~\ref{sec:gate_decomposition}. In the following, we numerically simulate the time evolution of the second Rényi entanglement asymmetry using Qulacs \cite{Suzuki:2020nbf}, a Python-based quantum circuit simulator, and investigate the quantum Mpemba effect.\footnote{
In this work, we do not explicitly simulate the whole quantum circuit of the SWAP test itself, since the corresponding expectation values and statistical errors are analytically known. Instead, we directly compute the quantities appearing on the left-hand sides of \eqref{eq:purity_ancilla} and \eqref{eq:purity_Q_ancilla} by state vector simulation.
}

%%%%%%%%%%%%%%%%%%%%%%%%%%%%%%%%%%%%
%%%%%%%%%%%%%%%%%%%%%%%%%%%%%%%%%%%%
\subsection{Quantum Mpemba effect in spin-product initial states as a benchmark}
\begin{figure}
    \centering
    \includegraphics[width=0.5\linewidth]{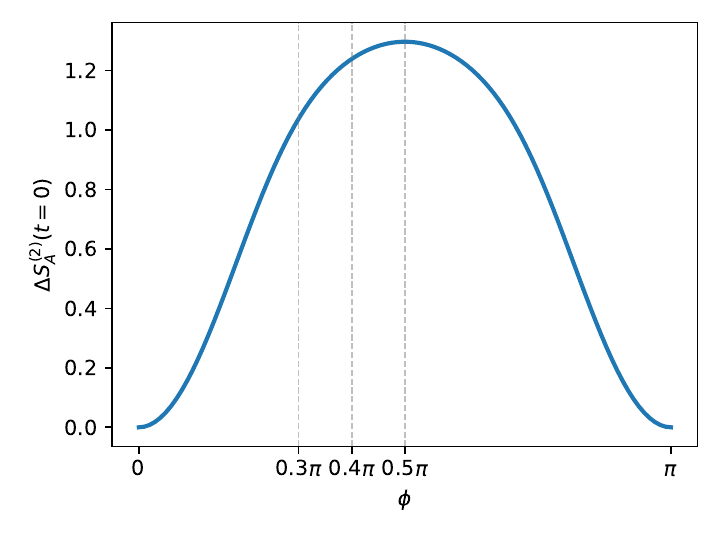}
    \caption{
    Initial second R\'enyi entanglement asymmetry $\Delta S_A^{(2)}(t=0)$ as a function of $\phi$ for the spin-product initial states \eqref{eq:initial_state_Ry}. The subsystem size is fixed to $N_A=4$. The asymmetry vanishes at $\phi=0$ and $\phi=\pi$ and reaches its maximum at $\phi=\pi/2$, reflecting the degree of symmetry breaking in the initial state. The dashed vertical lines indicate $\phi=0.3\pi$, $0.4\pi$, and $0.5\pi$, which are used in the time-evolution analysis.
    }
    \label{fig:EA_initial_Ry}
\end{figure}

%%%%%%%%%%%%%%%%%%%%%%%%%%%%%%%%%%%%
%%%%%%%%%%%%%%%%%%%%%%%%%%%%%%%%%%%%
We first consider spin-product initial states as benchmark examples for demonstrating the proposed measurement protocol and for developing an intuitive understanding of the symmetry-restoration dynamics. 
\begin{align}\label{eq:initial_state_Ry}
\left\{\,
\begin{aligned}
    \ket{\psi_1(\phi)}_{\mathrm F}
    &=
    e^{-\i \frac{\phi}{2}\sum_{n=1}^{N}Y_n}
    \ket{00\cdots 0},
    \\
    \ket{\psi_1(\phi)}_{\mathrm N}
    &=
    e^{-\i \frac{\phi}{2}\sum_{n=1}^{N}Y_n}
    \ket{0101\cdots 01}.
\end{aligned}
\right.
\end{align}
Because these states allow a simple characterization of the initial charge-sector distribution, they provide a useful setting in which to illustrate that the second R\'enyi entanglement asymmetry captures the expected quantum Mpemba behavior before turning to fermion-parity-preserving initial states. Here, the reference state is chosen to be the ferromagnetic state
$\ket{\Omega}=\ket{00\cdots 0}$ for
$\ket{\psi_1(\phi)}_{\mathrm F}$ and the N\'eel state
$\ket{\Omega}=\ket{0101\cdots 01}$ for
$\ket{\psi_1(\phi)}_{\mathrm N}$. For the present choices of the subsystem and the charge operator, the two states give the same initial second R\'enyi entanglement asymmetry as a function of $\phi$, as shown in Fig.~\ref{fig:EA_initial_Ry}. The angle $\phi$ controls the degree of initial symmetry breaking. In the following, we use
$\phi=0.3\pi$, $0.4\pi$, and $0.5\pi$
as representative values.

%%%%%%%%%%%%%%%%%%%%%%%%%%%%%%%%%%%%
\subsubsection{Ferromagnetic initial state}
%%%%%%%%%%%%%%%%%%%%%%%%%%%%%%%%%%%%

\begin{figure}
    \centering
    \includegraphics[width=0.5\linewidth]{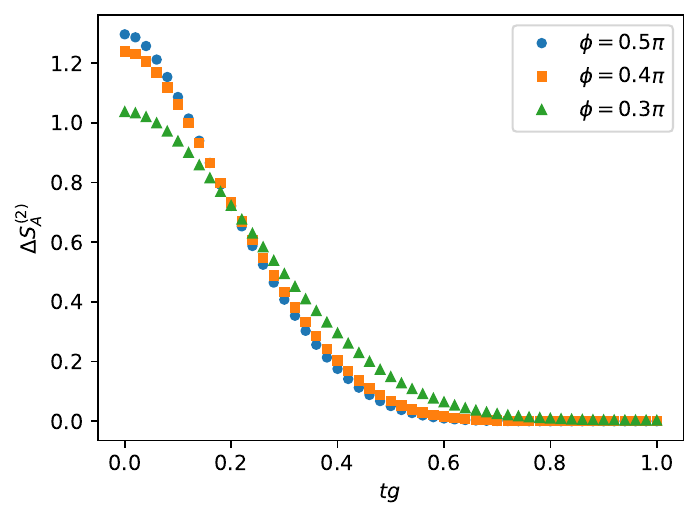}
    \caption{
    Time evolution of the second R\'enyi entanglement asymmetry $\Delta S_A^{(2)}(t)$ for the spin-product ferromagnetic initial state \eqref{eq:initial_state_Ry_ferro}. We compare the results for $\phi=0.3\pi$, $0.4\pi$, and $0.5\pi$. The parameters are fixed to $N=18$, $N_A=4$, $\theta=\pi/3$, $m/g=4$, and $ag=1/4$. For the Suzuki--Trotter decomposition, we use the time step $\delta t=0.02/g$.
    }
    \label{fig:QME_Ry_zero}
\end{figure}

We first investigate the quantum Mpemba effect for the spin-product ferromagnetic state:
\begin{align}\label{eq:initial_state_Ry_ferro}
    \ket{\psi_1(\phi)}_{\mathrm F}
    &=
    e^{-\i \frac{\phi}{2}\sum_{n=1}^{N}Y_n}
    \ket{00\cdots 0} \ .
    %,
\end{align}
In Fig.~\ref{fig:QME_Ry_zero}, we plot the time evolution of the second R\'enyi entanglement asymmetry for $\phi=0.3\pi$, $0.4\pi$, and $0.5\pi$. Although the three initial states have different nonzero values of the R\'enyi entanglement asymmetry, all three curves approach zero at sufficiently late times, indicating the restoration of the $U(1)$ symmetry. More importantly, among the three states considered here, the state with the largest initial R\'enyi entanglement asymmetry, $\phi=0.5\pi$, restores the symmetry faster than those with smaller initial asymmetries. The crossing of the curves therefore provides a signature of the quantum Mpemba effect.\footnote{
In the main text, we focus on the representative parameter set $\theta=\pi/3$, $m/g=4$, and $ag=1/4$ for the Schwinger model. We have confirmed qualitatively similar behavior of the second R\'enyi entanglement asymmetry for other parameter choices. Additional results for different values of $m/g$ and $\theta$ are presented in Appendix~\ref{sec:parameter_dependence}.
}
This benchmark demonstrates the application of our measurement protocol to quantum Mpemba dynamics in a lattice quantum field theory.

The initial distribution of the conserved charge provides a qualitative interpretation of this behavior \cite{Liu:2024kzv, Xiao:2026zia}. Since the Hamiltonian commutes with the total charge defined in \eqref{eq:charge_def}, each component of the initial state with a definite total charge remains in the same charge sector throughout the time evolution. In particular, neither the set of populated charge sectors nor the probability weight in each sector changes with time.

To make this structure explicit, we decompose the spin-product ferromagnetic state as
\begin{align}
    \ket{\psi_1(\phi)}_{\mathrm F}
    &=
    \prod_{n=1}^{N}
    \left( 
    \cos \frac{\phi}{2}\ket{0}_n + \sin \frac{\phi}{2}\ket{1}_n
    \right)\nonumber\\
    &=
    \sum_{q=0}^{N}
    \sqrt{p_q^{\mathrm F}(\phi)}
    \ket{\psi_q(\phi)}_{\mathrm F},
\label{eq:charge_sector_decomposition_Ferro}
\end{align}
where $\ket{\psi_q(\phi)}_{\mathrm F}$ is a normalized eigenstate with charge $q$ and $p_q^{\mathrm F}(\phi)$ is the probability weight of the charge-$q$ sector whose expression is given by
\begin{align}
    p_q^{\mathrm F}(\phi)
    &=
    \binom{N}{q}
    \cos^{2(N-q)}\left(\frac{\phi}{2}\right)
    \sin^{2q}\left(\frac{\phi}{2}\right).
\label{eq:weight_of_charge_sector_Ferro}
\end{align}
Correspondingly, the expectation value and variance of the total charge are obtained by
\begin{align}\label{eq:charge_distribution_Ferro}
\left\{\,
\begin{aligned}
    \langle Q \rangle_{\mathrm F}
    &=
    N\sin^2\left(\frac{\phi}{2}\right)\\
    (\Delta Q)^2_{\mathrm F}
    &=
    \frac{N}{4} \sin^2 \phi
\end{aligned}
\right.\ ,
\end{align}
where the variance is obtained by $(\Delta Q)^2_{\mathrm F} = \langle Q^2 \rangle_{\mathrm F} - \langle Q \rangle_{\mathrm F}^2$. Since the Hamiltonian has the $U(1)$ symmetry, the probability weight of each charge sector, expectation value, and variance are conserved under time evolution. 
In the range of $\phi\in [0,\pi/2]$, it is clear that both the expectation value and the variance increase with $\phi$. 

Because $[H,Q]=0$, the Hamiltonian and the total charge can be simultaneously diagonalized. We introduce a common eigenbasis satisfying
\begin{align}
    H\ket{q,\alpha}
    &=
    E_{q,\alpha}\ket{q,\alpha},
    &
    Q\ket{q,\alpha}
    &=
    q\ket{q,\alpha},
\end{align}
where $\alpha$ labels the energy eigenstates within the charge-$q$ sector. Each normalized charge-sector component can then be expanded as
\begin{align}
    \ket{\psi_q(\phi)}_{\mathrm F}
    &=
    \sum_{\alpha}
    c_{q,\alpha}^{\mathrm F}(\phi)\ket{q,\alpha},
    &
    c_{q,\alpha}^{\mathrm F}(\phi)
    &=
    \braket{q,\alpha|\psi_q(\phi)}_{\mathrm F},
\end{align}
with
\begin{align}
    \sum_{\alpha}
    \left|c_{q,\alpha}^{\mathrm F}(\phi)\right|^2
    =
    1.
\end{align}
The reduced density matrix therefore evolves as
\begin{align}
    \rho_A(t)
    &=
    \Tr_B\left[
    e^{-\i Ht}
    \ket{\psi_1(\phi)}_{\mathrm F}
    {}_{\mathrm F} \bra{\psi_1(\phi)}
    e^{\i Ht}
    \right]
    \nonumber\\
    &=
    \sum_{q,q'}
    \sqrt{
    p_q^{\mathrm F}(\phi)
    p_{q'}^{\mathrm F}(\phi)
    }
    \sum_{\alpha,\beta}
    c_{q,\alpha}^{\mathrm F}(\phi)
    c_{q',\beta}^{\mathrm F*}(\phi)
    e^{-\i(E_{q,\alpha}-E_{q',\beta})t}
    \Tr_B\left[
    \ket{q,\alpha}\bra{q',\beta}
    \right].
\label{eq:reduced_density_matrix_charge_resolved}
\end{align}

This expression shows that the reduced density matrix contains oscillatory contributions with frequencies determined by the energy differences $E_{q,\alpha}-E_{q',\beta}$. The destructive interference among contributions with different frequencies can suppress the off-diagonal blocks of $\rho_A(t)$ with respect to the subsystem charge. This dephasing leads to an effective restoration of the $U(1)$ symmetry at the subsystem level and is observed as the decay of the R\'enyi entanglement asymmetry.

\begin{figure}[t]
    \centering
    \begin{subfigure}[b]{0.3\textwidth}
        \centering
        \includegraphics[width=\linewidth]{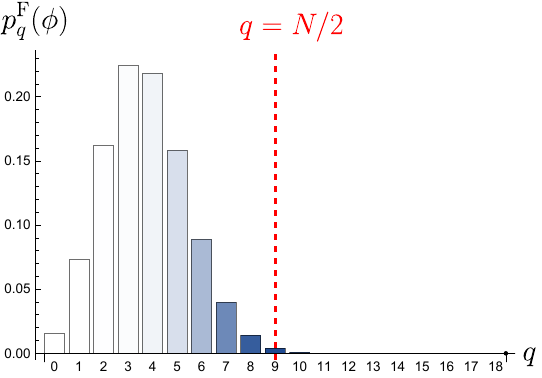}
        \caption*{$\phi = 0.3\pi$}
    \end{subfigure}
    \hfill
    \begin{subfigure}[b]{0.3\textwidth}
        \centering
        \includegraphics[width=\linewidth]{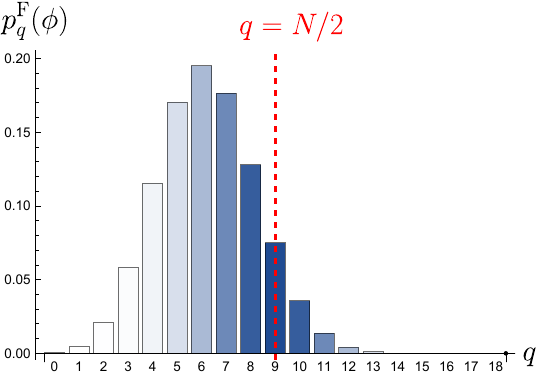}
        \caption*{$\phi = 0.4\pi$}
    \end{subfigure}
    \hfill
    \begin{subfigure}[b]{0.3\textwidth}
        \centering
        \includegraphics[width=\linewidth]{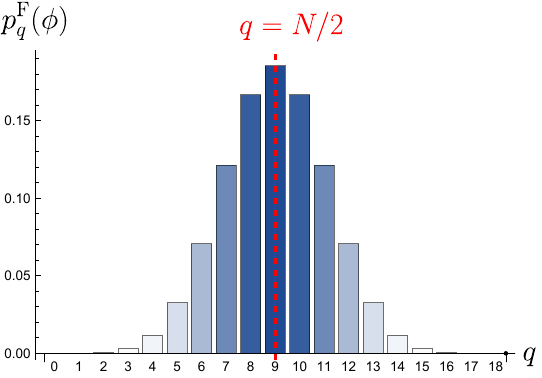}
        \caption*{$\phi = 0.5\pi$}
    \end{subfigure}
    \caption{
    Charge-sector distributions for the spin-product ferromagnetic state $\ket{\psi_1(\phi)}_{\mathrm F}$ with $N=18$. The horizontal axis denotes the total charge $q$, while the vertical axis denotes the probability $p_q^{\mathrm F}(\phi)$. The left, middle, and right panels correspond to $\phi=0.3\pi$, $0.4\pi$, and $0.5\pi$, respectively. The dimension of the charge-$q$ sector is $\binom{N}{q}$ and is maximal at $q=N/2$. The color shading schematically represents the dimension of the corresponding charge sector.
    }
    \label{fig:charge_distrubution_spin_product_ferro}
\end{figure}

The conserved initial charge distribution determines which total-charge sectors can participate in this process. The distributions $p_q^{\mathrm F}(\phi)$ for $\phi=0.3\pi$, $0.4\pi$, and $0.5\pi$ are shown in Fig.~\ref{fig:charge_distrubution_spin_product_ferro}. For $\phi=0.3\pi$, the distribution is concentrated in relatively low-charge sectors. As $\phi$ increases, it shifts toward the central region of the charge spectrum. At $\phi=0.5\pi$, the binomial distribution is centered at $q=N/2$, where the dimension of the fixed-charge subspace,
\begin{align}
    \dim\mathcal{H}_q
    =
    \binom{N}{q},
\end{align}
is maximal.

A larger charge-sector dimension does not, by itself, guarantee faster dephasing. The relaxation rate also depends on the energy spectrum within each sector, the overlaps $c_{q,\alpha}^{\mathrm F}(\phi)$, and the matrix elements that remain after tracing out subsystem $B$. Nevertheless, a broader distribution over larger charge sectors can allow a greater number of energy eigenstates and oscillation frequencies to contribute to the reduced dynamics. This provides a natural qualitative interpretation of the ordering observed in Fig.~\ref{fig:QME_Ry_zero}: the state with $\phi=0.3\pi$, whose charge distribution is concentrated in relatively small sectors, exhibits the slowest symmetry restoration, whereas the state with $\phi=0.5\pi$, whose distribution is centered around the largest charge sectors, exhibits the fastest restoration.

Thus, for the spin-product ferromagnetic state, the initial charge distribution provides an intuitive, although not complete, explanation of the observed quantum Mpemba effect. In the next subsection, we investigate the quantum Mpemba effect in the spin-product N\'eel state as a further benchmark. 

%%%%%%%%%%%%%%%%%%%%%%%%%%%%%%%%%%%%
\subsubsection{N\'eel initial state}
%%%%%%%%%%%%%%%%%%%%%%%%%%%%%%%%%%%%

\begin{figure}[t]
    \centering
    \begin{subfigure}[b]{0.3\textwidth}
        \centering
        \includegraphics[width=\linewidth]{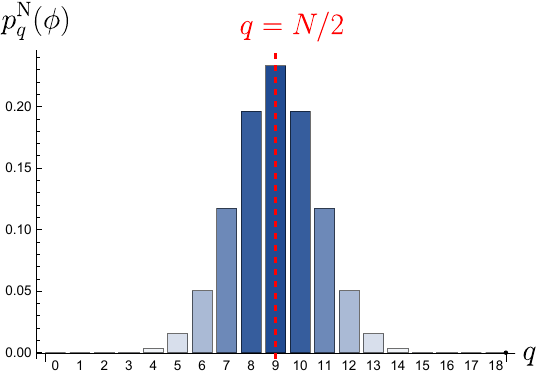}
        \caption*{$\phi = 0.3\pi$}
    \end{subfigure}
    \hfill
    \begin{subfigure}[b]{0.3\textwidth}
        \centering
        \includegraphics[width=\linewidth]{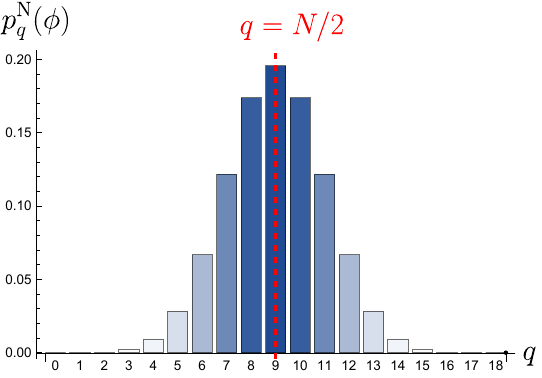}
        \caption*{$\phi = 0.4\pi$}
    \end{subfigure}
    \hfill
    \begin{subfigure}[b]{0.3\textwidth}
        \centering
        \includegraphics[width=\linewidth]{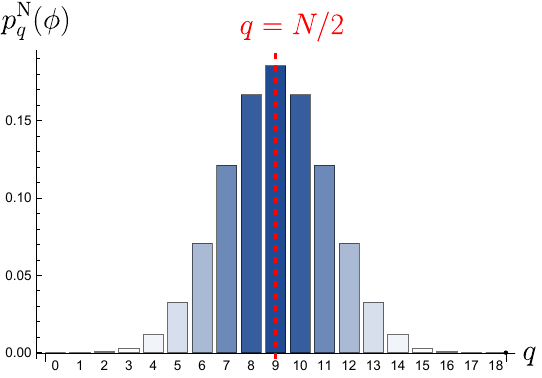}
        \caption*{$\phi = 0.5\pi$}
    \end{subfigure}
    \caption{
    Charge-sector distributions for the spin-product N\'eel initial state $\ket{\psi_1(\phi)}_{\mathrm N}$ with $N=18$. The horizontal axis denotes the total charge $q$, while the vertical axis denotes the probability $p_q^{\mathrm N}(\phi)$. The left, middle, and right panels correspond to $\phi=0.3\pi$, $0.4\pi$, and $0.5\pi$, respectively. All three distributions are centered at $q=N/2$, while their widths increase with $\phi$.
    }
    \label{fig:charge_distribution_spin_product_Neel}
\end{figure}

We next consider the second benchmark initial state, namely the
spin-product N\'eel state,
\begin{align}\label{eq:initial_state_Ry_neel}
    \ket{\psi_1(\phi)}_{\mathrm N}
    &=
    e^{-\i \frac{\phi}{2}\sum_{n=1}^{N}Y_n}
    \ket{0101\cdots 01}.
\end{align}
As in the ferromagnetic case, this state can be decomposed into sectors
of fixed total charge:
\begin{align}
    \ket{\psi_1(\phi)}_{\mathrm N}
    &=
    \sum_{q=0}^{N}
    \sqrt{p^{\mathrm N}_q(\phi)}
    \ket{\psi_q(\phi)}_{\mathrm N},
    \quad
    Q\ket{\psi_q(\phi)}_{\mathrm N}
    =
    q\ket{\psi_q(\phi)}_{\mathrm N},
    \quad
    \|\ket{\psi_q(\phi)}_{\mathrm N}\|=1 .
\label{eq:charge_sector_decomposition_Neel}
\end{align}
The probability weight of each charge sector is given by
\cite{Xiao:2026zia}
\begin{align}
    p^{\mathrm N}_q(\phi)
    &=
    \sum_{k=\max(0,q-N/2)}^{\min(q,N/2)}
    \binom{N/2}{k}
    \binom{N/2}{q-k}
    \cos^{N+2q-4k}\left(\frac{\phi}{2}\right)
    \sin^{N-2q+4k}\left(\frac{\phi}{2}\right).
\label{eq:weight_of_charge_sector_Neel}
\end{align}
The expectation value and variance of the total charge are
\begin{align}
\left\{\,
\begin{aligned}
    \langle Q\rangle_{\mathrm N}
    &=
    \frac{N}{2},
    \\
    (\Delta Q)^2_{\mathrm N}
    &=
    % \langle Q^2\rangle_{\mathrm N}
    % -
    % \langle Q\rangle_{\mathrm N}^2
    % &=
    \frac{N}{4}\sin^2\phi .
\end{aligned}
\right.
\label{eq:charge_distribution_Neel}
\end{align}
Thus, the mean charge is independent of $\phi$, whereas the width of the charge distribution increases monotonically with $\phi$ for $0\leq\phi\leq\pi/2$. As illustrated in Fig.~\ref{fig:charge_distribution_spin_product_Neel}, all three distributions are centered at $q=N/2$, where the dimension of the fixed-charge sector is maximal, while their widths increase with $\phi$.

\begin{figure}[t]
    \centering
    \includegraphics[width=0.5\linewidth]{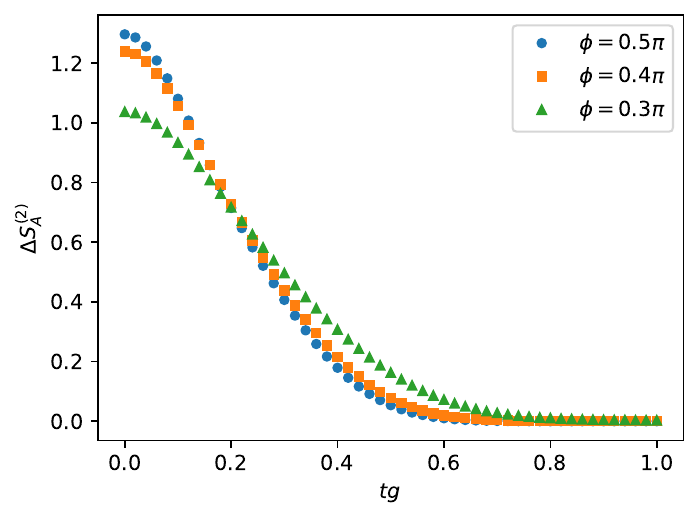}
    \caption{
    Time evolution of the second R\'enyi entanglement asymmetry $\Delta S_A^{(2)}(t)$ for the spin-product N\'eel initial state $\ket{\psi_1(\phi)}_{\mathrm N}$. We compare the results for $\phi=0.3\pi$, $0.4\pi$, and $0.5\pi$. The parameters are fixed to $N=18$, $N_A=4$, $\theta=\pi/3$, $m/g=4$, and $ag=1/4$. For the Suzuki--Trotter decomposition, we use the time step $\delta t=0.02/g$.
    }
    \label{fig:QME_Ry_neel}
\end{figure}

The spin-product N\'eel state also exhibits the quantum Mpemba effect, as shown in Fig.~\ref{fig:QME_Ry_neel}. Among the three initial states considered here, the state with $\phi=0.5\pi$ has the broadest charge distribution and exhibits the fastest symmetry restoration. A broader charge distribution involves a larger range of total-charge sectors and may therefore provide more channels contributing to the dephasing dynamics. The observed ordering of the relaxation rates is consistent with this charge-sector picture.

%%%%%%%%%%%%%%%%%%%%%%%%%%%%%%%%%%%%
%%%%%%%%%%%%%%%%%%%%%%%%%%%%%%%%%%%%
\subsection{Quantum Mpemba effect for fermion-parity-preserving initial states}
%%%%%%%%%%%%%%%%%%%%%%%%%%%%%%%%%%%%
%%%%%%%%%%%%%%%%%%%%%%%%%%%%%%%%%%%%
Having established the behavior of the protocol for simple benchmark states, we now turn to fermion-parity-preserving initial states that are compatible with the fermionic superselection rule. This allows us to test whether the quantum Mpemba behavior captured by the entanglement asymmetry persists in a physically admissible fermionic setting. In a fermionic theory, physical states cannot contain coherence between sectors with even and odd fermion parity \cite{Wick:1952nb,Vidal:2021xmv}. We therefore consider the following fermion-parity-preserving initial states:
\begin{align}\label{eq:initial_state_Ryy}
\left\{\,
\begin{aligned}
    \ket{\psi_2(\phi)}_{\mathrm F}
    &=
    e^{-\i \frac{\phi}{2}\sum_{n=1}^{N-1}Y_nY_{n+1}}
    \ket{00\cdots0}
    \\
    \ket{\psi_2(\phi)}_{\mathrm N}
    &=
    e^{-\i \frac{\phi}{2}\sum_{n=1}^{N-1}Y_nY_{n+1}}
    \ket{0101\cdots01}
\end{aligned}
\right.\ .
\end{align}
Since the unitary $e^{-\i \frac{\phi}{2}\sum_{n=1}^{N-1}Y_nY_{n+1}}$ commutes with the fermion-parity operator $P=(-1)^Q$, these states have definite global parity:
\begin{align}\label{eq:Ryy_state_parity}
\left\{\,
\begin{aligned}
    P\ket{\psi_2(\phi)}_{\mathrm F}
    &=
    +\ket{\psi_2(\phi)}_{\mathrm F},
    \\
    P\ket{\psi_2(\phi)}_{\mathrm N}
    &=
    (-1)^{N/2}\ket{\psi_2(\phi)}_{\mathrm N},
\end{aligned}
\right.
\end{align}
where we assume even $N$.

The global parity constraint also imposes a nontrivial structure on the reduced density matrix. Writing $P=P_A\otimes P_B$, with $P_A=(-1)^{Q_A}$, and using $[\rho(t),P]=0$, one finds
\begin{align}
    P_A\rho_A(t)P_A
    &=
    \Tr_B\left[
    (P_A\otimes I_B)\rho(t)(P_A\otimes I_B)
    \right]
    \nonumber\\
    &=
    \Tr_B\left[
    (I_A\otimes P_B)\rho(t)(I_A\otimes P_B)
    \right]
    =
    \rho_A(t).
\end{align}
Therefore,
\begin{align}\label{eq:Ryy_state_constraint}
    [\rho_A(t),P_A]=0
    \qquad
    \text{for all }t .
\end{align}
On the other hand, for generic symmetry-breaking angles $\phi$, the initial states in \eqref{eq:initial_state_Ryy} are not invariant under the full $U(1)$ symmetry, and hence
\begin{align}
    [\rho_A(0),Q_A]\neq0 .
\end{align}
Thus, these states preserve fermion parity while still breaking the continuous $U(1)$ symmetry.

To make the resulting constraint on the symmetry-breaking coherence explicit, we decompose the reduced density matrix into subsystem-charge blocks,
\begin{align}
    \rho_{A;q_A,q'_A}
    =
    \Pi_{q_A}\rho_A\Pi_{q'_A},
\end{align}
where $\Pi_{q_A}$ projects onto the eigenspace of $Q_A$ with charge
$q_A$. Since
$P_A\Pi_{q_A}=(-1)^{q_A}\Pi_{q_A}$, parity conservation implies
\begin{align}
    P_A\rho_{A;q_A,q'_A}P_A
    =
    (-1)^{q_A-q'_A}
    \rho_{A;q_A,q'_A}.
\end{align}
Combining this relation with
$[\rho_A(t),P_A]=0$, we obtain
\begin{align}\label{eq:parity_conservation_rho_A}
    \rho_{A;q_A,q'_A}=0,
    \qquad
    q_A-q'_A\ \text{odd}.
\end{align}
Hence, coherences between charge sectors with opposite fermion parity are forbidden, whereas coherences such as $q_A=0\leftrightarrow2$, $q_A=1\leftrightarrow3$, and $q_A=0\leftrightarrow4$ are allowed. This provides a clear distinction from the spin-product initial states, for which no such parity selection rule constrains the charge off-diagonal blocks.

\begin{figure}[t]
    \centering
    \centering
    \begin{subfigure}[b]{0.45\textwidth}
        \centering
        \includegraphics[width=\linewidth]{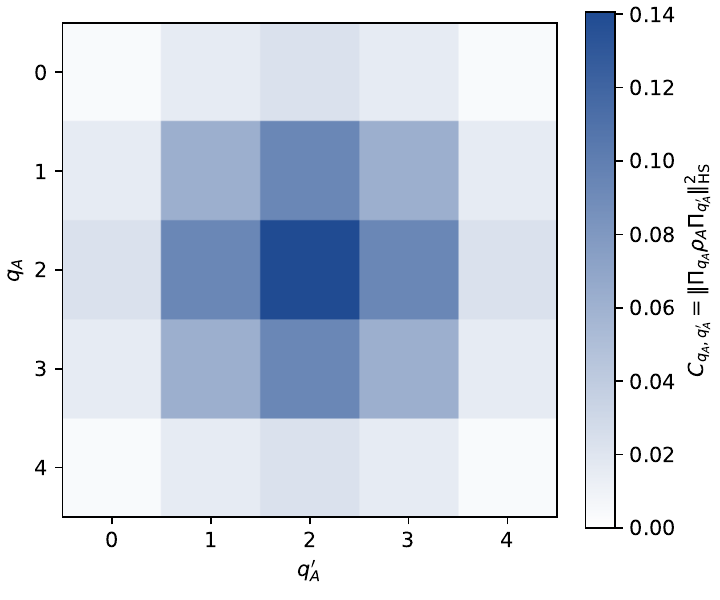}
        \caption*{Spin-product state}
    \end{subfigure}
    \hfill
    \begin{subfigure}[b]{0.45\textwidth}
        \centering
        \includegraphics[width=\linewidth]{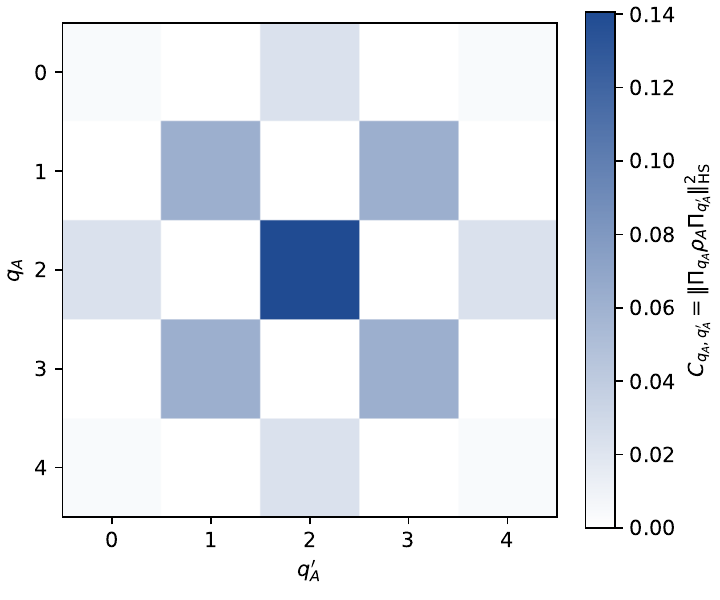}
        \caption*{Fermion-parity-preserving state}
    \end{subfigure}
    \caption{
    Charge-block coherence weights
    $C_{q_A,q'_A}=\|\Pi_{q_A}\rho_A\Pi_{q'_A}\|_{\mathrm{HS}}^2$
    at $t=0$ for the spin-product ferromagnetic initial state $\ket{\psi_1(\phi)}_{\mathrm F}$ (left) and the fermion-parity-preserving ferromagnetic initial state $\ket{\psi_2(\phi)}_{\mathrm F}$ (right). The two panels use the same color scale, with white corresponding to zero coherence. 
%    Off-diagonal blocks encode coherences between different subsystem charge sectors and therefore characterize the breaking of the subsystem $U(1)$ symmetry. 
    %In the fermion-parity-preserving state, all blocks with odd $q_A-q'_A$ vanish, producing the checkerboard structure required by fermion-parity conservation.
    We use $N_A=4$, and $\phi=0.5\pi$.
    }
    \label{fig:charge_block_structure}
\end{figure}

To visualize this structure, we define
\begin{align}
    C_{q_A,q'_A}
    =
    \left\|
    \Pi_{q_A}\rho_A\Pi_{q'_A}
    \right\|_{\mathrm{HS}}^2,
\end{align}
where $\|\cdot\|_{\mathrm{HS}}$ denotes the Hilbert--Schmidt norm. Fig.~\ref{fig:charge_block_structure} compares $C_{q_A,q'_A}$ for the spin-product and fermion-parity-preserving ferromagnetic initial states at $t=0$. 
Note that the off-diagonal blocks encode coherences between different subsystem charge sectors and therefore characterize the breaking of the subsystem $U(1)$ symmetry. 
The spin-product state exhibits coherence between both equal- and opposite-parity charge sectors. By contrast, the fermion-parity-preserving state displays the checkerboard structure implied by \eqref{eq:parity_conservation_rho_A}: all blocks with odd $q_A-q'_A$ vanish, while even-charge-difference coherences remain finite. The same selection rule holds for the N\'eel initial state.

\begin{figure}[t]
    \centering
    \begin{subfigure}[b]{0.48\textwidth}
        \centering
        \includegraphics[width=\linewidth]{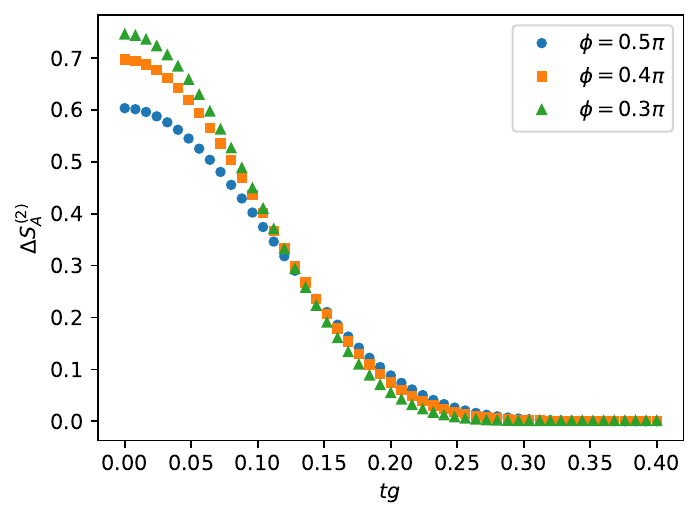}
        \caption*{Ferromagnetic}
    \end{subfigure}
    \hfill
    \begin{subfigure}[b]{0.48\textwidth}
        \centering
        \includegraphics[width=\linewidth]{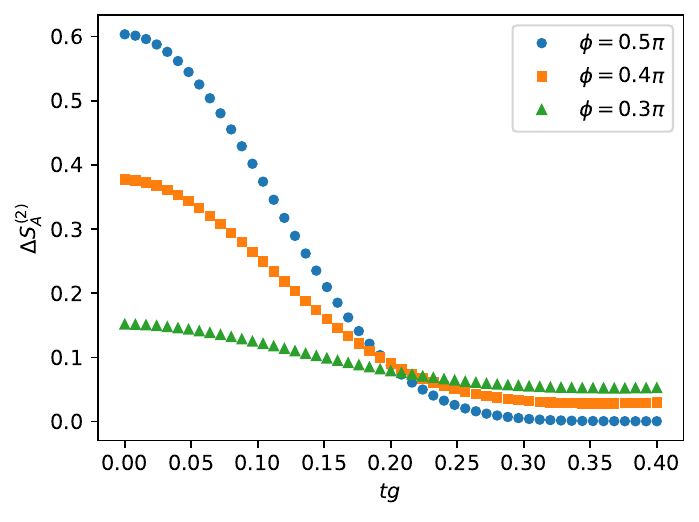}
        \caption*{N\'eel}
    \end{subfigure}
    \caption{
    Time evolution of the second R\'enyi entanglement asymmetry
    $\Delta S_A^{(2)}(t)$ for the fermion-parity-preserving initial
    states in \eqref{eq:initial_state_Ryy}.
    The left and right panels show the ferromagnetic and N\'eel
    families, respectively, for
    $\phi=0.3\pi$, $0.4\pi$, and $0.5\pi$.
    The parameters are fixed to
    $N=18$, $N_A=4$, $\theta=\pi/3$, $m/g=4$, and $ag=1/4$.
    The Suzuki--Trotter time step is $\delta t=0.02/g$.
    }
    \label{fig:QME_Ryy}
\end{figure}

Despite this additional constraint on the allowed charge coherences, the fermion-parity-preserving initial states still exhibit Mpemba-like relaxation dynamics. Fig.~\ref{fig:QME_Ryy} shows the time evolution of the R\'enyi entanglement asymmetry for the two families \eqref{eq:initial_state_Ryy}.
For the ferromagnetic family, the entanglement asymmetry approaches values close to zero at late times within the simulated time window, and the relaxation curves cross, providing a clear signature of the quantum Mpemba effect. A similar crossing is observed for the N\'eel family. In this case, however, the curves for $\phi=0.3\pi$ and $0.4\pi$ retain nonzero residual values at late times for $N=18$.

As discussed in Appendix~\ref{sec:parameter_dependence}, these residual late-time values exhibit a noticeable system-size dependence, suggesting that they may at least partly originate from finite-size effects. Additional results on the energy expectation values of the fermion-parity-preserving initial states are presented in Appendix~\ref{sec:energy}. 
Under a thermal interpretation of the late-time state, the ferromagnetic family is associated with a negative effective temperature, whereas the N\'eel family is associated with a positive effective temperature. 
%Within the parameter range studied here, we observe a tendency for initial states associated with smaller $|\beta_{\rm eff}|$ to exhibit faster symmetry restoration. 
For the fermion-parity-preserving ferromagnetic family, we observe a tendency for initial states associated with more negative $\beta_{\rm eff}$ to exhibit faster symmetry restoration within the parameter range studied here.
Our numerical results for the initial energies and the symmetry-restoration dynamics suggest a possible connection between the effective temperature and the restoration rate.
For the fermion-parity-preserving N\'eel family, the role of the effective temperature is hard to disentangle from that of the initial charge-sector distribution.
One possible interpretation is that the relaxation reflects competing effects associated with these two properties.
Since both vary with $\phi$, the present results do not allow us to isolate their respective contributions.
%Although this observation suggests a possible connection between the effective temperature and the restoration rate, the present results are not sufficient to establish such a relation conclusively.

The system sizes accessible to the present state-vector simulations, however, are not sufficient to establish the thermodynamic-limit behavior conclusively. Extending the analysis to substantially larger systems would require more efficient classical approaches, such as tensor-network methods \cite{Paeckel:2019yjf,Florio:2024aix,Papaefstathiou:2024zsu,Jeyaretnam:2024tkj}, or quantum hardware. We therefore leave a systematic thermodynamic-limit analysis for future work and turn in the next section to the quantum-computing resources required to implement the protocol.

%%%%%%%%%%%%%%%%%%%%%%%%%%%%%%%%%%%%
%%%%%%%%%%%%%%%%%%%%%%%%%%%%%%%%%%%%
%%%%%%%%%%%%%%%%%%%%%%%%%%%%%%%%%%%%
\section{Quantum Resource Estimation}
\label{sec:resource_estimation}
%%%%%%%%%%%%%%%%%%%%%%%%%%%%%%%%%%%%
%%%%%%%%%%%%%%%%%%%%%%%%%%%%%%%%%%%%
%%%%%%%%%%%%%%%%%%%%%%%%%%%%%%%%%%%%

In the previous section, we investigated the quantum Mpemba effect using a classical state-vector simulator. In this section, we estimate the quantum resources required to implement our protocol on a digital quantum computer. At the algorithmic level, there are two distinct sources of error:
\begin{itemize}
    \item Algorithmic error associated with the circuit implementation of the time evolution:
    \begin{align}
        \epsilon_{\text{alg}}
        =
        \left\|
        e^{-\i Ht}
        -
        U_{\text{circ}}(t)
        \right\|,
    \end{align}
    where $U_{\text{circ}}(t)$ denotes the time-evolution operator implemented by the quantum circuit and $\|\cdot\|$ denotes the operator norm.
    \item Statistical error associated with the SWAP-test measurement:
    \begin{align}
        \epsilon_{\text{stat}}
        =
        \sqrt{
        \frac{
        \langle Z^2\rangle_{\text{ancilla}}
        -
        \langle Z\rangle_{\text{ancilla}}^2
        }{N_{\text{shot}}}
        }.
    \end{align}
\end{itemize}

As discussed in Section~\ref{sec:QC_of_EA}, the statistical error is bounded by \eqref{eq:statistical_error1} (or equivalently \eqref{eq:statistical_error2}) and, for a fixed additive accuracy, does not explicitly depend on the quantum state or the total system size. In particular, a SWAP-test expectation value can be estimated to additive accuracy $\epsilon_{\text{stat}}$ using $\mathcal{O}(\epsilon_{\text{stat}}^{-2})$ measurements. We note that this statement concerns additive errors in the measured expectation values. Since the R\'enyi entanglement asymmetry is defined as the logarithm of the ratio of two purities, the number of measurements required to estimate it to a prescribed accuracy generally depends on the values of these purities, and hence on the initial state and evolution time. In particular, smaller purities require higher measurement precision, which is an inherent limitation of purity measurements. In the following, we therefore focus mainly on the per-shot circuit resources associated with the dynamical evolution.

The systematic error in the time evolution directly bounds the corresponding error in the purity. Let $P_{\text{exact}}$ and $P_{\text{circ}}$ denote the purities obtained from the exact and circuit-implemented time evolutions, respectively. One can show the following inequality.\footnote{
One can show \eqref{eq:purity_error} as follows. Since the purities $P_{\text{exact}}$ and $P_{\text{circ}}$ are given by the expectation value of the SWAP operator in the two-copy system:
%SWAP operator in two-copy system: 
$P_{\text{exact}}=\bra{\psi^{\otimes2}}e^{iH t\otimes 2} \text{SWAP}_A e^{-iH t\otimes 2} \ket{\psi^{\otimes 2}}$,
$P_{\text{circ}}=\bra{\psi^{\otimes2}}U_{\text{circ}}^{\dagger \otimes 2} \text{SWAP}_A U_{\text{circ}}^{\otimes 2}\ket{\psi^{\otimes2}}$,
we can evaluate the absolute value of the difference
%difference 
as follows.
\begin{align}
\begin{aligned}
   \left|P_{\text{exact}}-P_{\text{circ}} \right| 
   &\leq
   \left|
   \bra{\psi^{\otimes2}} e^{iH t\otimes 2} 
   \text{SWAP}_A
   \left( e^{-iH t\otimes 2} - U_{\text{circ}}^{\otimes 2 }\right)\ket{\psi^{\otimes 2}}
   \right|
   +
   \left|
   \bra{\psi^{\otimes 2}} \left( e^{iH t\otimes 2} - U_{\text{circ}}^{\dagger \otimes2} \right)
   \text{SWAP}_A
   U_{\text{circ}}^{\otimes2}\ket{\psi^{\otimes 2}}
   \right| \\
   &\leq 
   \left\| e^{-iH t\otimes 2} - U_{\text{circ}}^{\otimes 2}\right\| 
   +
   \left\|e^{iH t\otimes 2}-U_{\text{circ}}^{\dagger \otimes 2}\right\|\\
   &\leq
   4\left\| e^{-iH t}-U_{\text{circ}}\right\|,
%   &\leq    \left\| e^{iH t\otimes 2} - U_{\text{circ}}^{\otimes 2}\right\|    + \left\|e^{-iH t\otimes 2}-U_{\text{circ}}^{\dagger \otimes 2}\right\|\\  
%&\leq   4\left\| e^{iH t}-U_{\text{circ}}\right\|,
   \end{aligned}\nonumber
\end{align}
herein, inequalities $|A+B|\leq |A|+|B|$, $|\braket{\mathcal{O}}|\leq \left\|\mathcal{O}\right\|$, $\left\|\mathcal{O}_1\mathcal{O}_2\right\|\leq \left\|\mathcal{O}_1\right\|\left\|\mathcal{O}_2\right\|$, and operator norm of unitary operators $ \left\|\mathcal{U}\right\|=1$ are used. 
}
\begin{align}\label{eq:purity_error}
    \left|
    P_{\text{exact}}
    -
    P_{\text{circ}}
    \right|
    \leq
    4
    \left\|
    e^{-\i Ht}
    -
    U_{\text{circ}}(t)
    \right\|
    =
    4\epsilon_{\text{alg}}.
\end{align}
The required circuit resources therefore depend on how accurately the time-evolution operator is implemented. In general, this cost depends on the Hamiltonian, system size, simulation time, target accuracy, and, for an actual device, the native gate set and physical gate errors. We first estimate the algorithmic resources required for the Trotterized time evolution and then discuss noisy and fault-tolerant implementations.

Throughout this section, we quote the gate counts for a single copy of the Hamiltonian simulation. The second-R\'enyi SWAP-test protocol requires two copies of the state, and hence approximately twice as many gates in total. However, because the two copies can be evolved in parallel, this doubles the circuit width but does not increase the ideal depth of the time evolution. We explicitly include the two-copy structure when estimating the accumulated physical-gate error of the full protocol.

%%%%%%%%%%%%%%%%%%%%%%%%%%%%%%%%%%%%
%%%%%%%%%%%%%%%%%%%%%%%%%%%%%%%%%%%%
\subsection{Error-bound scaling}
\label{subsec:error_bound_scaling}
%%%%%%%%%%%%%%%%%%%%%%%%%%%%%%%%%%%%
%%%%%%%%%%%%%%%%%%%%%%%%%%%%%%%%%%%%

Let us first estimate the number of gates required to implement the time-evolution operator $e^{-\i Ht}$ using a Trotterized decomposition. Let $U(\delta t)$ denote one Trotter step, where $\delta t=t/M$ and $M$ is the total number of Trotter steps. The total Trotter error satisfies
\begin{align}\label{eq:systematic_error_Trotter}
    % \epsilon_{\text{alg}}
    % =
    \left\|
    e^{-\i Ht}
    -
    U^M(\delta t)
    \right\|
    \leq
    M
    \left\|
    e^{-\i H\delta t}
    -
    U(\delta t)
    \right\|.
\end{align}
We determine the required gate resources according to the following three steps:
\begin{itemize}
    \item[\textbf{STEP~1:}] Evaluate the error of a single Trotter step, $\left\|e^{-\i H\delta t}-U(\delta t)\right\|$.
    \item[\textbf{STEP~2:}] Determine the number of Trotter steps $M$ required to satisfy the target accuracy,
    \begin{align}\label{eq:M_def}
        M
        \left\|
        e^{-\i Ht/M}
        -
        U(t/M)
        \right\|
        \leq
        \epsilon_{\text{alg}}.
    \end{align}
    \item[\textbf{STEP~3:}] Count the elementary gates required for each Trotter step and combine this result with the value of $M$ obtained in STEP~2 to determine the total gate cost.
\end{itemize}

We first carry out STEP~1 by evaluating a theoretical upper bound on the one-step error \cite{Childs_2021}. To make the discussion concrete, we focus on the second-order product formula. We decompose the Hamiltonian as
\begin{align}
    H
    =
    \sum_{a=1}^{K}
    H_{a},
\end{align}
where the components $H_a$ are not required to commute with one another. In our implementation, each $H_a$ is itself composed of mutually commuting terms, namely, 
\begin{align}
    H_a = \sum_{i\in I_a} h^{(a)}_{i},\quad \left[ h^{(a)}_{i}, h^{(a)}_{i'} \right]=0,\quad \forall i, i' \in I_a\ ,
\end{align}
where $I_a$ labels the individual interactions in the $a$-th block. Each term $H_a$ can therefore be straightforwardly implemented as a product of elementary Pauli rotations. The second-order product formula is\footnote{Here, the product of operators is defined by
$\prod_{a = 1}^{K} U_a = U_K \cdots U_1$ and $\prod_{a = K}^{1} U_a = U_1 \cdots U_K$.
}
\begin{align}
    U(\delta t)
    =
    \prod_{a=1}^{K}
    e^{-\i\frac{\delta t}{2} H_a }
    \prod_{b=K}^{1}
    e^{-\i\frac{\delta t}{2}H_b}.
\end{align}
The corresponding one-step error can be bounded as
\begin{align}\label{eq:Trotter_error_1step}
    \left\|
    e^{-\i H\delta t}
    -
    U(\delta t)
    \right\|
    \leq
    \alpha_{\text{com}}\delta t^3,
\end{align}
where the commutator bound is given by \cite{Childs_2021}
\begin{align}\label{eq:commutator_bound}
    \alpha_{\text{com}}
    &=
    \frac{1}{24}
    \sum_{a=1}^{K-1}
    \left\|
    \left[
    H_{a},
    \left[
    H_{a},
    \sum_{b=a+1}^{K}
    H_{b}
    \right]
    \right]
    \right\|
    % \nonumber\\
    % &\quad
    +
    \frac{1}{12}
    \sum_{a=1}^{K-1}
    \left\|
    \left[
    \sum_{c=a+1}^{K}
    H_{c},
    \left[
    \sum_{b=a+1}^{K}
    H_{b},
    H_{a}
    \right]
    \right]
    \right\|.
\end{align}

A simpler analytical bound can be obtained by further bounding the nested commutators using the triangle inequality and the locality of the interactions. In contrast to the general commutator bound in \eqref{eq:commutator_bound}, the following estimate assumes that the Hamiltonian is composed of at most two-body interactions. We define
%Furthermore, the norms of nonvanishing nested commutators can be upper bounded using the triangle inequality $\|[A,B] \|\le 2 \|A\| \|B\|$. One can derive the following inequality as well.
\begin{align}
    \lambda_{\max}
    :=
    \max_j
    \sum_{a=1}^{K}
    \sum_{\substack{i\in I_a\\
    j\in {\rm supp}(h_i^{(a)})}}
    \left\|
    h_i^{(a)}
    \right\|,
\end{align}
which characterizes the maximum total interaction strength involving any single site, and
\begin{align}
    \gamma
    :=
    \sum_{a=1}^{K-1}
    \sum_{i\in I_a}
    \left\|
    h_i^{(a)}
    \right\|.
\end{align}
For the one- and two-body interactions considered here, one can derive the following upper bound.
\begin{align}\label{eq:worst-case_bound}
    \left\|
    e^{-\i H\delta t}
    -
    U(\delta t)
    \right\|
    \leq
    \frac{7}{3} \gamma \lambda_{\text{max}}^2 \delta t^3,
\end{align}
We refer to \eqref{eq:worst-case_bound} as the analytical worst-case bound. 
A derivation of this inequality for $k$-local Hamiltonians
%Hamiltonian 
is given in Appendix~\ref{sec:analytical_trotter_bound}.

For the lattice Schwinger model, the product formula used in \eqref{eq:Trotterization} corresponds to the ordering\footnote{
For a decomposition $H=\sum_{a=1}^{3} H_a$, the commutator bound generally depends on the ordering of the Hamiltonian components. Among the six possible permutations, the ordering used here gives the smallest bound for the representative parameter set employed in our numerical resource estimate. 
%We do not claim that this ordering is optimal for arbitrary Hamiltonian parameters.
}
\begin{align}
    (H_1,H_2,H_3)
    =
    (H_{\rm even},H_{\rm odd},H_Z+H_{ZZ}) .
\end{align}
For this ordering, the coefficients scale as $\gamma = \mathcal{O} \left( N \right)$, $\lambda_{\text{max}} = \mathcal{O} \left( N^2 \right)$. Therefore, the analytical worst-case bound gives
\begin{align}
    \left\|
    e^{-\i H\delta t}
    -
    U(\delta t)
    \right\|
    \leq
    \mathcal{O} \left( N^5 \delta t^3 \right).
\end{align}
Combining this result with \eqref{eq:M_def}, we obtain
\begin{align}
    M_{\text{worst}}
    =
    \mathcal{O}
    \left(
    N^{5/2}
    \frac{t^{3/2}}{\epsilon_{\text{alg}}^{1/2}}
    \right).
\end{align}

\begin{figure}[t]
    \centering
    \begin{subfigure}[b]{0.48\textwidth}
        \centering
        \includegraphics[width=\linewidth]{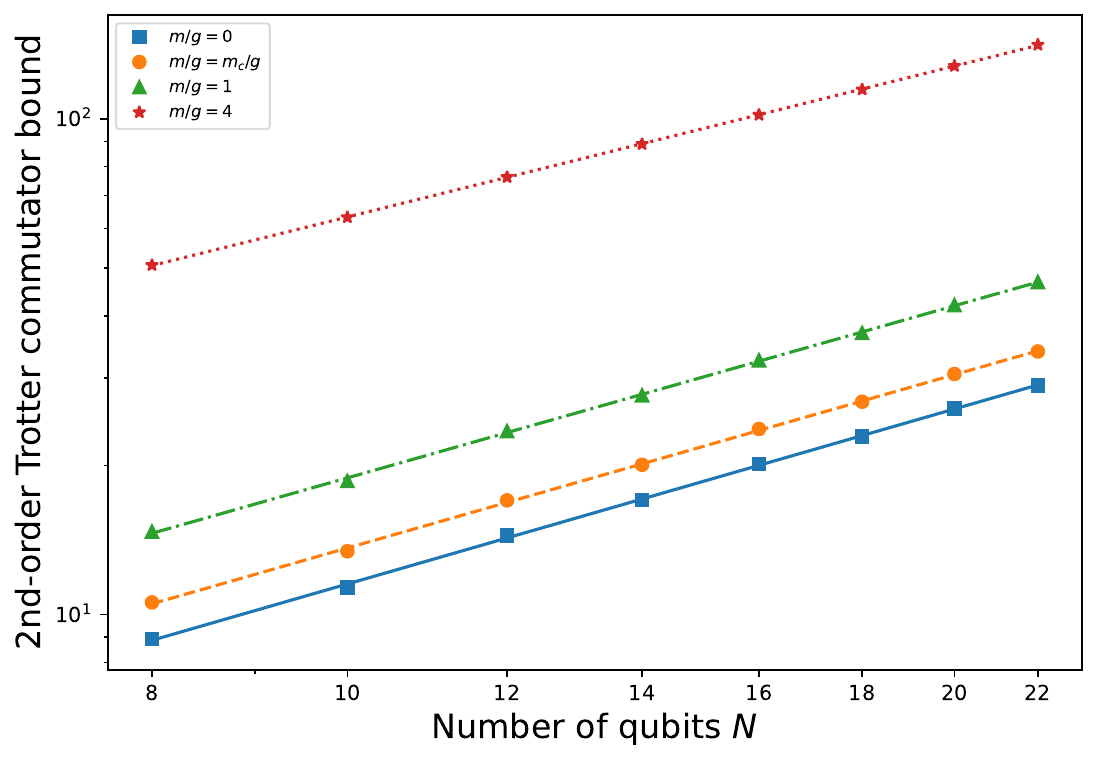}
    \end{subfigure}
    \hfill
    \begin{subfigure}[b]{0.48\textwidth}
        \centering
        \includegraphics[width=\linewidth]{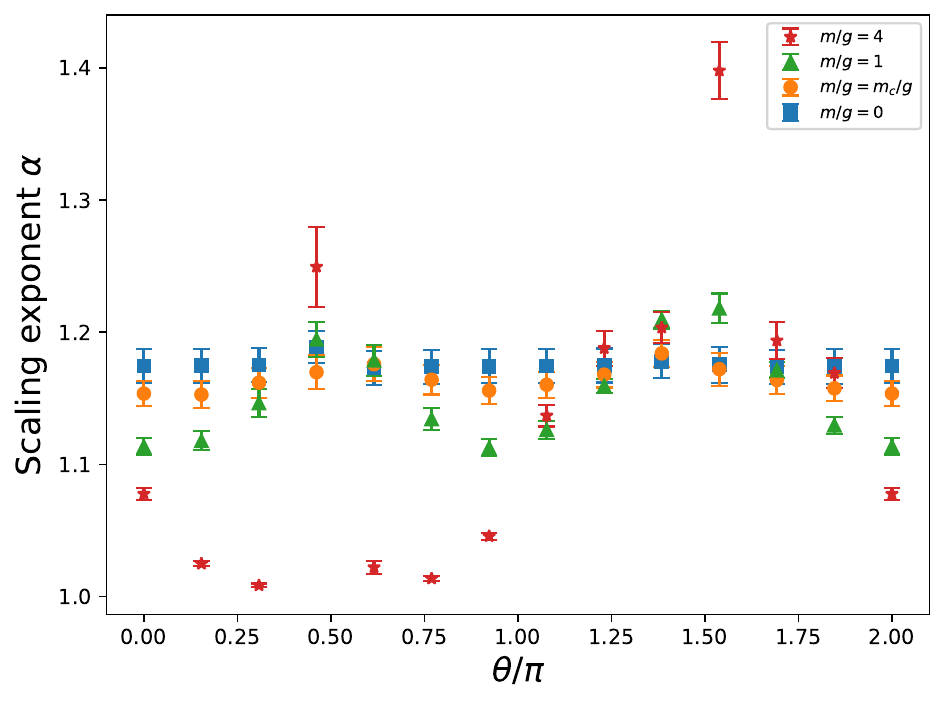}
    \end{subfigure}
    \caption{Commutator bound for the second-order product formula. 
    (Left) The commutator bound $\alpha_{\text{com}}$ defined in \eqref{eq:commutator_bound} as a function of the system size $N$ on a log-log scale. 
    The symbols show the numerical values, while the 
    %solid 
    lines show power-law fits of the form $\alpha_{\text{com}} = A N^\alpha$.
    %$\alpha_{\text{com}}\propto N^\alpha$. 
    Different colors correspond to different choices of $m/g$ with fixed $ag = 0.25$ and $\theta = \pi/3 $. 
    %$\theta/ = \pi/3 $. 
    %The particular values of fitting power scaling $\alpha$ are given in Table \ref{tab:fitting_value_com}. 
    The fitted values of $\alpha$ and $A$ are given in Table~\ref{tab:fitting_value_com}. 
    (Right) The effective scaling exponent $\alpha$ as a function of $\theta/\pi$ at fixed $ag = 0.25$. The fits are performed using $N=\{8,10,\cdots,22\}$ and the critical mass value is set at $m_c/g = 0.3335$. 
    }
    \label{fig:commutator_bound}
\end{figure}

The analytical worst-case estimate is obtained by repeatedly applying triangle and norm inequalities and is therefore generally conservative. For a tighter estimate, we numerically evaluate the commutator coefficient $\alpha_{\rm com}$ in \eqref{eq:commutator_bound} for the lattice Schwinger model. Fig.~\ref{fig:commutator_bound} shows $\alpha_{\rm com}$ as a function of the system size $N$.
Over the accessible finite-size range, the results are well described by an effective power law,
\begin{align}\label{eq:commutator_power_scaling}
    \alpha_{\text{com}}
    \simeq
    A N^{\alpha},
\end{align}
where $A$ and $\alpha$ denote the fitted coefficient and effective exponent, respectively, both of which depend on the parameter set.\footnote{
%\MH{Clarify the presence of $\theta$-dependence at $m=0$}
At $m=0$, $\alpha$ remains dependent on $\theta$ through the definition of $m_{\text{lat}}$ at finite lattice spacing $a$.
}
We numerically find approximately $1.0\lesssim\alpha$ over the parameter range considered here.\footnote{
The scaling of the commutator bound for the lattice Schwinger model without the topological term was previously investigated in Ref.~\cite{Nguyen_2022}, where an exponent $\alpha \simeq 1.1$ was obtained.
% \MH{Mention their values}
Here, we extend this analysis to include the $\theta$ dependence.
}
The fitted values of $A$ and $\alpha$ are summarized in Table~\ref{tab:fitting_value_com}. Since these fits are based on relatively small system sizes, we do not interpret the fitted values of $\alpha$ as rigorous asymptotic exponents.

\begin{table}[h]
    \centering
    \renewcommand{\arraystretch}{1.25}
    \begin{tabular}{|c|c|c|}
    \hline
    \textbf{Parameter} 
    & \textbf{Fitted $\alpha$} 
    & \textbf{Fitted $A$} \\
    \hline
    
    $m/g = 0$ 
    & $1.173\pm0.010$ 
    & $0.774\pm0.021$ \\
    \hline
    
    $m/g = m_c/g$ 
    & $1.162\pm0.009$ 
    & $0.938\pm0.024$ \\
    \hline
    
    $m/g = 1$ 
    & $1.154\pm0.009$ 
    & $1.324\pm0.030$ \\
    \hline
    
    $m/g = 4$ 
    & $1.011\pm0.002$ 
    & $6.176\pm0.034$ \\
    \hline

\end{tabular}
\caption{Fitted power $\alpha$ and coefficient $A$ for the commutator bound $\alpha_{\mathrm{com}}$, corresponding to the left panel of Fig.~\ref{fig:commutator_bound}. 
%Figs.~\ref{fig:commutator_bound}. 
The numerical data are fitted to the power-law form  $\alpha_{\mathrm{com}}=A N^\alpha$ for different values of $m/g$, with $ag=0.25$ and $\theta=\pi/3 $ fixed.}
\label{tab:fitting_value_com}
\end{table}

We next carry out STEP~2 and estimate the required number of Trotter steps. Substituting \eqref{eq:Trotter_error_1step} and $\delta t=t/M$ into \eqref{eq:systematic_error_Trotter}, we obtain
\begin{align}
    \epsilon_{\text{alg}}
    \leq
    \frac{\alpha_{\text{com}}t^3}{M^2}.
\end{align}
Therefore, a sufficient number of Trotter steps for a target error $\epsilon_{\text{alg}}$ is
\begin{align}\label{eq:the_number_of_Trotter_steps_commutator_bound}
    M_{\text{com}}
    &:=
    \left\lceil
    \alpha_{\text{com}}^{1/2}
    \frac{t^{3/2}}{\epsilon_{\text{alg}}^{1/2}}
    \right\rceil
    \sim
    \mathcal{O}
    \left(
    N^{\alpha/2}
    \frac{t^{3/2}}{\epsilon_{\text{alg}}^{1/2}}
    \right),
\end{align}
where the approximation follows from the effective scaling in \eqref{eq:commutator_power_scaling}. 

\begin{figure}[t]
    \centering
    \begin{subfigure}[b]{0.48\textwidth}
        \centering
        \includegraphics[width=\linewidth]{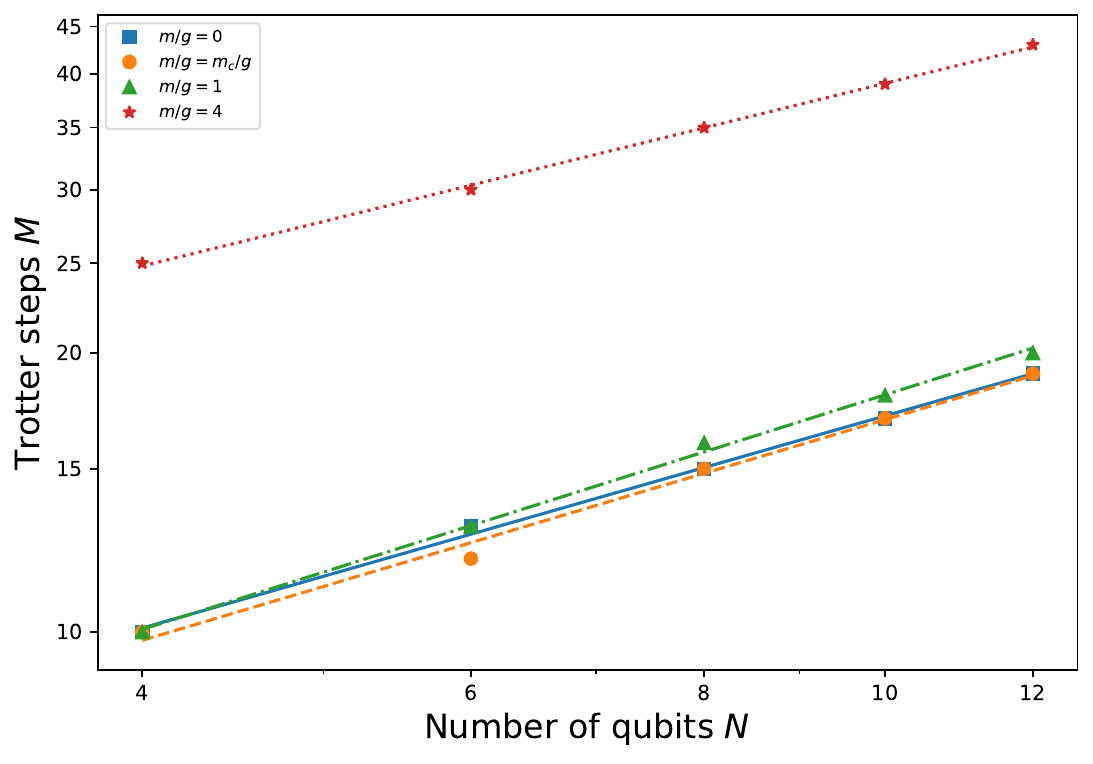}
    \end{subfigure}
    \hfill
    \begin{subfigure}[b]{0.48\textwidth}
        \centering
        \includegraphics[width=\linewidth]{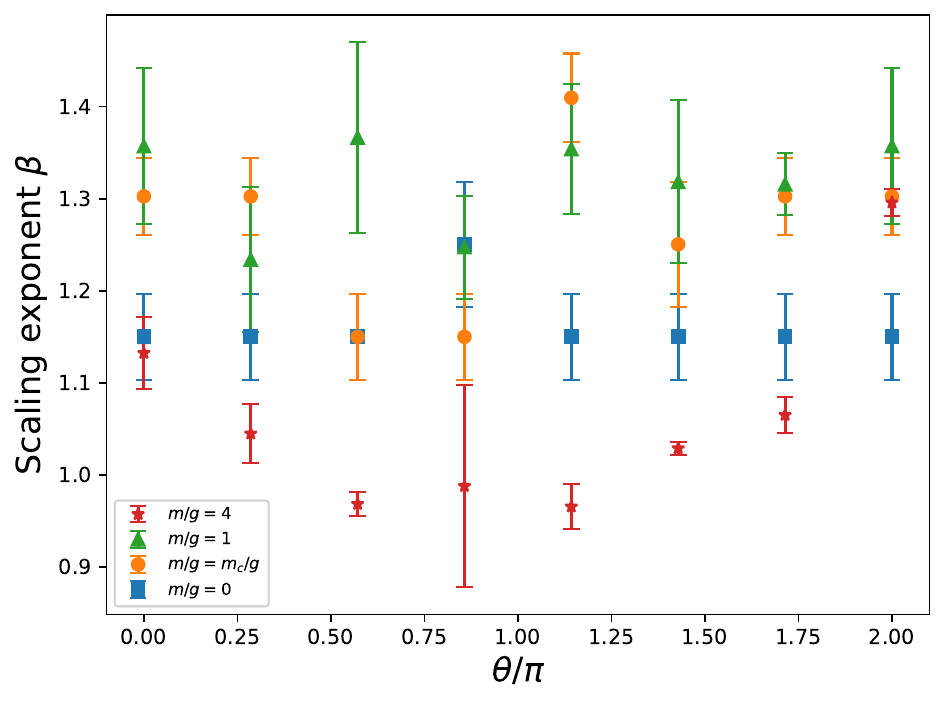}
    \end{subfigure}
    \caption{Empirical Trotter-step requirement defined in \eqref{eq:M_empirical}. We set $tg=1$, $ag = 0.25$ and $\epsilon_{\text{alg}}=0.01$ as representative values. (Left) $M_{\text{emp}}$ as a function of the system size $N$ on a log-log scale. 
    The symbols denote the numerical results, while the 
    %solid 
    lines show power-law fits of the form $M_{\text{emp}} = B N^{\beta/2}$. 
%    $M_{\text{emp}}\propto N^{\beta/2}$. 
    Different colors correspond to different choices of $m/g$ at fixed $\theta = \pi /3$. 
%    The particular values of fitting power scaling $\beta$ are given in Table \ref{tab:fitting_value_emp}. 
    The fitted values of $\beta$ and $B$ are given in Table~\ref{tab:fitting_value_emp}. 
    (Right) The effective scaling exponent $\beta$ as a function of $\theta/\pi$. 
    The fits 
    %on the right panel 
    are performed using $N=\{4,6,8,10,12\}$, and the critical mass value is set at $m_c/g = 0.3335$.  
    }
    \label{fig:empirical}
\end{figure}

Although the commutator bound provides a sufficient value of $M$, it can still substantially overestimate the number of steps required in practice. We further determine the required value of $M$ empirically by directly evaluating the total Trotter error. We define
\begin{align}\label{eq:M_empirical}
    M_{\text{emp}}
    =
    \min
    \left\{
    M\in\BZ_{+}
    \,\middle|\,
    \left\|
    e^{-\i Ht}
    -
    U^M(t/M)
    \right\|
    \leq
    \epsilon_{\text{alg}}
    \right\}.
\end{align}
For the small systems accessible to exact diagonalization, we evaluate the operator norm in \eqref{eq:M_empirical} directly. Fig.~\ref{fig:empirical} shows the system-size dependence of $M_{\text{emp}}$. Over the accessible range, the numerical results are again well fitted by an effective power law,
\begin{align}
    M_{\text{emp}}
    \simeq
    B N^{\beta/2},
\end{align}
where $B$ and $\beta/2$ denote the fitted coefficient and effective exponent, respectively. The fitted values of $B$ and $\beta$ are summarized in Table~\ref{tab:fitting_value_emp}. The fitted exponent $\beta$ is found to be comparable to the exponent $\alpha$ inferred from the commutator bound over the accessible system sizes. Thus, the empirical estimate does not necessarily exhibit a parametrically more favorable system-size scaling. Nevertheless, the absolute values of $M_{\rm emp}$ are substantially smaller than the sufficient Trotter-step counts obtained from the commutator bound, reflecting the conservativeness accumulated in the analytical bound. In the next subsection, we explicitly compare the corresponding gate counts obtained from the two estimates. As in the analysis of $\alpha_{\rm com}$, however, the fitted exponent $\beta$ is obtained from relatively small systems and should not be interpreted as a controlled asymptotic scaling exponent.

\begin{table}[h]
    \centering
    \renewcommand{\arraystretch}{1.25}
    \begin{tabular}{|c|c|c|}
    \hline
    \textbf{Parameter} 
    & \textbf{Fitted $\beta$} 
    & \textbf{Fitted $B$} \\
    \hline
    
    $m/g = 0$ 
    & $1.150\pm0.030$ 
    & $4.551\pm0.141$ \\
    \hline
    
    $m/g = m_c/g$ 
    & $1.196\pm0.062$ 
    & $4.275\pm0.271$ \\
    \hline
    
    $m/g = 1$ 
    & $1.272\pm0.035$ 
    & $4.169\pm0.151$ \\
    \hline
    
    $m/g = 4$ 
    & $0.991\pm0.019$ 
    & $12.491\pm0.248$ \\
    \hline

\end{tabular}
\caption{Fitted power $\beta$ and coefficient $B$ for the empirical number of Trotter steps $M_{\mathrm{emp}}$, corresponding to the left panel of Fig.~\ref{fig:empirical}.
%Figs.~\ref{fig:empirical}. 
The numerical data are fitted to the power-law form $M_{\mathrm{emp}}=B N^{\beta/2}$ for different values of $m/g$, with $ag=0.25$, $\theta=\pi / 3$ and $\epsilon_{\text{alg}}=0.01$ fixed.
}
\label{tab:fitting_value_emp}
\end{table}

%%%%%%%%%%%%%%%%%%%%%%%%%%%%%%%%%%%%
%%%%%%%%%%%%%%%%%%%%%%%%%%%%%%%%%%%%
\subsection{Gate-complexity scaling}
\label{subsec:gate_complexity}
%%%%%%%%%%%%%%%%%%%%%%%%%%%%%%%%%%%%
%%%%%%%%%%%%%%%%%%%%%%%%%%%%%%%%%%%%

We finally carry out STEP~3 by translating the required Trotter number into elementary gate counts. For noisy, non-fault-tolerant devices, two-qubit gates are typically among the dominant sources of circuit error, although their precise cost depends on the hardware platform and native gate set. We therefore use the CNOT count as a simple proxy for the two-qubit-gate cost.

For a single copy of the system, the total number of CNOT gates required for the second-order Suzuki--Trotter time evolution is
\begin{align}\label{eq:CNOT_count}
    G_{\text{CNOT}}^{(1)}
    =
    M(N^2+3N-2)
    +
    (N-1)(N-2),
    \qquad
    N=\text{even}.
\end{align}
%The detail of elementary gate decomposition is 
Details of elementary gate decomposition are given in Appendix~\ref{sec:gate_decomposition}. The leading large-$N$ behavior is therefore
\begin{align}
    G_{\text{CNOT}}^{(1)}
    =
    \mathcal{O}(MN^2).
\end{align}
Combining this with \eqref{eq:the_number_of_Trotter_steps_commutator_bound}, the commutator-bound estimate gives
\begin{align}
    G_{\text{CNOT}}^{(1)}
    =
    \mathcal{O}
    \left(
    N^{2+\alpha/2}
    \frac{t^{3/2}}{\epsilon_{\text{alg}}^{1/2}}
    \right),
\end{align}
while the analytical worst-case bound $\alpha_{\text{com}}=\mathcal{O}(N^5)$ gives
\begin{align}
    G_{\text{CNOT}}^{(1)}
    =
    \mathcal{O}
    \left(
    N^{9/2}
    \frac{t^{3/2}}{\epsilon_{\text{alg}}^{1/2}}
    \right).
\end{align}

The complete second-R\'enyi protocol also contains the controlled-SWAP operation used to evaluate $\Tr_A[\rho_A^2]$ and the $U_A$ operations required to evaluate $\Tr_A[\rho_{A,S}^2]$. These operations introduce an additional $\mathcal{O}(N_A)$ gate overhead in the implementation considered here. Since $N_A\leq N$, this overhead is subleading compared with the $\mathcal{O}(MN^2)$ cost of the time evolution and is therefore neglected in the leading resource estimates.

\begin{figure}[t]
    \centering
     \centering
        \begin{subfigure}[b]{0.49\textwidth}
        \centering
        \includegraphics[width=\linewidth]{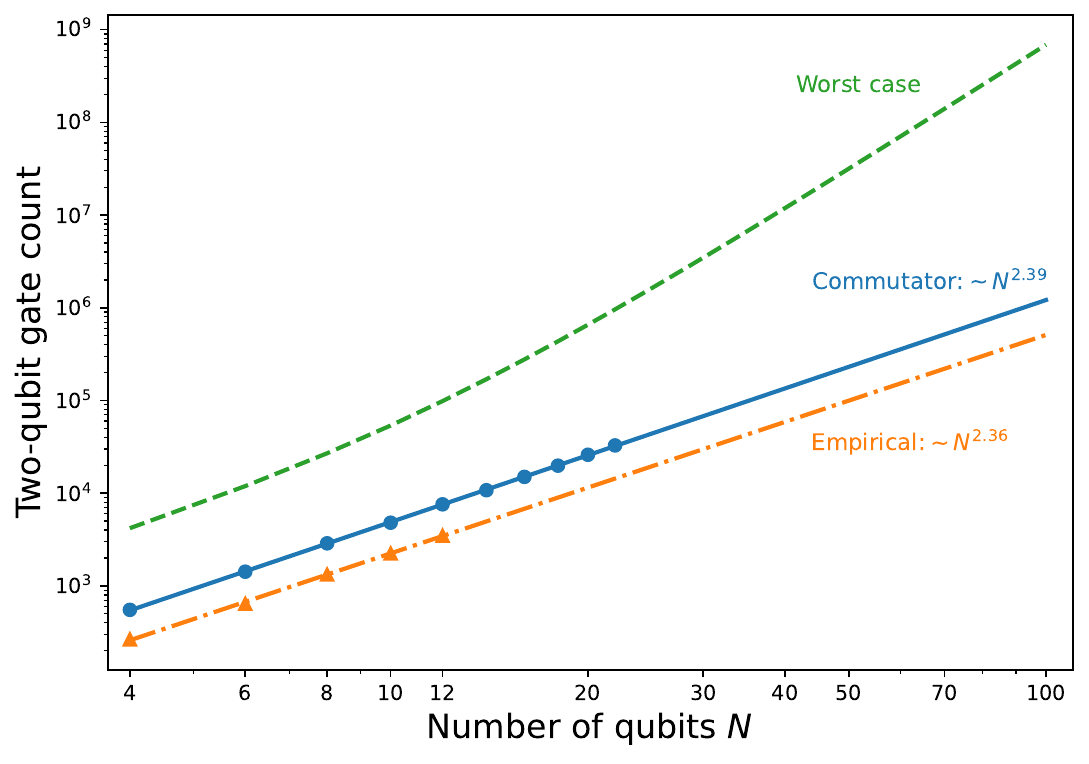}
        \caption*{$m=m_c$}
    \end{subfigure}
       \hfill
    \begin{subfigure}[b]{0.49\textwidth}
        \centering
        \includegraphics[width=\linewidth]{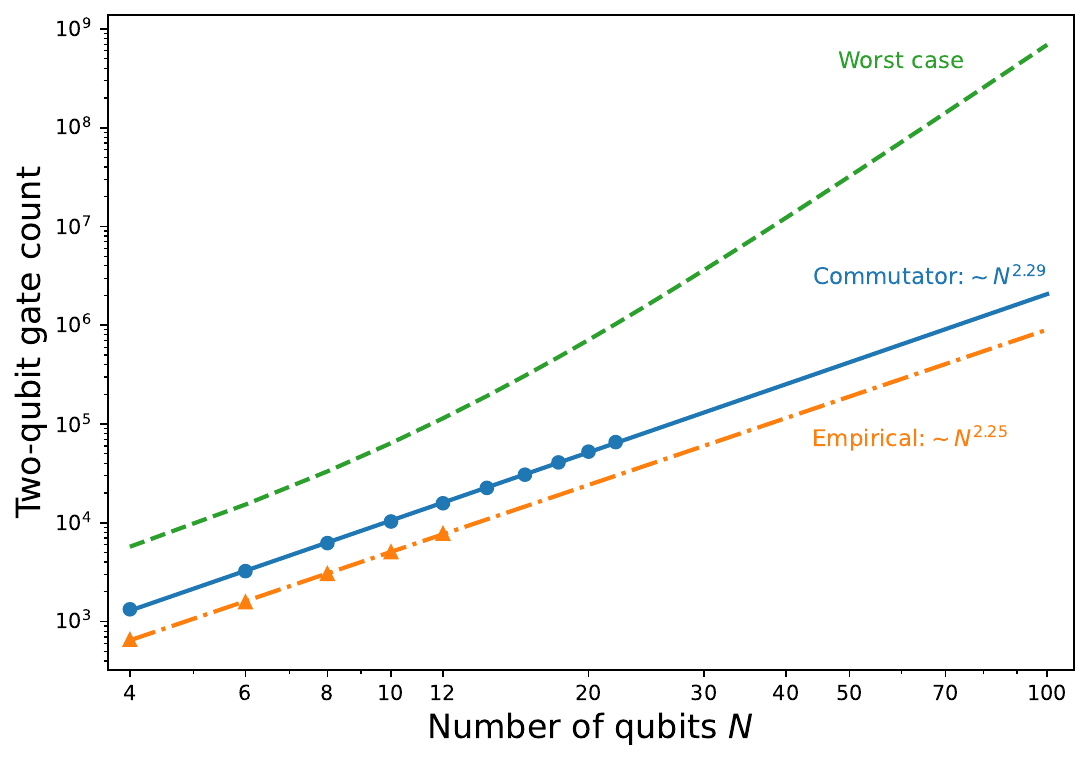}
        \caption*{$m/g=4$}
    \end{subfigure}    
    \caption{
    Estimated single-copy CNOT count required to implement the Trotterized time evolution with target error $\epsilon_{\text{alg}}=0.01$. The green, blue, and orange lines show the estimate obtained from the worst-case bound \eqref{eq:worst-case_bound}, the commutator bound \eqref{eq:the_number_of_Trotter_steps_commutator_bound}, and the empirical Trotter number \eqref{eq:M_empirical}, respectively. 
    %We compare the result with $m=m_c$ and $m/g=4$ in the left and right panels, respectively.
    %panel, which exhibit qualitatively similar behavior.
    The left and right panels show the results for $m=m_c$ and $m/g=4$, respectively.
    %The both
    Both cases exhibit qualitatively similar behavior.
    For the other parameters, we use $tg=1$, $\theta=\pi/3$, and $ag=1/4$ as a representative parameter set. The full second-R\'enyi protocol requires two such time-evolution circuits, which can be executed in parallel.
    }
    \label{fig:CNOT_count}
\end{figure}

Fig.~\ref{fig:CNOT_count} shows the single-copy CNOT count obtained using both the commutator-bound estimate and the empirical Trotter number. The empirical result is extrapolated from small system sizes and should therefore be regarded as an indicative estimate rather than a controlled prediction at large $N$. 

The CNOT count can also be used to obtain a rough indication of the physical gate fidelity required for an uncorrected device. Let $p_{\text{ph}}$ denote the physical error probability of a two-qubit gate and assume, for illustration, an independent stochastic-error model. Since the full SWAP-test protocol contains approximately $2G_{\text{CNOT}}^{(1)}$ two-qubit gates in the two parallel time-evolution circuits, the probability of at least one two-qubit-gate fault is
\begin{align}\label{eq:NISQ_fault_probability}
    P_{\text{fault}}
    =
    1-
    (1-p_{\text{ph}})^{2G_{\text{CNOT}}^{(1)}}
    \simeq
    2G_{\text{CNOT}}^{(1)}p_{\text{ph}},
\end{align}
where the final approximation holds for $2G_{\text{CNOT}}^{(1)}p_{\text{ph}}\ll1$. Thus, the heuristic condition
\begin{align}
    p_{\text{ph}}
    \ll
    \frac{1}{2G_{\text{CNOT}}^{(1)}}
\end{align}
provides a rough estimate of the required physical two-qubit-gate fidelity. For example, using the empirical Trotter-step estimate with $m/g=4$, $ag=0.25$, $\theta=\pi/3$, $tg=1$, and $\epsilon_{\rm alg}=0.01$, we obtain $G_{\rm CNOT}^{(1)}\simeq 3.1\times10^3$ for $N=8$, corresponding to the heuristic requirement $p_{\rm ph}\ll 1.6\times10^{-4}$. Recent trapped-ion quantum computers have achieved two-qubit gate error rates on the order of $10^{-3}$--$10^{-4}$ \cite{Ransford:2025ksn,Hughes:2025pem}, suggesting that a small-scale implementation is within reach of current NISQ technology, although the effect of accumulated gate errors must be taken into account. For a larger system with $N=100$, the estimated gate count increases to $G_{\rm CNOT}^{(1)}\simeq1.3\times10^6$, leading to $p_{\rm ph}\ll 3.9\times10^{-7}$ within the same error model.

Equation~\eqref{eq:NISQ_fault_probability} should not be interpreted as a device-independent bound on the error of the measured observable; it merely estimates the probability that at least one elementary gate fault occurs under a simple independent-noise model.

For actual noisy devices, the native elementary gate set, qubit connectivity, parallelization, coherent and correlated errors, and circuit compilation can substantially modify the practical resource requirement. In particular, for small-scale demonstrations of the quantum Mpemba effect, a device-specific optimization of the trade-off among the number of qubits, circuit depth, statistical error, and systematic error may lead to substantially more efficient implementations. We therefore leave a detailed hardware-specific analysis of noisy devices for future work.

For fault-tolerant quantum computing (FTQC), Clifford gates such as CNOT gates are generally expected to be considerably less expensive than non-Clifford operations and are often treated as a subleading resource in first-pass logical resource estimates. Non-Clifford gates can be implemented using magic states, whose preparation and distillation can constitute a substantial space--time overhead \cite{Bravyi:2004isx}. We therefore focus on the $T$-gate cost associated with synthesizing the arbitrary single-qubit rotations appearing in the Trotter circuit.

For a single copy of the time evolution, the number of $R_z$ rotations is
\begin{align}\label{eq:Rz_count}
    L_{R_z}^{(1)}
    =
    \frac{M}{2}
    \left(
    N^2+5N-2
    \right)
    +
    \frac{N^2}{2}
    -
    \frac{N}{2}
    +1\ ,
    \qquad
    N=\text{even}.
\end{align}
Again, we neglect the $\mathcal{O}(N_A)$ measurement-circuit overhead when estimating the leading time-evolution cost.

For a generic single-qubit $R_z$ rotation, deterministic ancilla-free synthesis over the Clifford+$T$ gate set requires a typical $T$ count
\begin{align}
    n_T(\delta)
    =
    3\log_2\left(\frac{1}{\delta}\right)
    +
    \mathcal{O}
    \left(
    \log\log\frac{1}{\delta}
    \right)
\end{align}
to achieve approximation error $\delta$ \cite{ross2016,selinger2014}.\footnote{Lower expected $T$ counts are possible in more permissive models that allow ancillary qubits and adaptive mid-circuit measurements, or probabilistic mixtures of Clifford+$T$ circuits \cite{Kliuchnikov:2022rqw,Morisaki:2025mwy}.}

The rotation-synthesis error must be included in the total algorithmic error budget. We therefore separate the algorithmic error into the Trotterization error and the rotation-synthesis error,
\begin{align}
    \epsilon_{\text{alg}}
    \lesssim
    \epsilon_{\text{Trot}}
    +
    \epsilon_{\text{syn}}.
\end{align}
For a single-copy circuit, a simple uniform allocation based on the triangle inequality assigns
\begin{align}
    \delta
    =
    \frac{\epsilon_{\text{syn}}}{L_{R_z}^{(1)}}
\end{align}
to each $R_z$ rotation. The corresponding single-copy $T$ count is therefore approximately
\begin{align}\label{eq:T_count}
    G_T^{(1)}
    &\simeq
    L_{R_z}^{(1)}
    \left[
    3\log_2
    \left(
    \frac{L_{R_z}^{(1)}}{\epsilon_{\text{syn}}}
    \right)
    +
    \mathcal{O}
    \left(
    \log\log
    \frac{L_{R_z}^{(1)}}{\epsilon_{\text{syn}}}
    \right)
    \right].
\end{align}
This estimate provides a transparent reference point rather than an optimized fault-tolerant compilation.

\begin{figure}[t]
    \centering
    \begin{subfigure}[b]{0.49\textwidth}
        \centering
        \includegraphics[width=\linewidth]{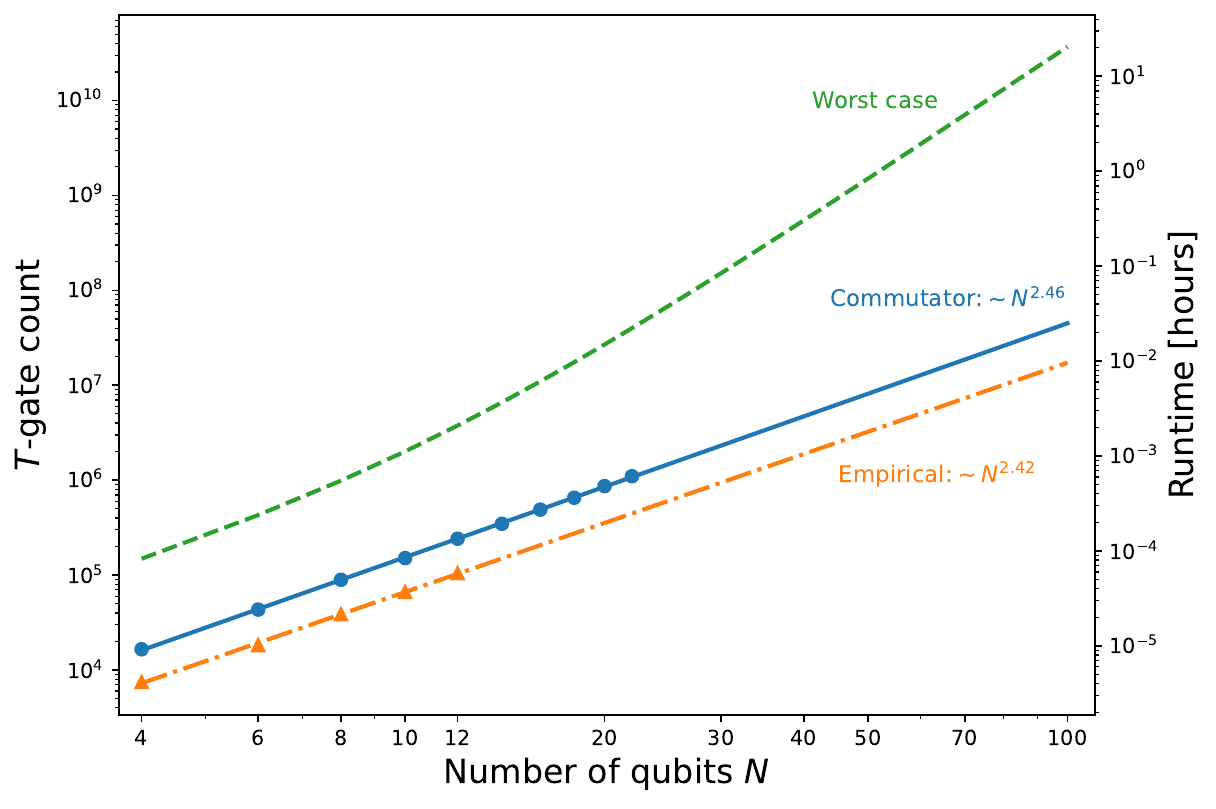}
        \caption*{$m=m_c$}
         \end{subfigure}
            \hfill
    \begin{subfigure}[b]{0.49\textwidth}
        \centering
        \includegraphics[width=\linewidth]{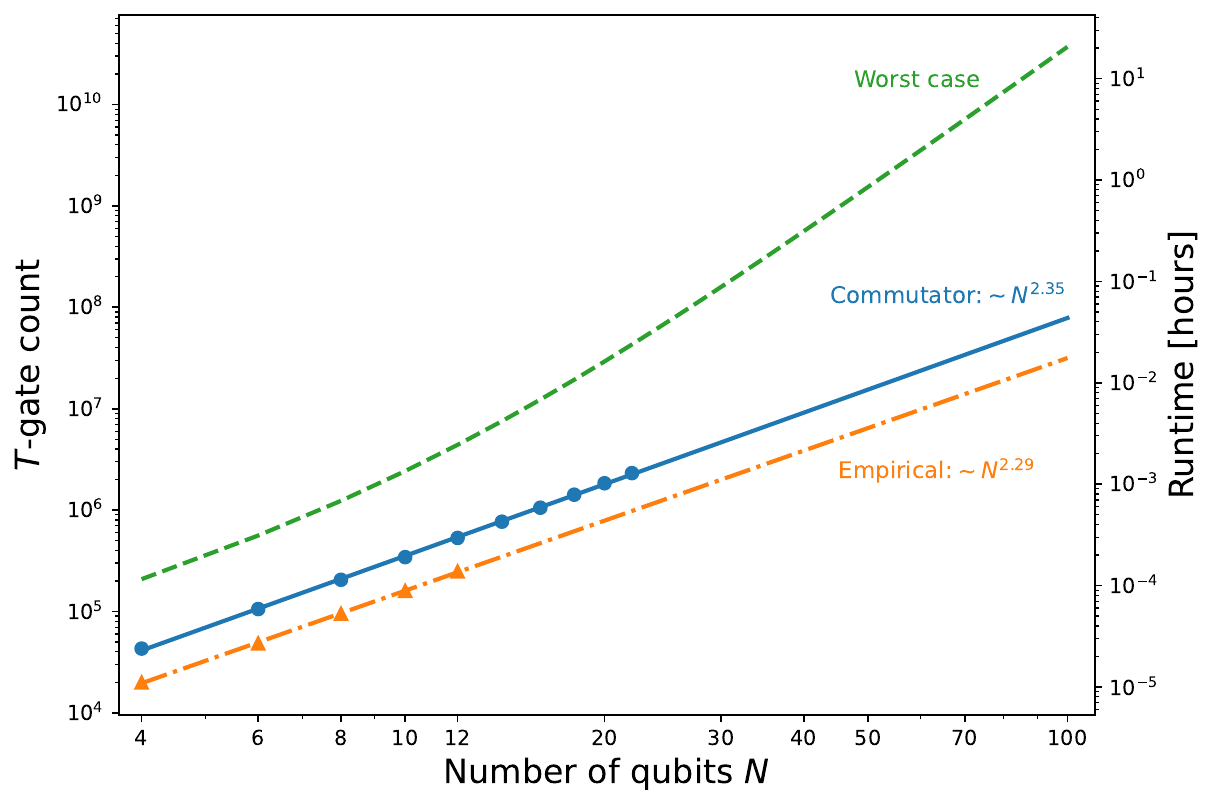}
        \caption*{$m/g=4$}
    \end{subfigure}
    \caption{
    Estimated single-copy $T$ count required to implement the Trotterized time evolution with target Trotter error $\epsilon_{\text{Trot}}=10^{-2}$ and rotation-synthesis error $\epsilon_{\text{syn}}=10^{-2}$. The left vertical axis shows the single-copy $T$ count, while the right vertical axis shows the corresponding runtime of the full two-copy protocol estimated from \eqref{eq:runtime}, assuming an aggregate logical $T$-gate throughput of $R_T=1\,\mathrm{MHz}$. The green, blue, and orange lines show the estimates obtained from the worst-case bound \eqref{eq:worst-case_bound}, the commutator bound \eqref{eq:the_number_of_Trotter_steps_commutator_bound}, and the empirical Trotter number \eqref{eq:M_empirical}, respectively. The left and right panels show the results for $m=m_c$ and $m/g=4$, respectively, and exhibit qualitatively similar behavior. The remaining parameters are fixed to $tg=1$, $\theta=\pi/3$, and $ag=1/4$ as a representative parameter set.
    }
    \label{fig:T_count}
\end{figure}

The $T$ count can be converted into an illustrative wall-clock time once a magic-state production and consumption rate is specified. Let $R_T$ denote the aggregate rate at which logical $T$ gates can be supplied and consumed by the computation.\footnote{This conversion is intended only as a throughput-based estimate. In an architecture-specific implementation, the achievable logical $T$-gate rate would depend on, among other factors, the number and layout of magic-state factories, the distillation protocol, the code distance, the surface-code cycle time, routing and lattice-surgery constraints, and the extent to which magic-state production and gate execution can be parallelized.} Since the complete SWAP-test protocol contains two copies, its total non-Clifford consumption is approximately $2G_T^{(1)}$. If both copies share a total available throughput $R_T$, the corresponding runtime is estimated as
\begin{align}\label{eq:runtime}
    t_{\text{run}}
    \simeq
    \frac{2G_T^{(1)}}{R_T}.
\end{align}

Fig.~\ref{fig:T_count} shows the corresponding single-copy $T$ count obtained from the commutator-bound and empirical estimates of $M$, together with the illustrative runtime obtained from the expression above. Here we use the representative choice $\epsilon_{\text{Trot}}=\epsilon_{\text{syn}}=10^{-2}$. For the runtime estimate, we assume an aggregate logical $T$-gate throughput of $R_T=1\,\mathrm{MHz}$ \cite{Litinski_2019,Sakamoto_2024}. This estimate should not be interpreted as a device-level prediction: it effectively assumes that the required non-Clifford gates can be supplied at the prescribed average rate and does not explicitly model magic-state factory sizing, $T$-depth constraints, scheduling, routing, or other architecture-dependent bottlenecks.

A fully architecture-specific fault-tolerant resource estimate would require specifying a hardware architecture and jointly optimizing the logical error rate, code distance, physical-qubit count, magic-state distillation factories, and circuit scheduling. In particular, the wall-clock time can depend strongly on how many magic-state factories are operated in parallel and whether their output rate is sufficient to saturate the non-Clifford demand of the circuit. Such an analysis is beyond the scope of the present work. Our aim here is instead to provide an algorithm-level resource estimate that makes explicit how the cost of implementing the real-time evolution required to study the quantum Mpemba effect scales with the system size, simulation time, and target accuracy.

%%%%%%%%%%%%%%%%%%%%%%%%%%%%%%%%%%%%
%%%%%%%%%%%%%%%%%%%%%%%%%%%%%%%%%%%%
\subsection{Early-FTQC resource estimation}
\label{earlyFTQC_resource}
%%%%%%%%%%%%%%%%%%%%%%%%%%%%%%%%%%%%
%%%%%%%%%%%%%%%%%%%%%%%%%%%%%%%%%%%%

Finally, we briefly discuss the resources required in an early fault-tolerant quantum computing (early-FTQC) regime. As a representative architecture, we consider STAR ver.~3 proposed in Ref.~\cite{toshio2026}. STAR ver.~3 is a partially fault-tolerant architecture in which Clifford operations are protected by quantum error correction, while analog non-Clifford rotations are implemented using STAR-magic mutation (SMM) and their residual errors are suppressed by error mitigation. Following Ref.~\cite{toshio2026}, we estimate the resources for Hamiltonian simulation using Time Evolution by Probabilistic Angle Interpolation (TE-PAI)~\cite{Kiumi:2024amh}.

TE-PAI constructs an unbiased estimator of the time-evolution superoperator by sampling random circuits, thereby avoiding the finite-step Trotter error considered in Section~\ref{subsec:error_bound_scaling}. For a Hamiltonian written as
\begin{align}
    H=\sum_{k=1}^{L}c_k\sigma_k,
    \qquad
    \lambda:=\sum_{k=1}^{L}|c_k|,
\end{align}
where $\sigma_k$ are Pauli strings, the leading resources are determined by the Hamiltonian $L_1$ norm $\lambda$, the evolution time $t$, and a tunable parameter $Q$ controlling the trade-off between circuit depth and sampling overhead. Choosing the TE-PAI rotation angle as $\Delta=2\arctan[Q/(2\lambda t)]$, the sampling overhead is $e^Q$, while the expected number of non-identity Pauli-rotation gates in a single-copy TE-PAI circuit is
\begin{align}\label{eq:TEPAI_gate_count}
    N_{\mathrm{gate}}^{(1)}
    =
    \frac{2(\lambda t)^2}{Q}+Q.
\end{align}
We refer to Refs.~\cite{Kiumi:2024amh,toshio2026} for the detailed construction of the TE-PAI estimator.

For the lattice Schwinger model considered here, the $L_1$ norm is given by
\begin{align}\label{eq:Schwinger_L1_bound}
\begin{aligned}
    \lambda
    =
    \frac{J}{12}(N-2)(N-1)N
    +
    \sum_{n=1}^{N-1}
    \left|w - (-1)^n \frac{m_{\text{lat}}}{2}\sin \theta \right|\\
    +
    \frac{1}{2}
    \sum_{n=1}^{N}
    \left|
    (-1)^n m_{\text{lat}}\cos \theta 
    -
    J \left(
    \left\lceil \frac{N-1}{2} \right\rceil - \left\lfloor \frac{n}{2} \right\rfloor
    \right)
    \right|
\end{aligned}
\end{align}

Thus, the leading behavior is $\lambda=\mathcal{O}(N^3)$. 
Since the second-R\'enyi
%R'enyi 
protocol requires two independently evolved copies, we independently sample a TE-PAI circuit for each copy. 
At the level of the two-copy Hamiltonian, the corresponding $L_1$ norm is twice the single-copy value. In our resource estimate, however, we treat the two copies as independent TE-PAI simulations and assign the same parameter $Q$ to each copy. The total expected gate count is therefore $2N_{\rm gate}^{(1)}$, while the combined TE-PAI sampling overhead is $e^{2Q}$.

We next incorporate the residual errors of the analog rotations in STAR ver.~3. Following Ref.~\cite{toshio2026}, the accumulated error of a single-copy TE-PAI circuit can be approximated in the relevant small-angle regime as $P_{\mathrm{total}}^{(1)}\simeq\alpha_{\mathrm{RUS}}\lambda t\,p_{\mathrm{ph}}$, where $p_{\mathrm{ph}}$ is the physical error rate and $\alpha_{\mathrm{RUS}}$ characterizes the repeat-until-success implementation of the analog rotations. For the two-copy protocol, we therefore use
\begin{align}\label{eq:Ptotal_two_copy}
    P_{\mathrm{total}}^{(2)}
    \simeq
    2\alpha_{\mathrm{RUS}}\lambda t\,p_{\mathrm{ph}}.
\end{align}
Since probabilistic error cancellation introduces a sampling overhead approximately proportional to $e^{4P_{\mathrm{total}}^{(2)}}$, we regard
\begin{align}\label{eq:STAR_feasibility}
    P_{\mathrm{total}}^{(2)}
    \lesssim1
\end{align}
as a rough criterion for a resource-efficient early-FTQC implementation. For example, taking $ag=1/4$, $m/g=4$, $tg=1$, $\theta=\pi/3$, $\alpha_{\mathrm{RUS}}=0.1$, and $p_{\mathrm{ph}}=10^{-3}$, this condition corresponds to $\lambda t \lesssim5\times10^3$, which allows approximately $N\lesssim 79$ qubits according to \eqref{eq:Schwinger_L1_bound}. This value should be interpreted as an indicative estimate rather than a sharp hardware limit.

To estimate the spacetime resources, we follow the STAR ver.~3 resource model of Ref.~\cite{toshio2026}. For a single $N$-qubit register, the fast-block layout requires
\begin{align}
    N_{\mathrm{patch}}^{(1)}
    =
    \left\lceil
    2N+\sqrt{8N}+11
    \right\rceil
\end{align}
surface-code patches, where each patch represents an encoded surface-code region used for a logical qubit or an ancillary workspace. To execute the two copies in parallel, we conservatively take $N_{\mathrm{patch}}^{(2)}=2N_{\mathrm{patch}}^{(1)}$. The logical error probability per surface-code cycle is approximated as
\begin{align}
    p_L(p_{\rm ph},d)
    =
    0.1
    \left(
    100p_{\rm ph}
    \right)^{(d+1)/2},
\end{align}
where $d$ is the code distance of the surface codes. 
In the STAR ver.~3 resource model, one lattice-surgery clock corresponds to $d$ surface-code cycles for code distance $d$. We denote by $C_{\rm SMM}$ the execution cost of one STAR-magic-mutation operation in units of these clocks and take $C_{\rm SMM}=3$, following Ref.~\cite{toshio2026}. Because the two copies are evolved in parallel, the circuit duration entering the logical-error estimate is approximately that of a single TE-PAI circuit, whereas the number of simultaneously allocated patches is doubled. We therefore choose the smallest code distance $d$ satisfying
\begin{align}\label{eq:STAR_code_distance_two}
    p_L(p_{\rm ph},d)^{-1}
    \geq
    100\,
    d\,
    N_{\rm gate}^{(1)}
    C_{\rm SMM}
    N_{\rm patch}^{(2)}.
\end{align}
Here, $dN_{\rm gate}^{(1)}C_{\rm SMM}$ estimates the number of surface-code cycles during the time evolution, while $N_{\rm patch}^{(2)}$ accounts for the number of code patches exposed to logical errors. The factor of $100$ is introduced so that the accumulated error from fault-tolerant Clifford operations remains below approximately $1\%$, following the conservative prescription of Ref.~\cite{toshio2026}. The resulting number of physical qubits is estimated as
\begin{align}
    N_{\rm phys}
    \simeq
    2d^2N_{\rm patch}^{(2)}.
\end{align}
The additional resources required for the controlled-SWAP and symmetry-resolving operations scale with $N_A$ and are neglected in this leading estimate.

To convert the circuit duration into physical time, we denote by $\tau_{\rm cyc}$ the duration of one surface-code cycle and take $\tau_{\rm cyc}=1\,\mu\mathrm{s}$ following Ref.~\cite{toshio2026}. Because the two copies are evolved in parallel, the single-shot runtime is approximated by
\begin{align}
    t_{\rm shot}
    =
    N_{\rm gate}^{(1)}
    C_{\rm SMM}
    d\,\tau_{\rm cyc}.
\end{align}
Including the TE-PAI sampling overhead and PEC overhead, the total runtime required to estimate the two-copy observable to additive precision $\epsilon_{\text{alg}}$ is then estimated as
\begin{align}\label{eq:STAR_total_runtime}
    t_{\rm total}
    =
    t_{\rm shot}
    \frac{
    e^{2Q}
    e^{4P_{\rm total}^{(2)}}
    }{\epsilon_{\text{alg}}^2}.
\end{align}
Here, $\epsilon_{\text{alg}}$ denotes the target additive precision of the TE-PAI estimator, rather than the operator-norm error used for deterministic Trotterized evolution.
For the numerical resource estimates shown below, we take $Q=1$ and $\epsilon_{\text{alg}}=0.01$.

\begin{figure}[t]
    \centering
    \begin{subfigure}[b]{0.48\textwidth}
        \centering
        \includegraphics[width=\linewidth]{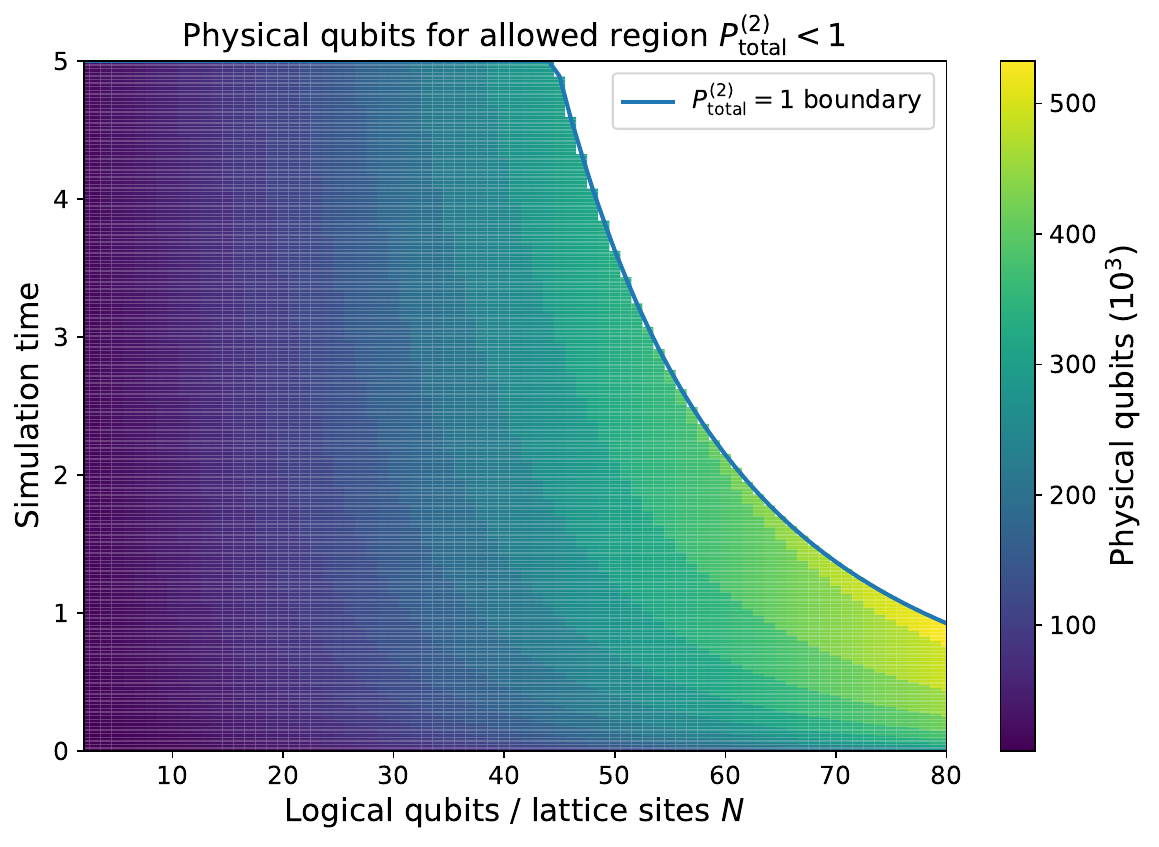}
        \caption*{(a) Number of physical qubits}
    \end{subfigure}
    \hfill
    \vspace{0.3cm}
    \begin{subfigure}[b]{0.48\textwidth}
        \centering
        \includegraphics[width=\linewidth]{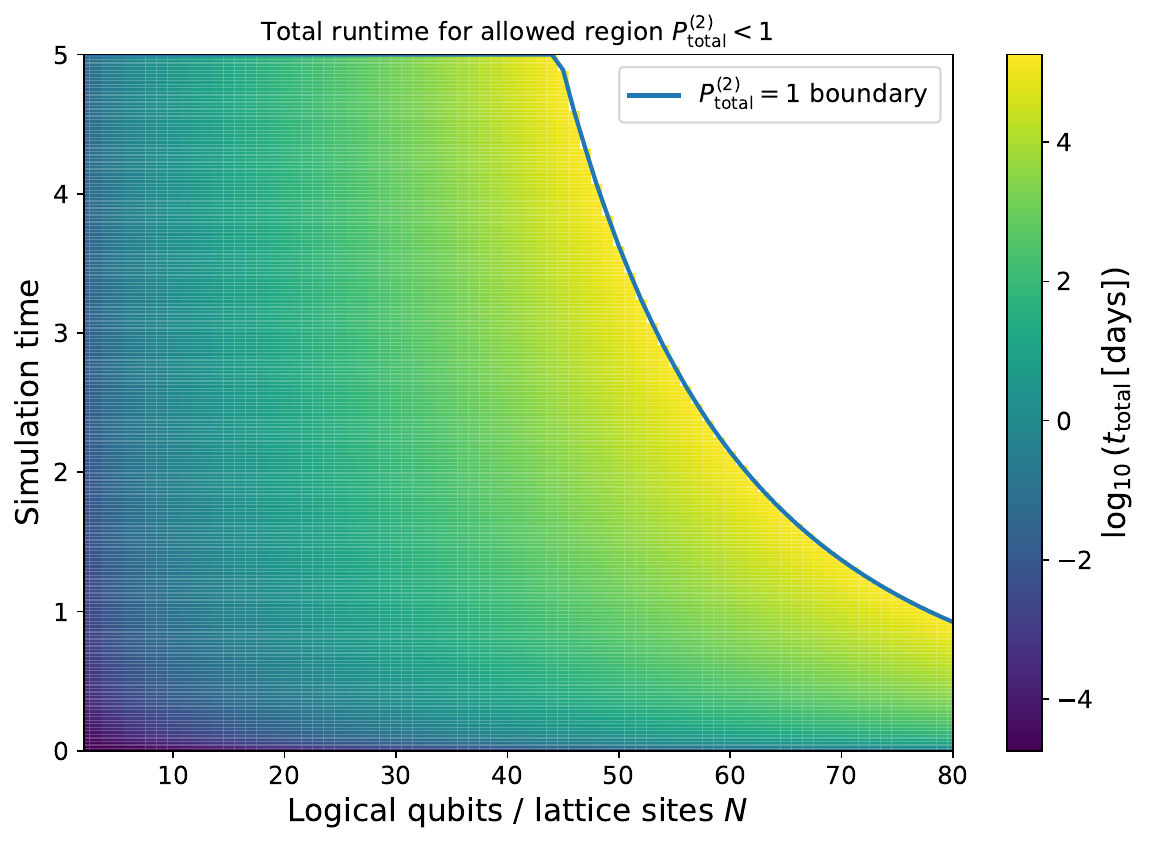}
        \caption*{(b) Total runtime}
    \end{subfigure}
    % \begin{subfigure}[b]{0.48\textwidth}
    %     \centering
    %     \includegraphics[width=\linewidth]{codedistance_Starv3.pdf}
    %     \caption*{(b)}
    % \end{subfigure}
    % \vspace{0.3cm}
    % \begin{subfigure}[b]{0.5\textwidth}
    %     \centering
    %     \includegraphics[width=\linewidth]{runtime_Starv3.pdf}
    %     \caption*{(c)}
    % \end{subfigure}
    \caption{
    Early-FTQC resource estimates for the two-copy Schwinger-model simulation using TE-PAI on STAR ver.~3. (a) Estimated number of physical qubits, and (b) total runtime including the TE-PAI sampling and PEC overheads. We use $Q=1$, $\epsilon_{\text{alg}}=0.01$, $C_{\mathrm{SMM}}=3$, $\tau_{\mathrm{cyc}}=1\,\mu\mathrm{s}$, $p_{\mathrm{ph}}=10^{-3}$, and $\alpha_{\mathrm{RUS}}=0.1$. The two TE-PAI circuits are independently sampled and assumed to be executed in parallel. Only parameter sets $(tg,N)$ satisfying $P_{\rm total}^{(2)}\lesssim1$ are shown, with the physical parameters fixed to $ag=1/4$, $m/g=4$, and $\theta=\pi/3$.
    }
    \label{fig:resource_estimation_Starv3}
\end{figure}

Fig.~\ref{fig:resource_estimation_Starv3} summarizes the resulting physical-qubit requirement and total runtime. These estimates employ a naive independent TE-PAI sampling scheme for the two copies and are intended as a simple indication of the resources required in an early-FTQC architecture rather than an optimized implementation.

Finally, we briefly compare the resource requirements of the three approaches considered in this section. For NISQ devices, current two-qubit gate error rates on the order of $10^{-3}$--$10^{-4}$ suggest that small-scale implementations may be feasible, while extending uncorrected simulations to $N=50$ would require $p_{\rm ph}\ll 2.2\times10^{-6}$ within our simplified noise model. In contrast, the early-FTQC approach based on STAR ver. 3 offers a possible route toward larger system sizes even with a relatively high physical error rate of $p_{\rm ph}=10^{-3}$, at the cost of substantial physical-qubit overhead. For example, at $N=50$ and $tg=1$, the estimated resource requirement is approximately $2.8\times10^5$ physical qubits under the assumptions used in Fig.~\ref{fig:resource_estimation_Starv3}. This system size is already challenging for direct classical state-vector simulations. Nevertheless, as discussed above, the sampling overhead associated with probabilistic error cancellation restricts the accessible system sizes to approximately $N\lesssim79$ for $p_{\rm ph}=10^{-3}$ and $tg=1$ under our chosen criterion. If large-scale fault-tolerant quantum computers become available, they could provide a further route toward simulations of larger systems. For illustration, using the empirical Trotter-step estimate at $N=50$, we obtain a single-copy T count of approximately $8.5\times10^6$, corresponding to an estimated single-shot runtime for the two-copy circuit of about $17$ seconds at an assumed logical T-gate throughput of $1$ MHz. These estimates illustrate the different resource constraints of the three approaches, rather than establishing a direct performance ranking.

%%%%%%%%%%%%%%%%%%%%%%%%%%%%%%%%%%%%
%%%%%%%%%%%%%%%%%%%%%%%%%%%%%%%%%%%%
%%%%%%%%%%%%%%%%%%%%%%%%%%%%%%%%%%%%
\section{Summary and Discussion}
\label{sec:summary}
%%%%%%%%%%%%%%%%%%%%%%%%%%%%%%%%%%%%
%%%%%%%%%%%%%%%%%%%%%%%%%%%%%%%%%%%%
%%%%%%%%%%%%%%%%%%%%%%%%%%%%%%%%%%%%

In this work, we constructed a quantum measurement protocol for estimating the second R\'enyi entanglement asymmetry by combining the SWAP test with a charge-resolving procedure based on randomized symmetry rotations. For a fixed additive accuracy $\epsilon$, the required number of measurements scales as $N_{\mathrm{shot}}=\mathcal{O}(\epsilon^{-2})$ without explicit dependence on the total system size. This favorable measurement complexity makes the protocol particularly attractive for studying symmetry-restoration dynamics in quantum many-body systems and lattice quantum field theories, although the circuit resources required for state preparation and real-time evolution generally grow with the system size. As a representative application, we investigated the quantum Mpemba effect in the lattice Schwinger model for both spin-product benchmark states and fermion-parity-preserving initial states using classical state-vector simulations. We further estimated the circuit resources required to implement the protocol on noisy quantum devices, fault-tolerant quantum computers, and a representative early-FTQC architecture. Taken together, these results connect the proposed measurement protocol to a concrete nonequilibrium field-theory application and provide quantitative resource estimates for its implementation on quantum hardware.

Several directions for future work are worth exploring:
\begin{itemize}

    \item {\bf Generalization to higher-order R\'enyi entanglement asymmetries and the $n\to1$ limit}\\
    In this work, we focused on the second R\'enyi entanglement asymmetry $\Delta S_A^{(2)}$. The protocol can be generalized to higher integer orders $\Delta S_A^{(n)}$ by introducing $n$ copies of the system and implementing the corresponding cyclic-permutation measurements. It would be interesting to investigate whether information obtained for several integer values of $n$ can be used to infer the von Neumann entanglement asymmetry $\Delta S_A$, defined in terms of the relative entropy. Such an analytic continuation to the $n\to1$ limit is generally nontrivial and requires further investigation.

    \item {\bf Hamiltonian-simulation algorithms beyond product formulas}\\
    In this work, we employed the Suzuki--Trotter decomposition to implement the real-time evolution. In the fault-tolerant regime, alternative Hamiltonian-simulation algorithms, including linear-combination-of-unitaries methods \cite{Berry:2014ivo} and approaches based on quantum singular value transformation \cite{Gilyen:2018khw}, can provide more favorable asymptotic complexity for suitable Hamiltonians and access models. These approaches typically require additional ancilla qubits and more elaborate circuit constructions. It would therefore be interesting to compare their practical resource requirements with the product-formula implementation considered here in the context of the quantum Mpemba effect.

    \item {\bf Large-volume and continuum limits of the lattice Schwinger model}\\
    From the field-theory perspective, it would be valuable to investigate the quantum Mpemba effect in the lattice Schwinger model at larger system sizes and toward the continuum limit $ag\to0$. The present state-vector simulations are limited to systems of up to $18$ qubits in our work. More scalable classical approaches, such as tensor-network methods, may provide access to substantially larger systems and longer real-time evolution \cite{Paeckel:2019yjf,Florio:2024aix,Papaefstathiou:2024zsu,Jeyaretnam:2024tkj}. Such studies would also help clarify the finite-size effects observed in the late-time entanglement asymmetry and determine whether the quantum Mpemba behavior persists in the large-volume and continuum limits.

    \item {\bf Extensions to other symmetries and quantum field theories}\\
    Although we focused on a global $U(1)$ symmetry in the lattice Schwinger model as an illustrative example, the general strategy of combining multi-copy measurements with symmetry-twirling procedures can potentially be extended to other symmetry groups and many-body models. The precise implementation depends on how the corresponding symmetry action and twirling operation can be realized on a quantum computer and is therefore model and symmetry dependent. Recent studies have also investigated connections between entanglement asymmetry and chiral anomalies \cite{Florio:2025xup,Sierra:2026eev}. Exploring the real-time dynamics of entanglement asymmetry in lattice models with anomalous or more general symmetry structures would be an interesting direction for future work.

\end{itemize}

\section*{Acknowledgments}
We thank Tomoya Hayata for valuable discussions and important comments, which helped us substantially improve the analysis presented in this work. 
We also thank Dongwook Ghim for discussion on the mass shift formula in the lattice Schwinger model. 
%We also thank Dongwook Ghim for kindly sharing his ongoing work on the mass shift formula in the lattice Schwinger model. 
H.F. also thanks Yutaro Akahoshi for introducing him
%me 
to K.F. and for providing the opportunity to collaborate. 
The work of H.F. was supported by a Grant-in-Aid for JSPS fellows (Grant No.~24KJ1637).
M.H. is supported by JST CREST Grant Number JPMJCR24I3, JSPS Grant-in-Aid for Transformative Research Areas (A) ``Extreme Universe'' JP21H05190 [D01], JSPS KAKENHI Grant Number JP22H01222 and the Royal Society grants ICA/R2/242058 and IEC/R3/243103.

%%%%%%%%%%%%%%%%%%%%%%%%%%%%%%%%%%%%%%
%%%%%%%%%%%%%%%%%%%%%%%%%%%%%%%%%%%%%%
%%%%%%%%%%%%%%%%%%%%%%%%%%%%%%%%%%%%%%
\appendix

%%%%%%%%%%%%%%%%%%%%%%%%%%%%%%%%%%%%%%
%%%%%%%%%%%%%%%%%%%%%%%%%%%%%%%%%%%%%%
%%%%%%%%%%%%%%%%%%%%%%%%%%%%%%%%%%%%%%
\appendix

%%%%%%%%%%%%%%%%%%%%%%%%%%%%%%%%%%%%%%
%%%%%%%%%%%%%%%%%%%%%%%%%%%%%%%%%%%%%%
\section{Decomposition into Elementary Gates}
\label{sec:gate_decomposition}
%%%%%%%%%%%%%%%%%%%%%%%%%%%%%%%%%%%%%%
%%%%%%%%%%%%%%%%%%%%%%%%%%%%%%%%%%%%%%

In this appendix, we provide the elementary-gate decompositions used in the quantum circuits and resource estimates discussed in the main text. We use the gate set
\begin{align}
\begin{aligned}
    H
    =
    \frac{1}{\sqrt{2}}
    \begin{pmatrix}
        1 & 1 \\
        1 & -1
    \end{pmatrix},
    \quad
    S
    &=
    \begin{pmatrix}
        1 & 0 \\
        0 & \i
    \end{pmatrix},
    \quad
    \mathrm{CNOT}
    =
    \begin{quantikz}[row sep=0.3cm, column sep=0.3cm]
        & \ctrl{1} & \qw \\
        & \targ{}  & \qw
    \end{quantikz}
    =
    \begin{pmatrix}
        1 & 0 & 0 & 0 \\
        0 & 1 & 0 & 0 \\
        0 & 0 & 0 & 1 \\
        0 & 0 & 1 & 0
    \end{pmatrix},
    \\
    R_Z(\alpha)
    &:=
    e^{-\i\alpha Z/2}
    =
    \begin{pmatrix}
        e^{-\i\alpha/2} & 0 \\
        0 & e^{\i\alpha/2}
    \end{pmatrix}.
\end{aligned}
\end{align}
Here, $H$, $S$, and CNOT are Clifford gates, whereas $R_Z(\alpha)$ is generally non-Clifford for an arbitrary angle $\alpha$. In a fault-tolerant Clifford+$T$ implementation, an arbitrary $R_Z(\alpha)$ rotation can be approximated to a prescribed accuracy using a sequence of Clifford and $T$ gates\cite{ross2016,selinger2014}.

The following decompositions are useful for implementing the initial-state preparation and real-time evolution:
\begin{align}
    R_Y(\alpha)
    &:=
    e^{-\i\alpha Y/2}
    =
    \begin{quantikz}
        & \gate{S^{\dagger}}
        & \gate{H}
        & \gate{R_Z(\alpha)}
        & \gate{H}
        & \gate{S}
        & \qw
    \end{quantikz},
    \label{eq:R_Y}
\end{align}
\begin{align}
    R_{ZZ}^{(n,m)}(\alpha)
    &:=
    e^{-\i\alpha Z_n Z_m/2}
    =
    \begin{quantikz}[row sep=0.3cm, column sep=0.3cm]
        \lstick{$n$} & \ctrl{1} & \qw                    & \ctrl{1} & \qw \\
        \lstick{$m$} & \targ{}  & \gate{R_Z(\alpha)}     & \targ{}  & \qw
    \end{quantikz},
    \label{eq:R_ZZ}
\end{align}
\begin{align}
    R_{YY}^{(n,m)}(\alpha)
    &:=
    e^{-\i\alpha Y_nY_m/2}
    =
    \begin{quantikz}[row sep=0.3cm, column sep=0.3cm]
        \lstick{$n$}
        & \gate{S^{\dagger}}
        & \gate{H}
        & \gate[2]{R_{ZZ}^{(n,m)}(\alpha)}
        & \gate{H}
        & \gate{S}
        & \qw \\
        \lstick{$m$}
        & \gate{S^{\dagger}}
        & \gate{H}
        &
        & \gate{H}
        & \gate{S}
        & \qw
    \end{quantikz},
    \label{eq:R_YY}
\end{align}
and
\begin{align}
    R_{XXYY}^{(n,m)}(\alpha)
    &:=
    e^{-\i\alpha(X_nX_m+Y_nY_m)/2}
    %\nonumber\\
    &=
    \begin{quantikz}[row sep=0.3cm, column sep=0.3cm]
        \lstick{$n$}
        & \ctrl{1}
        & \gate{H}
        & \ctrl{1}
        & \gate{R_Z(\alpha)}
        & \ctrl{1}
        & \gate{H}
        & \ctrl{1}
        & \qw \\
        \lstick{$m$}
        & \targ{}
        & \qw
        & \targ{}
        & \gate{R_Z(-\alpha)}
        & \targ{}
        & \qw
        & \targ{}
        & \qw
    \end{quantikz}.
    \label{eq:R_XXYY}
\end{align}
The Controlled-SWAP gate can be decomposed as
\begin{align}\label{eq:Controlled_SWAP}
    \text{Controlled-SWAP}
    =
    \begin{quantikz}%[row sep=0.3cm, column sep=0.3cm]
        &\ctrl{1} &\qw  \\
        &\swap{1} &\qw  \\
        &\targX{} &\qw
    \end{quantikz}
    =
    \begin{quantikz}%[row sep=0.3cm, column sep=0.3cm]
        &\qw       &\ctrl{1}  &\qw &\qw  \\
        &\targ{}   &\ctrl{1} &\targ{}  &\qw \\
        &\ctrl{-1}  &\targ{}   &\ctrl{-1} &\qw  
    \end{quantikz}\ ,
\end{align}
Here, $X_n$, $Y_n$, and $Z_n$ denote Pauli operators acting on the $n$th qubit.

For the initial states, we consider the spin-product and fermion-parity-preserving states
\begin{align}
\begin{aligned}
    \ket{\psi_1(\phi)}
    &=
    e^{-\i\frac{\phi}{2}\sum_{n=1}^{N}Y_n}
    \ket{\Omega}
    =
    \left[
        \prod_{n=1}^{N}
        R_Y^{(n)}(\phi)
    \right]
    \ket{\Omega},
    \\
    \ket{\psi_2(\phi)}
    &=
    e^{-\i\frac{\phi}{2}\sum_{n=1}^{N-1}Y_nY_{n+1}}
    \ket{\Omega}
    =
    \left[
        \prod_{n=1}^{N-1}
        R_{YY}^{(n,n+1)}(\phi)
    \right]
    \ket{\Omega}.
\end{aligned}
\end{align}
All single-qubit generators $Y_n$ commute with one another. Likewise $[Y_n Y_{n+1},Y_m Y_{m+1}]=0$ for arbitrary $n$ and $m$. Therefore, the two initial states can be prepared exactly by applying the decompositions in \eqref{eq:R_Y} and \eqref{eq:R_YY}, respectively.

For the real-time evolution, we employ the second-order product formula. Defining $\delta t=t/M$, a single Trotter step is written as
\begin{align}
    U_2(\delta t)
    &=
    e^{-\i\frac{\delta t}{2}(H_{ZZ}+H_Z)}
    e^{-\i\frac{\delta t}{2}H_{\mathrm{odd}}}
    e^{-\i\delta t H_{\mathrm{even}}}
    e^{-\i\frac{\delta t}{2}H_{\mathrm{odd}}}
    e^{-\i\frac{\delta t}{2}(H_{ZZ}+H_Z)},
    \label{eq:2nd_Trotter}
\end{align}
where
\begin{align}
\begin{aligned}
    H_{ZZ}
    &=
    \frac{J}{2}
    \sum_{1\le n<m\le N-1}
    (N-m) Z_nZ_m,
    \\
    H_Z
    &=
    \frac{m_{\mathrm{lat}}\cos\theta}{2}
    \sum_{n=1}^{N}(-1)^nZ_n
    -
    \frac{J}{2}
    \sum_{n=1}^{N-1}
    (n\bmod 2)
    \sum_{m=1}^{n}Z_m,
    \\
    H_{\mathrm{odd}}
    &=
    \frac{1}{2}
    \sum_{m=1}^{\lfloor N/2\rfloor}
    \left(
        w+\frac{m_{\mathrm{lat}}}{2}\sin\theta
    \right)
    \left(
        X_{2m-1}X_{2m}+Y_{2m-1}Y_{2m}
    \right),
    \\
    H_{\mathrm{even}}
    &=
    \frac{1}{2}
    \sum_{m=1}^{\lfloor(N-1)/2\rfloor}
    \left(
        w-\frac{m_{\mathrm{lat}}}{2}\sin\theta
    \right)
    \left(
        X_{2m}X_{2m+1}+Y_{2m}Y_{2m+1}
    \right).
\end{aligned}
\end{align}

Since all Pauli-$Z$ strings appearing in $H_{ZZ}+H_Z$ mutually commute, the corresponding exponential can be implemented exactly within each Trotter step using $R_Z$ and $R_{ZZ}$ rotations. Similarly, the interaction terms within $H_{\mathrm{odd}}$ mutually commute, as do those within $H_{\mathrm{even}}$, because they act on disjoint nearest-neighbor bonds. We therefore obtain
\begin{align}
\begin{aligned}
    e^{-\i\frac{\delta t}{2}H_{\mathrm{odd}}}
    &=
    \prod_{m=1}^{\lfloor N/2\rfloor}
    R_{XXYY}^{(2m-1,2m)}
    \left[
        \frac{\delta t}{2}
        \left(
            w+\frac{m_{\mathrm{lat}}}{2}\sin\theta
        \right)
    \right],
    \\
    e^{-\i\delta t H_{\mathrm{even}}}
    &=
    \prod_{m=1}^{\lfloor(N-1)/2\rfloor}
    R_{XXYY}^{(2m,2m+1)}
    \left[
        \delta t
        \left(
            w-\frac{m_{\mathrm{lat}}}{2}\sin\theta
        \right)
    \right].
\end{aligned}
\end{align}
Substituting \eqref{eq:R_XXYY} gives the corresponding decomposition into the elementary gate set.

When $M$ Trotter steps are concatenated, the last $(H_{ZZ}+H_Z)$ half-step of one Trotter step can be merged with the first half-step of the next. Consequently, the $(H_{ZZ}+H_Z)$ block is implemented $M+1$ times: the first and last applications have half angles, whereas the $M-1$ intermediate applications have full angles. The odd-bond and even-bond blocks are implemented $2M$ and $M$ times, respectively. Counting the elementary gates in these blocks gives the CNOT and $R_Z$ counts quoted in \eqref{eq:CNOT_count} and \eqref{eq:Rz_count} of the main text.

The gate counts above refer to a single copy of the Hamiltonian simulation. The second-R\'enyi protocol requires two copies, which doubles the total number of time-evolution gates, although the two copies can in principle be evolved in parallel. The additional SWAP-based measurement and the symmetry-resolving operation $( U_A )^x$ used in \eqref{eq:rho_A_S_U(1)}
%$U_A(\alpha)$ 
require only $\mathcal{O}(N_A)$ additional elementary gates. These contributions are therefore subleading compared with the $\mathcal{O}(MN^2)$ cost of the Hamiltonian simulation and are neglected in the leading resource estimates presented in the main text.

%%%%%%%%%%%%%%%%%%%%%%%%%%%%%%%%%%%%%%
%%%%%%%%%%%%%%%%%%%%%%%%%%%%%%%%%%%%%%
\section{Parameter Dependence of the Symmetry-Restoration Dynamics}
\label{sec:parameter_dependence}
%%%%%%%%%%%%%%%%%%%%%%%%%%%%%%%%%%%%%%
%%%%%%%%%%%%%%%%%%%%%%%%%%%%%%%%%%%%%%

In the main text, we used $N=18$, $N_A=4$, $\theta=\pi/3$, $m/g=4$, and $ag=1/4$ as a representative set of parameters for the lattice Schwinger model. In this appendix, we examine the dependence of the symmetry-restoration dynamics on the fermion mass, the $\theta$ angle, and the total system size. The purpose of this analysis is not to establish a detailed dynamical phase diagram, but rather to assess the robustness of the qualitative behavior discussed in the main text.

\begin{figure}
    \centering
    % TikZ code unchanged
    \begin{tikzpicture}
    %parameters
    \pgfmathsetmacro{\width}{5}
    \pgfmathsetmacro{\height}{2.5}
    %coordinates
    \coordinate (origin) at (0,0);
    \coordinate (p_middle) at (\width/2,\height/2);
    %axis
    \draw[->, line width = 1] ($(origin)+(0,-\height/5)$)--(0,\height*7/5) node[above] {$m/g$};
    \draw[->, line width = 1] ($(origin)+(-\width/6, 0)$)--(\width*7/6, 0) node[right] {$\theta$};
    \draw (origin) node[below left] {$O$};
    \draw (\width/2,0) node[below = 2mm] {$\pi$};
    \draw (\width,0) node[below = 1mm] {$2\pi$};
    %lines
    \draw[line width = 1] (\width/2,\height*5/4)--(p_middle);
    \draw[line width = 1, dashed] (\width/2,0)--(p_middle);
    \draw[line width = 1, dashed] (p_middle)--(0,\height/2) node[left] {$m_c/g$};
    %symbol
    \fill[red!80] (p_middle) circle [radius=0.1];
    \end{tikzpicture}
    \caption{
    Schematic zero-temperature phase structure of the continuum Schwinger model in the $\theta$--$m/g$ plane. The theory is periodic under $\theta\to\theta+2\pi$. The solid line at $\theta=\pi$ and $m>m_c$ denotes the first-order transition line, which terminates at the second-order critical point $(\theta,m/g)=(\pi,m_c/g)$, indicated by the red point.
    }
    \label{fig:phase_diagram}
\end{figure}
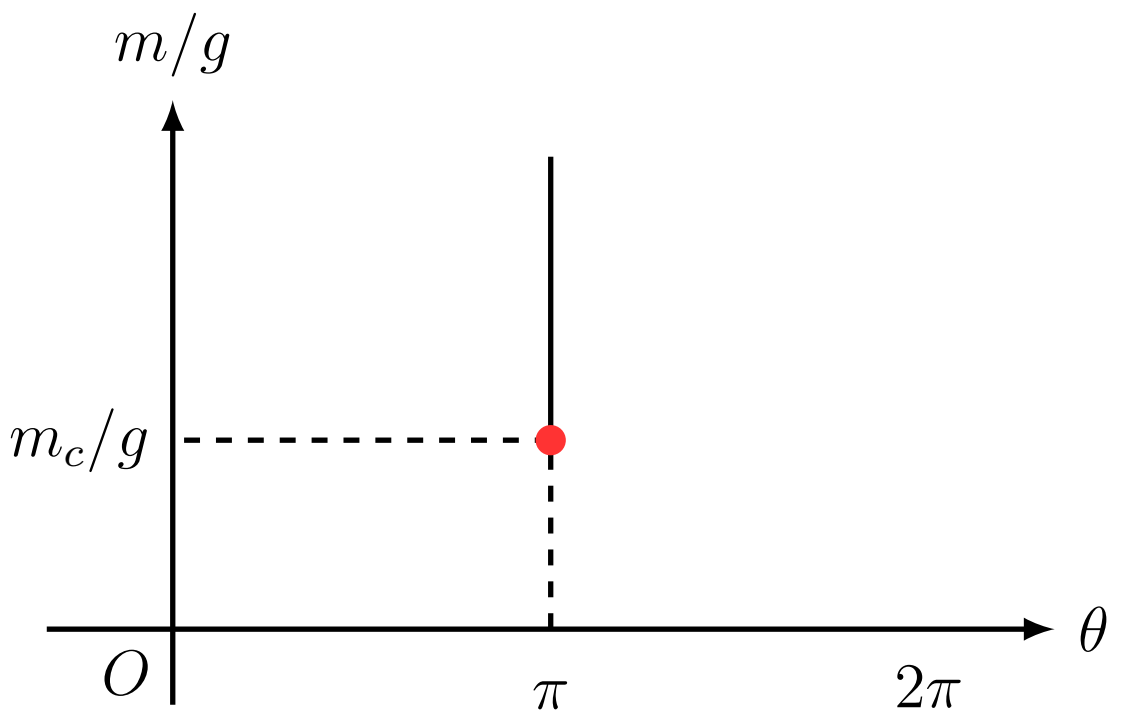

We first briefly recall the zero-temperature phase structure of the continuum Schwinger model. In the massless theory, the $\theta$ dependence can be removed by an anomalous chiral rotation, whereas for nonzero fermion mass the $\theta$ angle affects physical observables. The theory is $2\pi$ periodic in $\theta$, and $\theta=\pi$ is invariant under the corresponding discrete charge-conjugation transformation. At $\theta=\pi$, the massive Schwinger model exhibits a critical point at
\begin{align}
    \frac{m_c}{g}\simeq0.3335,
\end{align}
as determined by numerical studies \cite{Byrnes:2002nv,Cichy:2012rw,HAMER1982413}. For $m>m_c$, varying $\theta$ across $\pi$ gives a first-order transition associated with spontaneous breaking of the discrete symmetry, and the first-order line terminates at the second-order critical point $(\theta,m/g)=(\pi,m_c/g)$ \cite{Buyens:2017crb,Buyens:2015tea,Shimizu:2014fsa,Coleman:1975pw,Coleman:1976uz,Angelides:2023noe,Kaikov:2024acw}. A schematic representation of this continuum phase structure is shown in Fig.~\ref{fig:phase_diagram}.

\begin{figure}[t]
    \centering
    \begin{subfigure}[b]{0.4\textwidth}
    \centering
    \includegraphics[width=\linewidth]{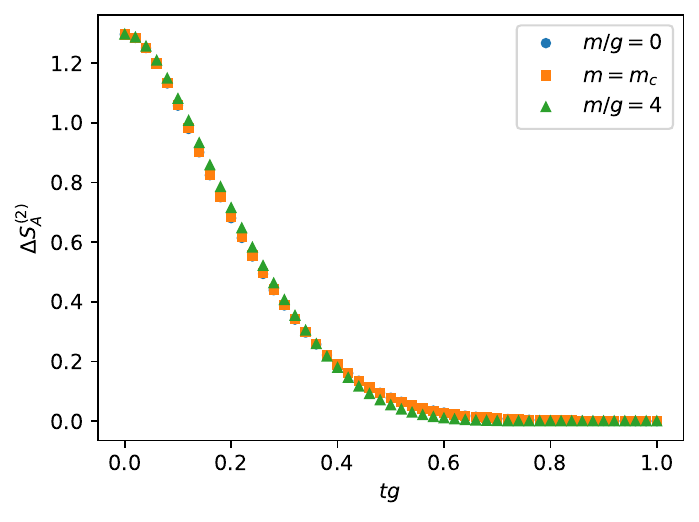}
    \caption*{$\theta=\pi/3$}
    \end{subfigure}
    \begin{subfigure}[b]{0.4\textwidth}
    \centering
    \includegraphics[width=\linewidth]{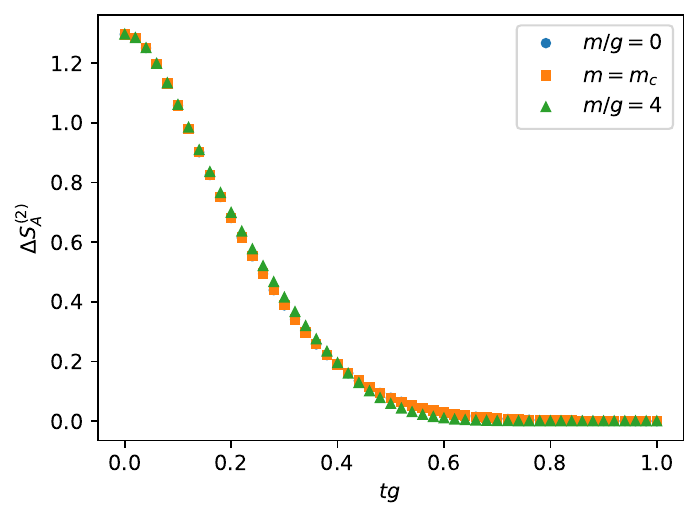}
    \caption*{$\theta=\pi$}
    \end{subfigure}
    \vspace{0.5cm}
    \begin{subfigure}[b]{0.4\textwidth}
    \centering
    \includegraphics[width=\linewidth]{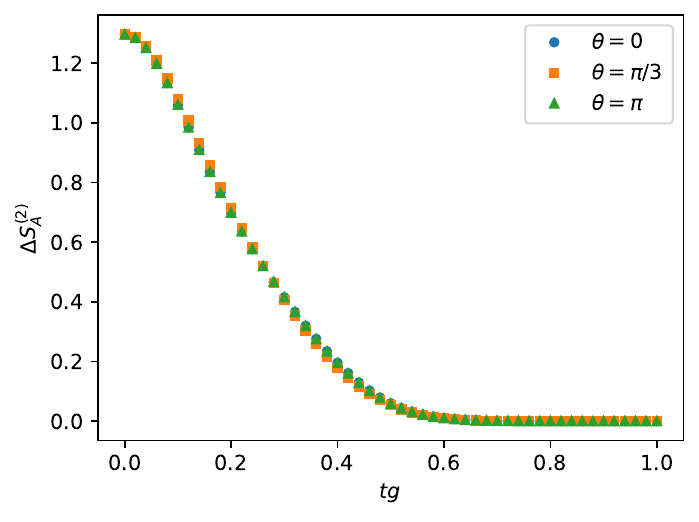}
    \caption*{$m/g=4$}
    \end{subfigure}
    \begin{subfigure}[b]{0.4\textwidth}
    \centering
    \includegraphics[width=\linewidth]{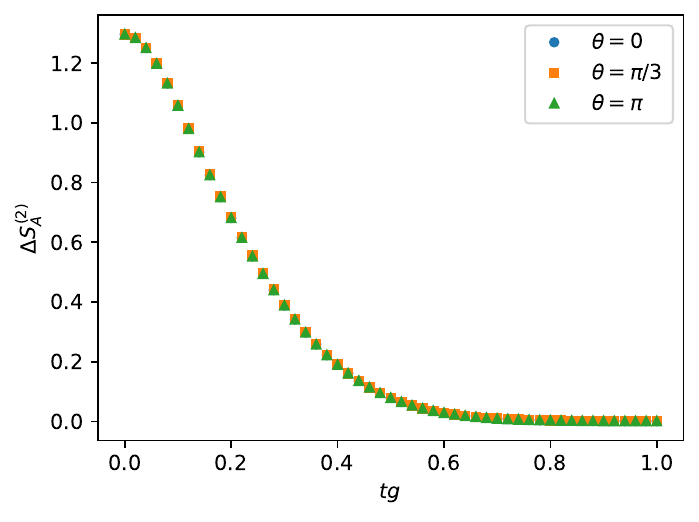}
    \caption*{$m=m_c$}
%    \caption*{$m_c/g=0.3335$}
    \end{subfigure}
    \caption{
    Dependence of the symmetry-restoration dynamics on the fermion mass and the $\theta$ angle for the spin-product N\'eel initial state \eqref{eq:initial_state_Ry_neel}. The vertical and horizontal axes denote the second R\'enyi entanglement asymmetry $\Delta S_A^{(2)}(t)$ and the dimensionless time $tg$,
    %the real time, 
    respectively. 
    The upper panels show the mass dependence at fixed $\theta=\pi/3$ (left) and $\theta=\pi$ (right), while the lower panels show the $\theta$ dependence at fixed $m/g=4$ (left) and $m=m_c$ (right).
    %$m_c/g=0.3335$ (right).      
    The remaining parameters are fixed to $N=18$, $N_A=4$, $\phi=0.5\pi$, $ag=1/4$, and $\delta t=0.02/g$.
    }
    \label{fig:parameter_dependence}
\end{figure}

We next examine whether the symmetry-restoration dynamics considered in this work exhibits a pronounced dependence on these parameters. Fig.~\ref{fig:parameter_dependence} shows the time evolution of the second R\'enyi entanglement asymmetry for the spin-product N\'eel initial state \eqref{eq:initial_state_Ry_neel}. The upper panels show the mass dependence at fixed $\theta=\pi/3$ and $\theta=\pi$, whereas the lower panels show the $\theta$ dependence at fixed $m/g=4$ and $m/g=0.3335$. Within the system sizes, lattice spacing, and evolution times considered here, we do not observe a pronounced qualitative change in the symmetry-restoration dynamics as either $m$ or $\theta$ is varied. In particular, the qualitative decay of $\Delta S_A^{(2)}(t)$ persists for parameter choices corresponding to the vicinity of the continuum transition line and critical endpoint.

This observation should, however, be interpreted with caution. The phase structure summarized in Fig.~\ref{fig:phase_diagram} refers to the ground state of the continuum theory in the infinite-volume limit, whereas the present simulations describe nonequilibrium dynamics on a finite lattice at nonzero lattice spacing. In particular, the continuum value $m_c/g\simeq0.3335$ should be regarded only as a reference parameter at finite $N$ and $a$, rather than as the exact critical point of the lattice system simulated here. Therefore, the absence of a sharp dynamical signature near these parameter values does not imply that the continuum vacuum phase structure is irrelevant to symmetry restoration. A systematic investigation of the continuum and thermodynamic limits would be required to establish such a connection. For the purposes of the present work, Fig.~\ref{fig:parameter_dependence} indicates that the representative parameter set adopted in the main text is sufficient to illustrate the qualitative symmetry-restoration behavior observed in our finite-size simulations.

\begin{figure}[t]
    \centering
    \begin{subfigure}[b]{0.48\textwidth}
    \centering
    \includegraphics[width=\linewidth]{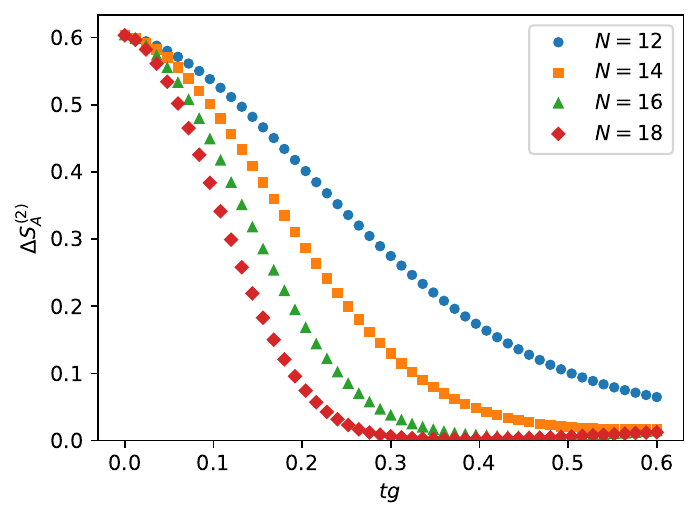}
    \caption*{$\phi=0.5\pi$}
    \end{subfigure}
    \hfill
    \begin{subfigure}[b]{0.48\textwidth}
    \centering
    \includegraphics[width=\linewidth]{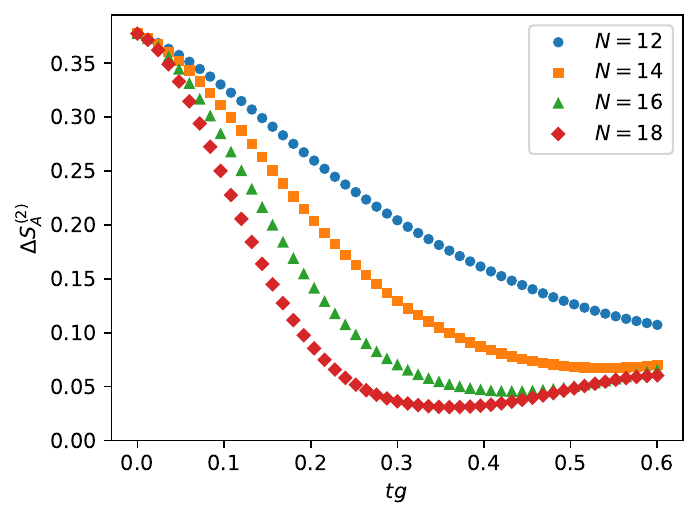}
    \caption*{$\phi=0.4\pi$}
    \end{subfigure}
    \caption{
    Time evolution of the second R\'enyi entanglement asymmetry $\Delta S_A^{(2)}(t)$ for total system sizes $N=12,14,16,$ and $18$. We use the fermion-parity-preserving N\'eel initial states in \eqref{eq:initial_state_Ryy}. The left and right panels correspond to $\phi=0.5\pi$ and $\phi=0.4\pi$, respectively. The remaining parameters are fixed to $N_A=4$, $\theta=\pi/3$, $m/g=4$, $ag=1/4$, and $\delta t=0.02/g$.
    }
\label{fig:N_dependence}
\end{figure}

We finally examine the dependence on the total system size. Fig.~\ref{fig:N_dependence} shows the time evolution of the second R\'enyi entanglement asymmetry for $N=12,14,16,$ and $18$, with the subsystem size fixed to $N_A=4$. Here, we use the fermion-parity-preserving N\'eel initial states defined in \eqref{eq:initial_state_Ryy}. For $\phi=0.5\pi$, increasing $N$ leads to a faster decrease of the entanglement asymmetry over the simulated time window, and the late-time values approach zero more closely for the larger systems. For $\phi=0.4\pi$, a nonzero residual value and a weak late-time increase remain visible at the accessible system sizes. Nevertheless, the minimum value of $\Delta S_A^{(2)}(t)$ decreases as $N$ is increased from $12$ to $18$.

The systematic decrease of the residual entanglement asymmetry with increasing $N$ suggests that the nonzero late-time value observed for the fermion-parity-preserving states is at least partly a finite-volume effect. However, the accessible range $N\le18$ is too limited to establish the thermodynamic-limit behavior or to conclude that the residual value vanishes as $N\to\infty$. The weak late-time increase may likewise reflect finite-size recurrences in the closed-system dynamics. We therefore regard the results in Fig.~\ref{fig:N_dependence} as evidence for appreciable finite-volume dependence rather than as a finite-size scaling analysis.

%%%%%%%%%%%%%%%%%%%%%%%%%%%%%%%%%%%%%%
%%%%%%%%%%%%%%%%%%%%%%%%%%%%%%%%%%%%%%
\section{Energy Expectation Values of the Fermion-Parity-Preserving Initial States}
\label{sec:energy}
%%%%%%%%%%%%%%%%%%%%%%%%%%%%%%%%%%%%%%
%%%%%%%%%%%%%%%%%%%%%%%%%%%%%%%%%%%%%%

For completeness, we examine the energy expectation values of the fermion-parity-preserving initial states introduced in \eqref{eq:initial_state_Ryy}. Since constant terms are omitted from the spin Hamiltonian \eqref{eq:Hamiltonian_spin}, the Hamiltonian is traceless, and the energy expectation value of the infinite-temperature state is
\begin{align}
    \langle H\rangle_{\beta=0}
    =
    \frac{\Tr H}{\dim\mathcal H}
    =
    0.
\end{align}
If the late-time state can be described by a thermal ensemble, an initial energy below this infinite-temperature value is associated with a positive effective inverse temperature, whereas an energy above it is associated with a negative effective inverse temperature.

\begin{figure}[t]
    \centering
    \begin{subfigure}[b]{0.48\textwidth}
        \centering
        \includegraphics[width=\linewidth]{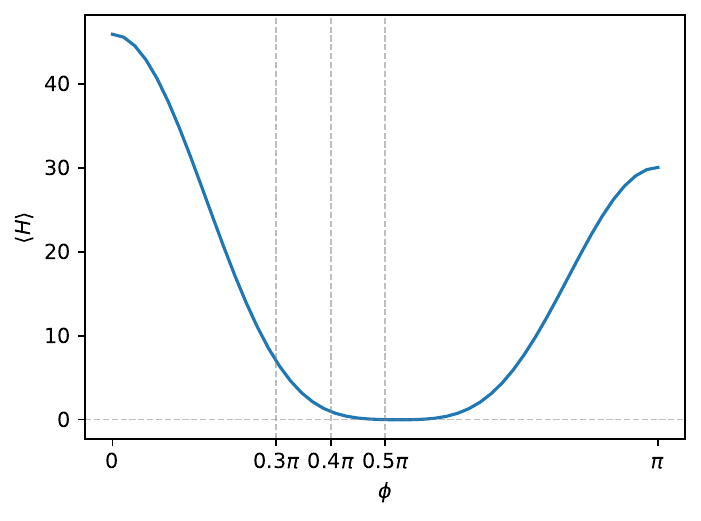}
        \caption*{Ferromagnetic}
    \end{subfigure}
    \hfill
    \begin{subfigure}[b]{0.48\textwidth}
        \centering
        \includegraphics[width=\linewidth]{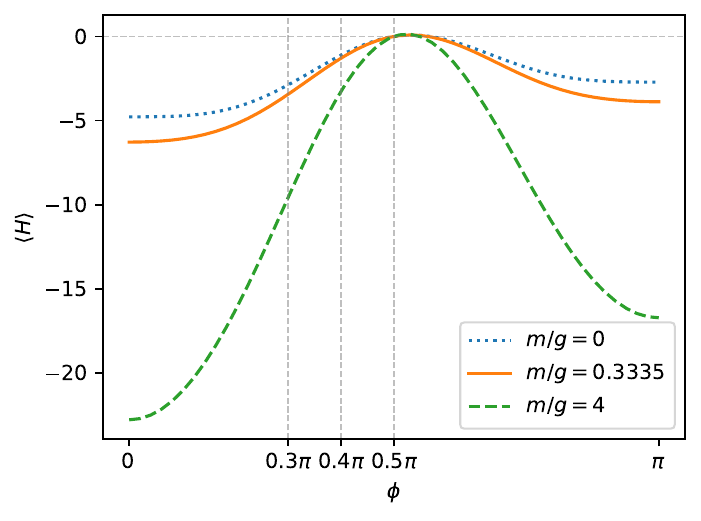}
        \caption*{N\'eel}
    \end{subfigure}
    \caption{
    Energy expectation values of the fermion-parity-preserving initial states in \eqref{eq:initial_state_Ryy} as functions of $\phi$. The left and right panels correspond to the ferromagnetic and N\'eel families, respectively. For the ferromagnetic family, the energy expectation value is independent of the fermion mass $m$, so the curves for different masses coincide. The remaining parameters are fixed to $N=18$, $\theta=\pi/3$, and $ag=1/4$.
    }
    \label{fig:Energy_Ryy}
\end{figure}

Fig.~\ref{fig:Energy_Ryy} shows the energy expectation values of the fermion-parity-preserving ferromagnetic and N\'eel initial states as functions of $\phi$. For the representative parameters used in the main text, the N\'eel family lies on the positive-temperature side of the infinite-temperature reference point, whereas the ferromagnetic family lies on the negative-temperature side. In both families, increasing $\phi$ from $0.3\pi$ to $0.5\pi$ brings the energy closer to the infinite-temperature value.
Under the thermal interpretation above, this corresponds to a decrease in $|\beta_{\rm eff}|$,
%with $\phi=0.5\pi$ associated with the highest effective temperature and $\phi=0.3\pi$ with the lowest 
with $\phi=0.5\pi$ and $\phi=0.3\pi$ corresponding to the smallest and largest $|\beta_{\rm eff}|$, respectively, among the three representative values.
%This ordering parallels the tendency observed in the main text that states associated with smaller $|\beta_{\rm eff}|$ exhibit faster symmetry restoration. 
The possible connection between the effective temperature and the symmetry-restoration dynamics, including its interplay with the initial charge-sector distribution, is discussed in the main text. 
We emphasize, however, that this observation provides only a phenomenological thermal interpretation of the conserved energy and does not establish thermalization or a causal relation between the effective temperature and the symmetry-restoration timescale.

%%%%%%%%%%%%%%%%%%%%%%%%%%%%%%%%%%%%%%
%%%%%%%%%%%%%%%%%%%%%%%%%%%%%%%%%%%%%%
\section{An Analytical Bound for the Second-Order Product Formula}
\label{sec:analytical_trotter_bound}
%%%%%%%%%%%%%%%%%%%%%%%%%%%%%%%%%%%%%%
%%%%%%%%%%%%%%%%%%%%%%%%%%%%%%%%%%%%%%

In this appendix, we derive the analytical bound used in \eqref{eq:worst-case_bound}. We start from the general commutator bound for the second-order product formula and further bound the nested commutators by exploiting the locality of the individual interaction terms.

Consider a Hamiltonian decomposed into $K$ blocks,
\begin{align}
    H=\sum_{a=1}^{K}H_a.
\end{align}
The second-order product formula is
\begin{align}
    U(\delta t)
    =
    \prod_{a=1}^{K}
    e^{-\i\frac{\delta t}{2}H_a}
    \prod_{b=K}^{1}
    e^{-\i\frac{\delta t}{2}H_b}.
\end{align}
As discussed in Sec.~\ref{subsec:error_bound_scaling}, its one-step error satisfies
\begin{align}\label{eq:commutator_bound_appendix}
\begin{aligned}
    \left\|
    e^{-\i H\delta t}
    -
    U(\delta t)
    \right\|
    &\leq
    \frac{(\delta t)^3}{24}
    \sum_{a=1}^{K-1}
    \left\|
    \left[
    H_a,
    \left[
    H_a,
    \sum_{b=a+1}^{K}H_b
    \right]
    \right]
    \right\|\\
    &\quad+
    \frac{(\delta t)^3}{12}
    \sum_{a=1}^{K-1}
    \left\|
    \left[
    \sum_{c=a+1}^{K}H_c,
    \left[
    \sum_{b=a+1}^{K}H_b,
    H_a
    \right]
    \right]
    \right\|.
\end{aligned}
\end{align}
The bound \eqref{eq:commutator_bound_appendix} itself is valid for a general decomposition of the Hamiltonian. To obtain a simpler locality-based estimate, however, we impose an additional assumption on the individual interaction terms.

We assume that the Hamiltonian can be decomposed into individual interactions $h_i$ satisfying
\begin{align}
    \left|
    \mathrm{supp}(h_i)
    \right|
    \leq k,
\end{align}
where $\mathrm{supp}(\mathcal{O})$ denotes the minimal set of sites on which the operator $\mathcal{O}$ acts nontrivially, and $k$ is independent of the system size.
% We assume that the Hamiltonian can be decomposed into individual interactions $h_i$ satisfying
% \begin{align}
% \left|
% \mathrm{supp}(h_i)
% \right|
% \leq s,
% \end{align}
% where $s$ is independent of the system size. 
Thus, each $h_i$ acts nontrivially on at most $k$ sites. No assumption of geometric locality is required. In particular, $k=2$ includes the one- and two-body interactions relevant to the lattice Schwinger model considered in this work.

In our implementation, each block $H_a$ is itself composed of mutually commuting individual interactions. We may therefore refine the product-formula decomposition into individual interactions without changing the product formula. To distinguish this refined decomposition from the original block decomposition, we write
\begin{align}
    H
    =
    \sum_{\mu=1}^{L}\widetilde H_\mu,
\end{align}
where
\begin{align}
    \widetilde H_\mu
    =
    h_{i_\mu},
    \qquad
    \mu=1,\ldots,L-1,
\end{align}
are individual $k$-body interactions, while the final component $\widetilde H_L$ may contain a sum of such interactions. The latter can be expanded into individual terms when evaluating the commutators below. We define
\begin{align}\label{eq:gamma_appendix}
    \gamma
    :=
    \sum_{\mu=1}^{L-1}
    \left\|
    h_{i_\mu}
    \right\|,
\end{align}
and
\begin{align}\label{eq:lambda_max_appendix}
    \lambda_{\max}
    :=
    \max_x
    \sum_{i:\,x\in\mathrm{supp}(h_i)}
    \left\|
    h_i
    \right\|.
\end{align}
Here, $\lambda_{\max}$ characterizes the maximum total interaction strength involving any single site.

We first establish a useful locality bound. Let $\mathcal{O}$ be an operator whose support contains at most $r$ sites. A nonvanishing commutator $[h_i,\mathcal{O}]$ requires the supports of $h_i$ and $\mathcal{O}$ to overlap. Therefore,
\begin{align}
    \sum_{i:\,[h_i,\mathcal{O}]\neq0}
    \left\|
    h_i
    \right\|
    &\leq
    \sum_{x\in\mathrm{supp}(\mathcal{O})}
    \sum_{i:\,x\in\mathrm{supp}(h_i)}
    \left\|
    h_i
    \right\|
    \nonumber\\
    &\leq
    r\lambda_{\max}.
    \label{eq:locality_lemma}
\end{align}

We now apply this bound to the two nested-commutator contributions in \eqref{eq:commutator_bound_appendix}. For a fixed $\mu$, we define
\begin{align}
    B_\mu
    :=
    \sum_{\nu=\mu+1}^{L}\widetilde H_\nu.
\end{align}
Expanding $B_\mu$ into individual interaction terms and using the triangle inequality together with
\begin{align}
    \|[A,B]\|
    \leq
    2\|A\|\|B\|,
\end{align}
we obtain
\begin{align}
    \left\|
    \left[
    h_{i_\mu},
    \left[
    h_{i_\mu},
    B_\mu
    \right]
    \right]
    \right\|
    \leq
    4
    \left\|
    h_{i_\mu}
    \right\|^2
    \sum_{j:\,[h_j,h_{i_\mu}]\neq0}
    \left\|
    h_j
    \right\|.
\end{align}
Since $h_{i_\mu}$ acts on at most $k$ sites, \eqref{eq:locality_lemma} gives
\begin{align}
    \sum_{j:\,[h_j,h_{i_\mu}]\neq0}
    \left\|
    h_j
    \right\|
    \leq
    k\lambda_{\max}.
\end{align}
Moreover, $\|h_{i_\mu}\|\leq\lambda_{\max}$, and hence
\begin{align}
    \left\|
    \left[
    h_{i_\mu},
    \left[
    h_{i_\mu},
    B_\mu
    \right]
    \right]
    \right\|
    \leq
    4k
    \left\|
    h_{i_\mu}
    \right\|
    \lambda_{\max}^2.
    \label{eq:first_nested_bound}
\end{align}

We next consider the second nested commutator. Expanding both occurrences of $B_\mu$ into individual interaction terms yields
\begin{align}
    \left\|
    \left[
    B_\mu,
    \left[
    B_\mu,
    h_{i_\mu}
    \right]
    \right]
    \right\|
    \leq
    4
    \left\|
    h_{i_\mu}
    \right\|
    \sum_{j:\,[h_j,h_{i_\mu}]\neq0}
    \left\|
    h_j
    \right\|
    \sum_{k:\,[h_k,[h_j,h_{i_\mu}]]\neq0}
    \left\|
    h_k
    \right\|.
\end{align}
For a nonvanishing commutator $[h_j,h_{i_\mu}]$, the supports of $h_j$ and $h_{i_\mu}$ must overlap on at least one site. Since both interactions act on at most $k$ sites,
\begin{align}
    \left|
    \mathrm{supp}(h_j)
    \cup
    \mathrm{supp}(h_{i_\mu})
    \right|
    \leq
    2k-1.
\end{align}
Furthermore, taking a commutator cannot enlarge the support beyond the union of the supports of the original operators,
\begin{align}
    \mathrm{supp}
    \left(
    [h_j,h_{i_\mu}]
    \right)
    \subseteq
    \mathrm{supp}(h_j)
    \cup
    \mathrm{supp}(h_{i_\mu}),
\end{align}
and therefore
\begin{align}
    \left|
    \mathrm{supp}
    \left(
    [h_j,h_{i_\mu}]
    \right)
    \right|
    \leq
    2k-1.
\end{align}
Applying \eqref{eq:locality_lemma} once more gives
\begin{align}
    \sum_{k:\,[h_k,[h_j,h_{i_\mu}]]\neq0}
    \left\|
    h_k
    \right\|
    \leq
    (2k-1)\lambda_{\max}.
\end{align}
Together with
\begin{align}
    \sum_{j:\,[h_j,h_{i_\mu}]\neq0}
    \left\|
    h_j
    \right\|
    \leq
    k\lambda_{\max},
\end{align}
we obtain
\begin{align}
    \left\|
    \left[
    B_\mu,
    \left[
    B_\mu,
    h_{i_\mu}
    \right]
    \right]
    \right\|
    \leq
    4k(2k-1)
    \left\|
    h_{i_\mu}
    \right\|
    \lambda_{\max}^2.
    \label{eq:second_nested_bound}
\end{align}

Substituting \eqref{eq:first_nested_bound} and \eqref{eq:second_nested_bound} into the commutator bound, we find
\begin{align}
    \left\|
    e^{-\i H\delta t}
    -
    U(\delta t)
    \right\|
    &\leq
    \left[
    \frac{4k}{24}
    +
    \frac{4k(2k-1)}{12}
    \right]
    \gamma
    \lambda_{\max}^2
    (\delta t)^3
    \nonumber\\
    &=
    \frac{k(4k-1)}{6}
    \gamma
    \lambda_{\max}^2
    (\delta t)^3.
    \label{eq:s_local_second_order_bound}
\end{align}
Thus, for a Hamiltonian composed of interactions acting on at most $k$ sites, the second-order product formula admits the locality-based bound \eqref{eq:s_local_second_order_bound}. For the one- and two-body interactions considered in the main text, $k=2$, and this reduces to
\begin{align}\label{eq:worst-case_bound_p=2_appendix}
    \left\|
    e^{-\i H\delta t}
    -
    U(\delta t)
    \right\|
    \leq
    \frac{7}{3}
    \gamma
    \lambda_{\max}^2
    (\delta t)^3,
\end{align}
which is the analytical worst-case bound used in \eqref{eq:worst-case_bound}.

For the lattice Schwinger Hamiltonian considered in this work, the explicit expressions for $\gamma$ and $\lambda_{\text{max}}$ associated with the refined interaction decomposition of the block ordering $(H_1,H_2,H_3)=(H_{\rm even},H_{\rm odd},H_Z+H_{ZZ})$ are
\begin{align}\label{eq:gamma_schwinger}
    \gamma
    &=
    \left| \left(
    w - \frac{m_{\text{lat}}}{2}\sin \theta
    \right)\right|
    \cdot
    \left\lfloor \frac{N-1}{2} \right\rfloor
    +
    \left| \left(
    w + \frac{m_{\text{lat}}}{2}\sin \theta
    \right)\right|
    \cdot
    \left\lfloor \frac{N}{2} \right\rfloor
\end{align}
\begin{align}\label{eq:lambda_max_schwinger}
\begin{aligned}
    \lambda_{\text{max}}
    =
    \max_{n\in \{1,\cdots ,N\}} 
    &\left\{ \quad
    \frac{J}{4} (N-n)(N+n-3) \right. \\
    &\quad+
    \left| w - (-1)^{n-1} \frac{m_{\text{lat}}}{2}\sin \theta \right|
    \left( 1-\delta_{n,1} \right)
    +
    \left| w - (-1)^n \frac{m_{\text{lat}}}{2}\sin \theta \right|
    \left( 1-\delta_{n,N} \right) \\
    &\left. \quad+
    \frac{1}{2}
    \left| 
    m_{\text{lat}}\cos \theta 
    -
    J(-1)^n \left(
    \left\lceil \frac{N-1}{2} \right\rceil - \left\lfloor \frac{n}{2} \right\rfloor
    \right)\ 
    \right|\quad
    \right\} \ ,
\end{aligned}
%\ ,
\end{align}
Substituting these expressions into \eqref{eq:worst-case_bound_p=2_appendix} gives the worst-case estimates shown in Figs.~\ref{fig:CNOT_count} and \ref{fig:T_count}. For fixed parameters of the Hamiltonian, \eqref{eq:gamma_schwinger} and \eqref{eq:lambda_max_schwinger} imply
\begin{align}
    \gamma=\mathcal{O}(N),
    \qquad
    \lambda_{\text{max}}=\mathcal{O}(N^2).
\end{align}
The former scaling follows from the $\mathcal{O}(N)$ hopping interactions entering $\gamma$. For $\lambda_{\text{max}}$, a given site participates in $\mathcal{O}(N)$ long-range $ZZ$ interactions whose coefficients can themselves scale as $\mathcal{O}(N)$, leading to an accumulated interaction strength of $\mathcal{O}(N^2)$. Consequently, the $k=2$ bound gives the one-step scaling $\mathcal{O}(N^5\delta t^3)$ used in the main text.

\bibliographystyle{utphys}
\bibliography{EA_QC}

\end{document}